\documentclass[usenatbib]{mnras}
\usepackage{graphicx}
\usepackage{rotating}
\usepackage{multirow}
\usepackage{amssymb}
\usepackage{amsmath}
\usepackage{hyperref}
\usepackage{breakurl}
\usepackage{color}
\usepackage{multicol}
\def\apj{ApJ}
\def\apjl{ApJ}
\def\mnras{MNRAS}

\def\aap{A\&A}
\def\aj{AJ}
\def\apjs{APJS}

\def\araa{ARAA}
\def\ssr{SSR}

\newcommand*{\Mwl}{$M_{\rm WL}$}
\newcommand*{\Lxrass}{$L_{X,\rm RASS}$}
\newcommand*{\Lx}{$L_{X,\rm ce}$}
\newcommand*{\Tx}{$T_{X,\rm ce}$}
\newcommand*{\Mgas}{$M_{\rm gas}$}
\newcommand*{\Yx}{$Y_{X}$}
\newcommand*{\Ysza}{$Y_{\rm SZA}$}
\newcommand*{\Ypl}{$Y_{\rm Pl}$}
\newcommand*{\Lk}{$L_{K,\rm tot}$}
\newcommand*{\Lbcg}{$L_{K,\rm BCG}$}
\newcommand*{\richness}{$\lambda$}

\hypersetup{draft}

\title[LoCuSS: galaxy cluster scaling relations]
{LoCuSS: Scaling relations between galaxy cluster mass, gas, and stellar content}

\author[S. L. Mulroy et al.]
{Sarah L. Mulroy,$^{1}$\thanks{E-mail: smulroy@star.sr.bham.ac.uk}
Arya Farahi,$^{2,3}$
August E. Evrard,$^{2,4}$
Graham P. Smith,$^{1}$ \and
Alexis Finoguenov,$^{5,6}$
Christine O'Donnell,$^{7}$
Daniel P. Marrone,$^{7}$
Zubair Abdulla,$^{8,9}$ \and
Herv\'e Bourdin,$^{10,11}$
John E. Carlstrom,$^{8,9,12,13}$
Jessica D\'emocl\`es,$^{14}$
Chris P. Haines,$^{15}$ \and
Rossella Martino,$^{11}$
Pasquale Mazzotta,$^{10,11}$
Sean L. McGee$^{1}$
and
Nobuhiro Okabe$^{16,17,18}$
\vspace{2mm}\\
	$^{1}$  School of Physics and Astronomy, University of Birmingham, Edgbaston, Birmingham, B15 2TT, UK \\
    $^{2}$  Department of Physics and Michigan Center for Theoretical Physics, University of Michigan, Ann Arbor, MI 48109, USA \\
    $^{3}$  McWilliams Center for Cosmology, Department of Physics, Carnegie Mellon, University, Pittsburgh, PA 15213, USA \\
    $^{4}$  Department of Astronomy, University of Michigan, Ann Arbor, MI 48109, USA \\
    $^{5}$  Max-Planck-Institute for Extraterrestrial Physics, Giessenbachstrasse, D-85741 Garching, Germany \\
    $^{6}$  Department of Physics, University of Helsinki, FI-00014 Helsinki, Finland \\
    $^{7}$  Steward Observatory, University of Arizona, 933 North Cherry Avenue, Tucson, AZ 85721, USA \\
    $^{8}$  Kavli Institute for Cosmological Physics, University of Chicago, Chicago, IL 60637, USA \\
    $^{9}$  Department of Astronomy and Astrophysics, University of Chicago, Chicago, IL 60637, USA \\
    $^{10}$ Harvard Smithsonian Centre for Astrophysics, 60 Garden Street, Cambridge, MA 02138, USA \\
    $^{11}$ Dipartimento di Fisica, Universit\`a degli Studi di Roma `Tor Vergata', via della Ricerca Scientifica 1, I-00133 Roma, Italy \\
    $^{12}$ Department of Physics, University of Chicago, Chicago, IL 60637, USA \\
    $^{13}$ Enrico Fermi Institute, University of Chicago, Chicago, IL 60637, USA \\
    $^{14}$ Service d’Astrophysique AIM, CEA-Saclay, F-91191 Gif sur Yvette, France \\
    $^{15}$ INAF - Osservatorio Astronomico di Brera, Via Brera 28, I-20122 Milano, Italy \\
    $^{16}$ Department of Physical Science, Hiroshima University, 1-3-1, Kagamiyama, Higashi-Hiroshima, Hiroshima 739-8526, Japan \\
    $^{17}$ Hiroshima Astrophysical Science Center, Hiroshima University, 1-3-1, Kagamiyama, Higashi-Hiroshima, Hiroshima 739-8526, Japan \\
    $^{18}$ Core Research for Energetic Universe, Hiroshima University, 1-3-1, Kagamiyama, Higashi-Hiroshima, Hiroshima 739-8526, Japan
}
\begin{document}

\date{Accepted. Received; in original form}

\pagerange{\pageref{firstpage}--\pageref{lastpage}} \pubyear{2018}

\maketitle

\label{firstpage}

\begin{abstract}
We present a simultaneous analysis of galaxy cluster scaling relations between weak-lensing mass and multiple cluster observables, across a wide range of wavelengths, that probe both gas and stellar content.
Our new hierarchical Bayesian model simultaneously considers the selection variable alongside all other observables in order to explicitly model intrinsic property covariance and account for selection effects.
We apply this method to a sample of 41 clusters at $0.15<z<0.30$, with a well-defined selection criteria based on RASS X-ray luminosity, and observations from \textit{Chandra}/\textit{XMM}, SZA, \textit{Planck}, UKIRT, SDSS and Subaru.
These clusters have well-constrained weak-lensing mass measurements based on Subaru/Suprime-Cam observations, which serve as the reference masses in our model. We present 30 scaling relation parameters for 10 properties.
All relations probing the intracluster gas are slightly shallower than self-similar predictions, in moderate tension with prior measurements, and the stellar fraction decreases with mass.
K-band luminosity has the lowest intrinsic scatter with a 95th percentile of 0.16, while the lowest scatter gas probe is gas mass with a fractional intrinsic scatter of $0.16 \pm 0.03$.
We find no distinction between the core-excised X-ray or high-resolution Sunyaev-Zel'dovich relations of clusters of different central entropy, but find with modest significance that higher entropy clusters have higher stellar fractions than their lower entropy counterparts. We also report posterior mass estimates from our likelihood model.
\end{abstract}


\section{Introduction}

Galaxy clusters form at rare peaks in the Universe's density distribution, and as such are rich laboratories for both cosmology and astrophysics \citep[e.g.][]{Allen2011,Kravtsov2012}. For cosmological purposes, counts and clustering of galaxy clusters are direct results of the late time growth of structure, and their measurement provides tests of cosmological parameters complementary to those of the cosmic microwave background (CMB) or supernovae \citep[e.g.][]{Weinberg2013}. The spatial abundance of galaxy clusters is a strong function of system mass, so such tests require accurate calibration of the absolute mass scale of halos as well as the statistical relationship between mass and observable properties. This requirement has motivated a significant effort to find and calibrate observable quantities which correlate with halo mass, so-called cluster scaling relations \citep[e.g.][]{Giodini2013}.

On the astrophysics side, galaxy clusters are a unique environment within which the majority of the baryon content is observable, either in stellar material or in hot intracluster gas \citep[e.g.][]{Gonzalez2013,Chiu2016}. The properties of the stellar and gas content of clusters is the result of a wide range of physical effects, including cooling, star formation, feedback, and accretion-driven processes such as shocks, tidal stripping and turbulence.  Thus the observable properties of gas and stellar material and their scaling with respect to the total cluster mass can give direct insight into the physics of these processes.

Ideally we would constrain the scaling relation of an observable with the `true' mass of the cluster, however in practise this is not measurable. A popular method of mass measurement uses X-ray properties together with the simplifying assumption of hydrostatic equilibrium \citep[e.g.][]{Mathews1978, Sarazin1988, Vikhlinin2006, Martino2014}. More recently, significant progress has been made in using the weak-lensing signal to probe the mass of galaxy clusters. When carefully accounting for systematic effects, these masses are thought to be, on average, close to unbiased with respect to the true mass \citep[e.g.][]{Becker&Kravtsov2011,Oguri2011,Bahe2012}, although \citet{Henson2017} report a 10 per cent mean bias that declines at very high masses. Crucially, these measurements do not rely on the assumption of hydrostatic equilibrium.

An often overlooked requirement for calibrating robust scaling relations is a clear understanding of the cluster sample selection and inclusion of the selection in the subsequent statistical analysis. As each observable has a non-zero scatter in its relation with mass, selection based on anything but true mass can bias the derived relations relative to those of the underlying halo population.  The latter are often characterized by cosmological simulations \citep[e.g.][]{LeBrun2017}. Cluster samples are commonly selected from optical, X-ray or Sunyaev-Zel'dovich (SZ) surveys \citep[e.g.][]{REFLEX,Rozo2009,Bleem2015}, and constraints on population model parameters are ultimately limited by both understanding of the selection function and sample size.

The 41 clusters in this work are particularly well studied over a wide range of wavelengths \citep[e.g.][]{Marrone2012,Martino2014,Mulroy2014,Haines2015,Okabe2016}. Combined with a well-defined selection function, they provide the first cluster sample with which to simultaneously constrain scaling relations for X-ray, SZ and optical observables. We report here the mean behaviours --- slopes, intercepts, and intrinsic scatter --- as well as correlations with the \Lxrass \ selection variable for 10 properties. The full covariance matrix is presented in a companion paper \citep{Farahi:inprep}.

In Section \ref{sec:data} we describe our cluster sample, its selection, and the wide range of multiwavelength data that we use in this paper. In Section \ref{sec:selfsimilar} we derive the expected scaling relations for a self-similar model, and in Section \ref{sec:regression} we describe our hierarchical Bayesian method to fit the scaling relations. We present our results in Section \ref{sec:results}, discuss these results and compare to the literature in Section \ref{sec:disc}, and conclude in Section \ref{sec:summary}. We assume $\Omega_{\rm M}=0.3$, $\Omega_\Lambda=0.7$ and $H_0=70\,{\rm km\,s^{-1}\,Mpc^{-1}}$. In this cosmology, at the average cluster redshift of $\langle z \rangle=0.22$, 1 arcsec corresponds to a projected physical scale of 3.55 kpc.  We employ a spherical mass and radius convention, $M_{500}$ and $r_{500}$, based on a mean enclosed density of 500 times the critical density evaluated in the above cosmology.

\section{Data}\label{sec:data}

\subsection{Sample}\label{sec:sample}
\begin{figure}
    \centering
    \includegraphics[width=\linewidth]{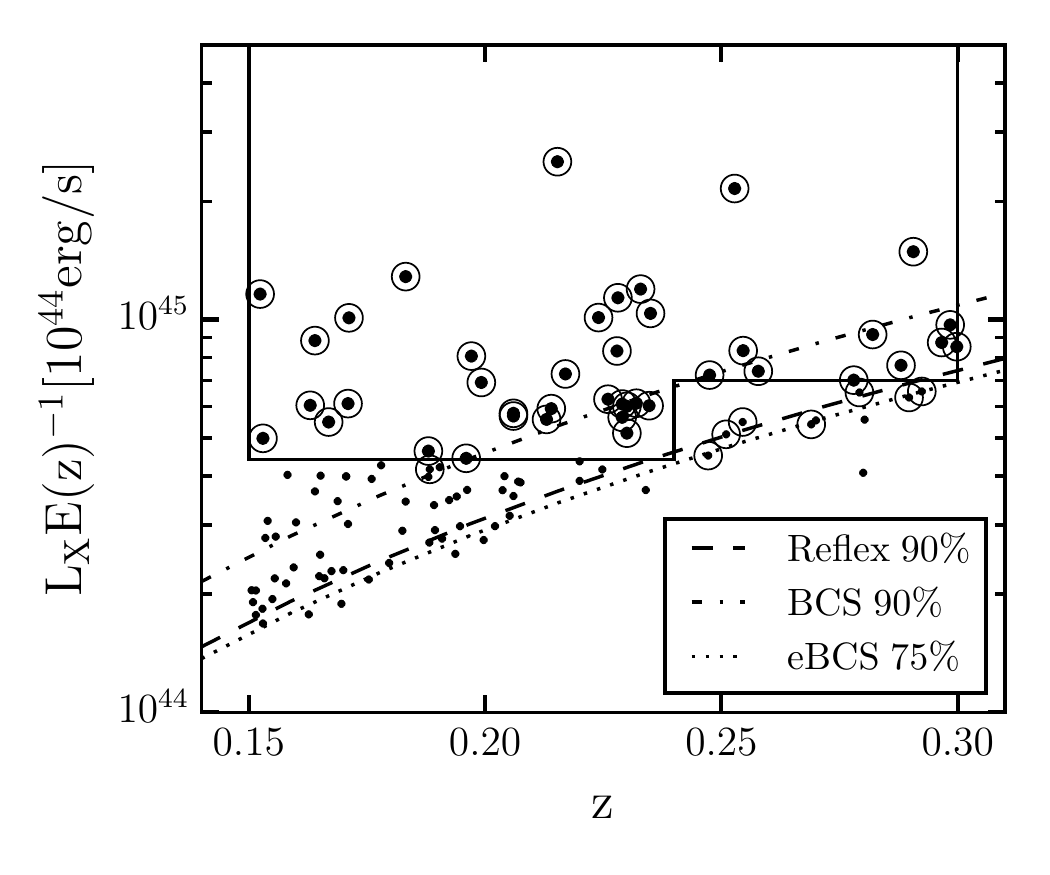}
  \caption{The $L_{X,\rm RASS}E(z)^{-1} -$ redshift distribution of the LoCuSS clusters. The large points show the 41 clusters passing the selection criteria and therefore used in this work, while the circles show the LoCuSS ``High-$L_X$'' clusters. The straight lines show the selection criteria, and the curves show the completeness limits for (e)BCS \citep{BCS,EBCS} and REFLEX \citep{REFLEX}.}
  \label{fig:sample}
\end{figure}

We study a sample of 41 X-ray luminous clusters from the ``High-$L_X$'' sample of the Local Cluster Substructure Survey (LoCuSS\footnote{\url{http://www.sr.bham.ac.uk/locuss}}), which was selected from the ROSAT All Sky Survey catalogues (RASS, \citealt{BCS,EBCS,REFLEX}). These are all the clusters satisfying clearly defined selection criteria: $n_{H} < 7 \times 10^{20}\rm cm^{-2}$; $-25^{\circ}<\delta<+65^{\circ}$; and an X-ray luminosity threshold of $L_{X,RASS}E(z)^{-1}>4.4\times 10^{44}\rm erg/s$ for clusters between $0.15<z \le 0.24$, and $L_{X,RASS}E(z)^{-1}>7.0 \times 10^{44}\rm erg/s$ for clusters between $0.24<z<0.30$ (Table \ref{tab:sample} \& Fig. \ref{fig:sample}), where $E(z)\equiv H(z)/H_{0}=\sqrt{\Omega_{\rm M}(1+z)^{3} + \Omega_\Lambda}$ is the evolution of the Hubble parameter. Therefore the only physical selection variable for this sample of galaxy clusters is the RASS X-ray luminosity, \Lxrass.

The \Lxrass \ measurements cover the soft-band X-ray [0.1$-$2.4]keV, and are taken from the ROSAT Brightest Cluster Sample and its low flux extension for objects in the northern hemisphere (BCS, \citealt{BCS}; eBCS, \citealt{EBCS}), and the ROSAT-ESO Flux Limited X-ray galaxy cluster survey for objects mostly in the southern hemisphere \citep[$\delta < 2.5^{\circ}$, REFLEX,][]{REFLEX}. For the clusters in the overlap between surveys (Abell0267: BCS, REFLEX and Abell2631: eBCS, REFLEX) we average the luminosities and errors. RASS luminosities are not core-excised due to the angular resolution of the instrument, and so are sensitive to the presence, or absence, of a cool core. We explore the effects of core treatment in Section \ref{sec:scatter}.

We observed this sample of clusters at X-ray, optical, near-infrared, and millimetre wavelengths over the period 2005$-$2014, building up a unique and comprehensive dataset.  The main facilities that we used are \emph{Chandra}, \emph{XMM-Newton}, Suprime-Cam on the Subaru telescope, Hectospec on the Multiple Mirror Telescope (MMT), WFCAM on the United Kingdom Infrared Telescope (UKIRT), and the Sunyaev-Zel'dovich Array (SZA).  The total investment of telescope time amounts to several million seconds. The following wavelength-specific sections describe the measurements of galaxy cluster weak-lensing masses and observable properties used in this article, with citations providing more complete details of their respective observations. The measurements are listed in Tables \ref{tab:sample} and \ref{tab:obs}, and summarized in Table \ref{tab:MOR}.

\begin{table*}
\caption{Cluster sample.}\label{tab:sample}
	\begin{center}
			\begin{tabular}{ l c c c c c c }
		      \hline
			Name & $\rm{RA}$ & $\rm{Dec}$ & Redshift & $L_{X,\rm RASS}$ & $M_{\rm WL}$ & $M_{\rm post}$ \\
			& $\alpha$ [J2000] & $\delta$ [J2000] & $z$ & $10^{44}\rm erg/s$ & $10^{14}M_{\odot}$ & $10^{14}M_{\odot}$ \\
		      \hline
Abell2697 & 0.7990 & -6.0860 & 0.2320 & $6.88^{+0.85}_{-0.85}$ & $6.61^{+1.20}_{-1.21}$ & $5.98^{+0.57}_{-0.53}$ \\
Abell0068 & 9.2785 & 9.1566 & 0.2546 & $9.47^{+2.61}_{-2.61}$ & $6.82^{+1.11}_{-1.01}$ & $6.38^{+0.52}_{-0.50}$ \\
Abell0115 & 14.0012 & 26.3424 & 0.1971 & $8.90^{+2.13}_{-2.13}$ & $5.39^{+1.62}_{-1.49}$ & $6.13^{+0.90}_{-0.82}$ \\
Abell0141 & 16.3864 & -24.6466 & 0.2300 & $5.76^{+0.90}_{-0.90}$ & $4.56^{+0.92}_{-0.86}$ & $5.01^{+0.68}_{-0.59}$ \\
Abell0209 & 22.9689 & -13.6112 & 0.2060 & $6.29^{+0.65}_{-0.65}$ & $12.34^{+1.64}_{-1.50}$ & $10.67^{+0.96}_{-0.86}$ \\
Abell0267 & 28.1748 & 1.0072 & 0.2300 & $6.74^{+1.42}_{-1.42}$ & $5.60^{+0.91}_{-0.85}$ & $5.48^{+0.55}_{-0.52}$ \\
Abell0291 & 30.4296 & -2.1966 & 0.1960 & $4.88^{+0.56}_{-0.56}$ & $4.46^{+1.02}_{-0.95}$ & $2.99^{+0.37}_{-0.33}$ \\
Abell0521 & 73.5287 & -10.2235 & 0.2475 & $8.18^{+1.36}_{-1.36}$ & $5.39^{+0.99}_{-0.93}$ & $5.62^{+0.63}_{-0.56}$ \\
Abell0586 & 113.0845 & 31.6335 & 0.1710 & $6.64^{+1.30}_{-1.30}$ & $7.21^{+1.60}_{-1.40}$ & $6.62^{+0.75}_{-0.68}$ \\
Abell0611 & 120.2367 & 36.0566 & 0.2880 & $8.86^{+2.53}_{-2.53}$ & $9.11^{+1.67}_{-1.56}$ & $6.42^{+0.70}_{-0.63}$ \\
Abell0697 & 130.7398 & 36.3666 & 0.2820 & $10.57^{+3.28}_{-3.28}$ & $7.71^{+1.54}_{-1.43}$ & $9.61^{+1.06}_{-1.02}$ \\
ZwCl0857.9+2107 & 135.1536 & 20.8946 & 0.2347 & $6.79^{+1.76}_{-1.76}$ & $2.07^{+0.99}_{-1.08}$ & $1.40^{+0.34}_{-0.29}$ \\
Abell0750 & 137.3024 & 10.9745 & 0.1630 & $6.59^{+1.40}_{-1.40}$ & $6.15^{+1.71}_{-1.35}$ & $6.19^{+1.10}_{-0.98}$ \\
Abell0773 & 139.4726 & 51.7271 & 0.2170 & $8.10^{+1.35}_{-1.35}$ & $10.07^{+1.07}_{-1.00}$ & $9.69^{+0.66}_{-0.61}$ \\
Abell0781 & 140.1075 & 30.4941 & 0.2984 & $11.29^{+2.82}_{-2.82}$ & $4.75^{+1.72}_{-1.89}$ & $7.07^{+1.45}_{-1.25}$ \\
ZwCl0949.6+5207 & 148.2048 & 51.8849 & 0.2140 & $6.60^{+1.15}_{-1.15}$ & $4.97^{+1.13}_{-1.04}$ & $3.06^{+0.40}_{-0.36}$ \\
Abell0907 & 149.5917 & -11.0640 & 0.1669 & $5.95^{+0.49}_{-0.49}$ & $11.52^{+1.95}_{-1.67}$ & $7.86^{+0.96}_{-0.84}$ \\
Abell0963 & 154.2652 & 39.0471 & 0.2050 & $6.39^{+1.19}_{-1.19}$ & $6.96^{+1.11}_{-1.03}$ & $5.77^{+0.63}_{-0.55}$ \\
ZwCl1021.0+0426 & 155.9152 & 4.1863 & 0.2906 & $17.26^{+2.93}_{-2.93}$ & $5.32^{+0.87}_{-0.82}$ & $5.57^{+0.64}_{-0.57}$ \\
Abell1423 & 179.3223 & 33.6110 & 0.2130 & $6.19^{+1.34}_{-1.34}$ & $4.44^{+0.89}_{-0.81}$ & $3.97^{+0.47}_{-0.42}$ \\
Abell1451 & 180.8199 & -21.5484 & 0.1992 & $7.63^{+1.63}_{-1.63}$ & $8.17^{+1.04}_{-0.96}$ & $7.87^{+0.75}_{-0.67}$ \\
ZwCl1231.4+1007 & 188.5728 & 9.7662 & 0.2290 & $6.32^{+1.58}_{-1.58}$ & $4.61^{+1.44}_{-1.47}$ & $5.02^{+0.77}_{-0.72}$ \\
Abell1682 & 196.7083 & 46.5593 & 0.2260 & $7.02^{+1.37}_{-1.37}$ & $8.52^{+1.06}_{-0.99}$ & $7.84^{+0.75}_{-0.68}$ \\
Abell1689 & 197.8730 & -1.3410 & 0.1832 & $14.07^{+1.13}_{-1.13}$ & $12.57^{+1.53}_{-1.40}$ & $12.00^{+0.97}_{-0.90}$ \\
Abell1763 & 203.8337 & 41.0012 & 0.2279 & $9.32^{+1.33}_{-1.33}$ & $15.80^{+2.16}_{-1.94}$ & $13.70^{+1.40}_{-1.23}$ \\
Abell1835 & 210.2588 & 2.8786 & 0.2528 & $24.48^{+3.35}_{-3.35}$ & $10.97^{+1.56}_{-1.44}$ & $11.03^{+0.93}_{-0.84}$ \\
Abell1914 & 216.4860 & 37.8165 & 0.1712 & $10.98^{+1.11}_{-1.11}$ & $7.83^{+1.35}_{-1.24}$ & $8.30^{+0.86}_{-0.81}$ \\
ZwCl1454.8+2233 & 224.3131 & 22.3428 & 0.2578 & $8.41^{+2.10}_{-2.10}$ & $3.74^{+1.46}_{-1.44}$ & $2.98^{+0.46}_{-0.42}$ \\
Abell2009 & 225.0813 & 21.3694 & 0.1530 & $5.37^{+0.99}_{-0.99}$ & $6.39^{+1.45}_{-1.25}$ & $4.73^{+0.54}_{-0.48}$ \\
RXCJ1504.1-0248 & 226.0313 & -2.8047 & 0.2153 & $28.07^{+1.49}_{-1.49}$ & $6.54^{+1.48}_{-1.32}$ & $6.19^{+0.95}_{-0.79}$ \\
Abell2111 & 234.9188 & 34.4243 & 0.2290 & $6.83^{+1.65}_{-1.65}$ & $5.09^{+1.39}_{-1.21}$ & $5.84^{+0.76}_{-0.67}$ \\
Abell2204 & 248.1956 & 5.5758 & 0.1524 & $12.50^{+1.34}_{-1.34}$ & $9.92^{+1.82}_{-1.59}$ & $10.11^{+1.01}_{-0.94}$ \\
Abell2219 & 250.0827 & 46.7114 & 0.2281 & $12.73^{+1.37}_{-1.37}$ & $8.65^{+1.34}_{-1.29}$ & $10.76^{+1.02}_{-0.93}$ \\
RXJ1720.1+2638 & 260.0420 & 26.6257 & 0.1640 & $9.57^{+1.07}_{-1.07}$ & $4.94^{+1.38}_{-1.17}$ & $4.55^{+0.65}_{-0.58}$ \\
Abell2261 & 260.6133 & 32.1326 & 0.2240 & $11.31^{+1.55}_{-1.55}$ & $10.75^{+1.30}_{-1.20}$ & $10.41^{+0.92}_{-0.83}$ \\
RXCJ2102.1-2431 & 315.5411 & -24.5335 & 0.1880 & $5.07^{+0.55}_{-0.55}$ & $3.71^{+0.87}_{-0.79}$ & $3.03^{+0.41}_{-0.37}$ \\
RXJ2129.6+0005 & 322.4165 & 0.0894 & 0.2350 & $11.66^{+2.92}_{-2.92}$ & $3.46^{+1.14}_{-1.22}$ & $4.02^{+0.57}_{-0.53}$ \\
Abell2390 & 328.4034 & 17.6955 & 0.2329 & $13.43^{+3.14}_{-3.14}$ & $10.53^{+1.52}_{-1.41}$ & $10.36^{+1.08}_{-0.96}$ \\
Abell2537 & 347.0926 & -2.1921 & 0.2966 & $10.17^{+1.45}_{-1.45}$ & $8.57^{+2.03}_{-1.82}$ & $7.77^{+0.99}_{-0.89}$ \\
Abell2552 & 347.8887 & 3.6349 & 0.2998 & $9.94^{+2.84}_{-2.84}$ & $7.16^{+1.88}_{-1.69}$ & $7.36^{+0.88}_{-0.78}$ \\
Abell2631 & 354.4155 & 0.2714 & 0.2779 & $8.07^{+2.11}_{-2.11}$ & $5.61^{+1.58}_{-1.78}$ & $5.66^{+0.72}_{-0.66}$ \\
		      \hline
		\end{tabular}
	\end{center}
{\footnotesize}
\end{table*}

\subsection{Gravitational Weak-Lensing Masses}\label{sec:mwl}

We use weak-lensing masses from \citet{Okabe2016} (as tabulated in their table 2), who calculate masses by fitting an NFW \citep*{NFW1997} mass profile to the shear profile obtained from Subaru/Suprime-Cam observations. We use $M_{500}$ values, defined as the mass within radius $r_{500}$, the radius within which the average density is $500 \times \rho_{\rm crit}(z)$, the critical density of the Universe. We adopt these weak-lensing determined radii, $r_{500,\rm WL}$, as the radii within which we measure the other aperture-integrated properties in this work (except \Yx \ and \richness). The systematic biases in the ensemble calibration of the weak-lensing mass calculations are controlled at $\sim$4 per cent level, based on careful selection of red background galaxies, extensive tests of both faint galaxy shape measurement methods and mass profile fitting methods \citep{Okabe2016}. The measurement errors on $M_{500}$ include contributions from shape noise, photometric redshift uncertainties and uncorrelated large-scale structure.  In our analysis below, we assume these weak-lensing masses to be unbiased in the mean with respect to true halo mass.  

\subsection{X-Ray Observables}\label{sec:xray}

We use X-ray measurements of the intracluster medium (ICM) described in \citet{Martino2014}, where most clusters were observed with the \emph{XMM-Newton} EPIC or \emph{Chandra} ACIS-I detectors, except for Abell0611 and ZwCl0949.6+5207 which were observed with the \emph{Chandra} ACIS-S detectors. We note that emission measure profiles were robust to X-ray telescope cross-calibration issues for the selected energy band, as shown in \citet{Martino2014}.

We consider bolometric [0.7$-$10]keV core-excised luminosity \Lx \ and the average gas temperature \Tx \ within an annulus of [0.15$-$1]$r_{500,\rm WL}$ to avoid the measurements being contaminated by potentially stochastic cool-core emission. The error bars in \Lx \ include marginalization over $T_X$. The gas mass \Mgas \ is measured within $r_{500,\rm WL}$. We also measure the integrated pressure proxy \Yx \ \citep{Kravtsov2006} for all but the two clusters with ACIS-S observations. Defined as the product of gas mass and average temperature, it is the X-ray analogue of the SZ parameter described in Section \ref{sec:sz}.

Both the luminosity and the \Yx \ parameter derive from spherically symmetric templates of the
X-ray emission measure per unit volume, [$\rm n_p n_e$](r), that were projected along the
line of sight, radially averaged and fitted to radial profiles of the soft [0.5$-$2]keV 
X-ray surface brightness. The bolometric estimate of \Lx \
derives from an extrapolation of the soft surface brightness
assuming the spectral energy distribution of the ICM to correspond to a
redshifted isothermal plasma with average temperature, $T$.

We estimate the \Yx \ parameter following the established methods based on its original definition \citep{Kravtsov2006} to ensure comparability with the literature. For each cluster we iterate about an existing \Yx$-M_{500}$ scaling relation, yielding a characteristic radius $r_{500}$, different from the weak-lensing $r_{500,\rm WL}$ radius within which the other X-ray observables are measured. For clusters observed with \emph{XMM-Newton} we use the relation of \citet{Arnaud2010}, and for those observed with \emph{Chandra} we use the relation of \citet{Vikhlinin2009}. Both relations are calibrated using hydrostatic mass estimates in a nearby cluster sample. The gas masses were computed from spherical integrals of the gas density profiles $\rm n_p$(r), and the gas temperatures correspond to spectroscopic measurements within projected [0.15$-$0.75]$r_{500}$ and [0.15$-$1]$r_{500}$, following the prescription of the relevant scaling relation study. We note that any bias in the assumed scaling relations would be a source of error for our \Yx \ measurements.

\subsection{Millimetre Observables -- Sunyaev-Zel'dovich Effect}\label{sec:sz}

The SZ effect is caused by the inverse Compton scattering of CMB photons by hot electrons, in this case in the ICM. These interactions boost the photon energy by $\sim \rm k_{B}T/m_{e}c^2$, leading to a characteristic 
distortion of the CMB spectrum in the direction of galaxy clusters. The CMB intensity is decreased below $\sim$220~GHz and increased above, in proportion 
to the `Comptonization' parameter, $Y$, which is an integral of the product of the electron 
density and temperature through the cluster. This integral of thermal pressure 
in the ICM, which is roughly in hydrostatic equilibrium with the gravitational 
potential well, should therefore be closely related to cluster mass \citep{Carlstrom2002,Arnaud2010,Marrone2012}.

\subsubsection{Sunyaev-Zel'dovich Array}\label{sec:sza}

One of the SZ measurement datasets employed in this paper is based on observations with the 
SZA, an interferometer comprising eight 3.5-metre antennas 
observing at 27$-$35~GHz. During the period of these observations, 2006$-$2014, the SZA 
initially observed from the floor of the Owens Valley, near Big Pine, CA, 
and later was relocated to the nearby Cedar Flat site of the Combined Array for 
Research in Millimeter-wave Astronomy (CARMA). For all observations presented here, 
the SZA antennas observed as an 8-element array, rather than in concert with other 
CARMA antennas as in, e.g., \citet{Plagge13}. The SZA was configured with six 
antennas in a compact configuration to maximize sensitivity to the large-scale cluster signal, 
with the remaining two antennas placed as `outriggers' to discriminate the emission 
from point-like radio sources from the SZ signature of clusters. The resolution of the 
compact array was approximately 2 arcmin, while baselines to the outrigger antennas 
yield a resolution closer to 20 arcsec. 

Observations with the SZA consist of roughly 6-h observing segments in which 
the antennas alternated between point-like calibrator sources and the cluster targets 
on $\sim$20-min cycles. 
The data were reduced using a MATLAB pipeline described in \citet{Muchovej07} 
to flag for weather and technical issues and to calibrate the data. Absolute 
calibration was established from observations of Mars and sometimes Jupiter. 

A Markov Chain Monte Carlo (MCMC) code was used to simultaneously
fit galaxy cluster and point source models to the data.
Point sources were identified from peaks in the flux density in 
long-baseline observations. Many of these sources were coincident with 1.4~GHz sources 
identified in the NRAO VLA Sky Survey \citep[NVSS;][]{NVSS} 
and/or the VLA Faint Images of the Radio Sky at Twenty-Centimeters 
\citep[FIRST;][]{FIRST}, and any sources in these catalogs that lie within 
2 arcmin of the cluster centre were automatically included as model components 
even if they were not obviously detected to prevent them from biasing the SZ signal. 
The SZ signal for each cluster was modeled as a generalized NFW pressure profile 
\citep{Nagai07} using the parameters determined 
by \cite{Planck13} from a joint fit to SZ and X-ray profiles of 62 massive clusters.  
These parameters include a concentration parameter, $c_{500} = 1.81$, the ratio of 
$r_{500}$ to the scale radius ($r_{\rm s}$) of the pressure profile. 
The weak-lensing values of 
$r_{500}$ and their uncertainties were used to define a Gaussian prior 
for the value of the scale radius, $r_{\rm s} = r_{500}/c_{500}$.

We are able to measure \Ysza \ for 30 of the 41 clusters, finding that the fields for nine are contaminated and that two clusters (RXCJ2102.1-2431 and ZwCl0857.9+2107) are non-detections. The two non-detections are near the low end of the sample weak-lensing mass distribution. The contaminated clusters contain 30 GHz sources that are not point-like at the 20 arcsec resolution of the SZA long baselines. In such cases, the interferometric measurement cannot cleanly distinguish between emission from spatially extended radio sources and the spatially extended SZ effect signal, which appears as `negative' emission. The degeneracy between extended radio source emission and cluster SZ signal makes the SZ measurements unreliable.

\subsubsection{Planck}\label{sec:planck}

We also calculate the $Y$ parameter from the six \textit{Planck} High Frequency maps \citep{Planck2016_maps} using a template fitting program similar to the method described in section 2.3 of \citet{Bourdin2017}. The maps are high-pass filtered to remove large-scale (1 deg) signals from the cosmic infrared background, SZ background, and instrumental offsets. On cluster scales, we subtract a spatially and spectrally variable model of the CMB and galactic thermal dust anisotropies.

An \citet{Arnaud2010} pressure profile template was fit to the residual flux within $5r_{500,\rm WL}$ using $\chi^2$ minimization, from which we calculate the cylindrical signal within $r_{500,\rm WL}$. While we use the brightest cluster galaxy (BCG) coordinates as the cluster centres, the \textit{Planck} team identify clusters as peaks in the signal map with a signal-to-noise above 4, and as such identify 38 of the 41 clusters in our sample, while we measure all 41. For the 38, our flux measurements are on average 10 per cent higher than those measured by the Matched Multi-Filter 3 (MMF3) algorithm \citep{Planck2016_y}, which we attribute to offsets of 1$-$2 arcmin in the cluster positions.

\subsubsection{Difference between $Y$ measurements}\label{sec:ycomp}

The SZA and \textit{Planck} estimates of cluster $Y$ parameters can be expected to be tightly correlated, but for several reasons they should not be perfectly so. Of principal importance in explaining differences in $Y$ is the difference in the angular scales probed by the two measurements. The SZA interferometric observations are absolutely insensitive to scales larger than 2$-$3 arcmin, set by the closest antenna pairs in the array, while the \textit{Planck} measurements are unable to capture details finer than $\sim$5$-$10 arcmin owing to the intrinsic resolution of the \textit{Planck} High Frequency maps.

The \textit{Planck} data necessarily infer the SZ signal within $r_{500,\rm WL}$ from a resolution element that is several times larger by assuming that a fixed pressure profile applies to all clusters and explains the observed, profile-integrated SZ signal detected in its large beam. The SZA interferometer, on the other hand, measures a range of spatial frequencies (the Fourier transform of the signal) with the greatest sensitivity to scales finer than $r_{500,\rm WL}$, and must use an assumed profile to fill in the missing spatial frequencies and estimate the signal that would be detected in an aperture of this larger size. Even when assuming the same profile, the two methods are sensitive to different deviations from the profile, from large scales for \textit{Planck} and fine scales for SZA, and are unlikely to agree perfectly. The SZA measurements suggest some significant deviations from the assumed inner shape of the profile for many clusters, manifesting as very different core radii for the pressure profile, but for consistency with the \textit{Planck} data we place a prior probability on the core radius based on the weak-lensing $r_{500,\rm WL}$ that reduces these differences. An additional difference, though one that would be a constant factor of $\sim$ 1.2 \citep{Arnaud2010} between \textit{Planck} and SZA for all clusters if they all had the same pressure profile, is the use of a cylindrical integration for the \textit{Planck} $Y$ and a spherical one for SZA. These integration choices are made to be consistent with the literature and to better accommodate the systematics of the two measurements.

\subsection{Optical and Infrared Observables}\label{sec:opt}

We also use optical and near-infrared observations of the member galaxies, calculating the $K$-band luminosity of the BCG, the total cluster $K$-band luminosity, and the optical richness.

\subsubsection{Near-Infrared Luminosity}\label{sec:nir}

To investigate the stellar content of the clusters, we use near-infrared data from WFCAM on UKIRT, where we observed in J and K band to depths of K$\sim$19 and J$\sim$21 \citep{Haines2009}. We lack this data for Abell2697. From these data we calculate both the $K$-band luminosity of the BCG, \Lbcg, and the total $K$-band luminosity of the cluster members, \Lk.

We analyse the data similar to \citet{Mulroy2014}. We convert from apparent $K$-band magnitude to rest-frame luminosity using a k-correction consistent with \citet{Mannucci2001} and the absolute $K$-band Vega magnitude of the sun, $M_{K,\odot}=3.39$ \citep{Johnson1966}. For the total luminosity, we select cluster members as galaxies lying along a ridge line in $(J-K)/K$ space. We select those within $r_{500,\rm WL}$ of the cluster centre down to a magnitude of $K\le K^{\ast}(\rm z)+2.5$, basing $K^{\ast}(\rm z)$ on \citet{Lin2006} and choosing this limit because $2 < K-K^{\ast} <2.5$ is the faintest 0.5mag width bin for which the average $K$-band magnitude error is $<$0.1 for all clusters. To account for the background we perform this same calculation on a control field \citep[the UKIDSS-DXS Lockman Hole and XMM-LSS fields;][]{Lawrence2007} within 40 apertures of radius $r_{500,\rm WL}$, subtracting the average from \Lk \ and adding the standard deviation to the measurement error. The other component of the measurement error is calculated by propagating the error on the weak-lensing radius. Note that the uncertainties in \citet{Mulroy2014} include a term calculated using bootstrap resampling of the members that we do not include here, because we are interested in the individual cluster measurement error and not the statistical properties of an ensemble of galaxies.

We note that the consistency found in \citet{Mulroy2014} between colour-magnitude selected luminosity and spectroscopically confirmed luminosity indicates the accuracy of colour-magnitude member selection in $(J-K)/K$ space, due to the sensitivity of near-infrared data to old stars and its relative insensitivity to recent star formation.

\subsubsection{Richness}\label{sec:richness}

We calculate the richness, $\lambda$, defined in \citet{Rozo2009} and improved in \citet{Rykoff2012}, for the 33 cluster overlap between our sample and the Sloan Digital Sky Survey \citep[SDSS,][]{Gunn1998,Doi2010,Alam2015}. This matched filter richness estimator is defined as the sum of the membership probabilities of all the galaxies, and was constructed as a low scatter optical mass proxy through extensive tests on the maxBCG cluster catalog \citep{Koester2007}.

For all potential cluster members, their membership probability is calculated considering their clustercentric radius, g-r colour and i-band magnitude. The richness estimator is the sum of these probabilities integrated down to $M^{\ast} + 1.75$, while the measurement error is derived from the variance. The corresponding radius is not equivalent to an overdensity radius such as $r_{500}$, but rather scales deterministically as $\lambda^{0.2}$. The mean radius for our sample is $1.4 \, \rm Mpc$. While the scale misalignment with respect to the other measures may add some additional variance, we retain the algorithm's choice so as to preserve consistency with other redMaPPer applications \citep{Rykoff2012, Rykoff2016}. We find good agreement between our values and redMaPPer values: $\left\langle \lambda_{\rm LoCuSS} / \lambda_{\rm redMaPPer} \right\rangle = 0.99 \pm 0.26$.

From a purely statistical point of view, \richness \ is simply another label tagged to each cluster.  We leave it to future work to identify physically meaningful, minimum variance estimators of these labels.

\begin{table*}
\caption{Cluster observables.}\label{tab:obs}
	\begin{center}
		\tabcolsep=1.0mm
			\begin{tabular}{ l c c c c c c c c c c }
		      \hline
			Name & $L_{X,\rm ce}$ & $k_BT_{X,\rm ce}$ & $M_{\rm gas}$ & $Y_X$ & $Y_{\rm SZA}D_{A}^2$ & $Y_{\rm Pl}D_{A}^2$ & $L_{K,\rm BCG}$ & $L_{K,\rm tot}$ & $\lambda$ \\
			& $10^{44}\rm erg/s$ & $\rm keV$ & $10^{14}M_{\odot}$ & $10^{14}M_{\odot}\rm keV$ & $10^{-5}\rm Mpc^2$ & $10^{-5}\rm Mpc^2$ & $10^{12}L_{\odot} $ & $10^{12}L_{\odot} $ & \\
		      \hline

Abell2697 & $11.68^{+0.44}_{-0.44}$ & $6.99^{+0.48}_{-0.38}$ & $0.90^{+0.07}_{-0.07}$ & $5.42^{+0.40}_{-0.40}$ & $7.61^{+0.78}_{-0.80}$ & $9.18^{+0.48}_{-0.48}$ & -- & -- & $91.44^{+4.37}_{-4.37}$ \\
Abell0068 & $9.91^{+0.59}_{-0.59}$ & $7.66^{+0.77}_{-0.62}$ & $0.80^{+0.04}_{-0.04}$ & $5.59^{+2.24}_{-2.24}$ & $9.38^{+1.21}_{-1.16}$ & $10.34^{+0.59}_{-0.59}$ & $1.01^{+0.01}_{-0.01}$ & $12.46^{+1.92}_{-2.26}$ & $93.04^{+4.60}_{-4.60}$ \\
Abell0115 & $10.68^{+0.40}_{-0.40}$ & $5.93^{+0.39}_{-0.32}$ & $0.87^{+0.15}_{-0.15}$ & $5.88^{+0.57}_{-0.57}$ & -- & $12.44^{+0.48}_{-0.48}$ & $0.74^{+0.00}_{-0.00}$ & $14.77^{+1.72}_{-2.13}$ & -- \\
Abell0141 & $5.10^{+0.88}_{-0.88}$ & $4.78^{+1.34}_{-0.83}$ & $0.60^{+0.05}_{-0.05}$ & $2.95^{+0.95}_{-0.95}$ & -- & $8.42^{+0.38}_{-0.38}$ & $0.64^{+0.00}_{-0.00}$ & $15.02^{+1.22}_{-1.48}$ & -- \\
Abell0209 & $13.59^{+1.02}_{-1.02}$ & $6.39^{+1.05}_{-0.77}$ & $1.44^{+0.08}_{-0.08}$ & $8.80^{+1.68}_{-1.68}$ & $10.79^{+0.96}_{-0.96}$ & $19.33^{+0.53}_{-0.53}$ & $0.90^{+0.00}_{-0.00}$ & $20.51^{+2.01}_{-1.98}$ & -- \\
Abell0267 & $6.34^{+2.88}_{-2.88}$ & $8.03^{+2.83}_{-1.81}$ & $0.70^{+0.05}_{-0.05}$ & $7.21^{+2.91}_{-2.91}$ & $6.47^{+0.61}_{-0.62}$ & $6.47^{+0.61}_{-0.61}$ & $1.44^{+0.01}_{-0.01}$ & $12.71^{+1.76}_{-2.79}$ & $96.38^{+4.03}_{-4.03}$ \\
Abell0291 & $3.37^{+0.08}_{-0.08}$ & $4.03^{+0.32}_{-0.29}$ & $0.47^{+0.04}_{-0.04}$ & $1.44^{+0.09}_{-0.09}$ & $2.57^{+0.57}_{-0.49}$ & $3.04^{+0.47}_{-0.47}$ & $0.55^{+0.00}_{-0.00}$ & $7.79^{+0.98}_{-0.96}$ & $53.86^{+2.75}_{-2.75}$ \\
Abell0521 & $15.33^{+1.09}_{-1.09}$ & $6.72^{+0.33}_{-0.29}$ & $1.08^{+0.09}_{-0.09}$ & $7.27^{+0.39}_{-0.39}$ & $5.34^{+0.60}_{-0.62}$ & $12.72^{+0.58}_{-0.58}$ & $0.95^{+0.01}_{-0.01}$ & $14.17^{+2.14}_{-2.69}$ & -- \\
Abell0586 & $6.20^{+0.54}_{-0.54}$ & $5.56^{+1.10}_{-0.79}$ & $0.73^{+0.06}_{-0.06}$ & $3.63^{+0.69}_{-0.69}$ & $10.29^{+1.34}_{-1.27}$ & $5.30^{+0.44}_{-0.44}$ & $0.81^{+0.00}_{-0.00}$ & $18.30^{+1.96}_{-2.36}$ & $105.96^{+4.38}_{-4.38}$ \\
Abell0611 & $12.00^{+0.94}_{-0.94}$ & $11.96^{+2.50}_{-2.40}$ & $0.69^{+0.05}_{-0.05}$ & -- & $8.47^{+0.78}_{-0.84}$ & $11.67^{+0.67}_{-0.67}$ & $1.33^{+0.01}_{-0.01}$ & $13.61^{+2.66}_{-2.82}$ & $100.90^{+4.64}_{-4.64}$ \\
Abell0697 & $22.55^{+2.29}_{-2.29}$ & $11.06^{+2.16}_{-1.83}$ & $1.22^{+0.10}_{-0.10}$ & $16.21^{+3.55}_{-3.55}$ & $16.35^{+1.51}_{-1.50}$ & $26.41^{+0.62}_{-0.62}$ & $1.50^{+0.01}_{-0.01}$ & $13.15^{+2.61}_{-2.62}$ & $147.28^{+5.13}_{-5.13}$ \\
ZwCl0857.9+2107 & $4.50^{+0.19}_{-0.19}$ & $3.97^{+0.15}_{-0.46}$ & $0.34^{+0.07}_{-0.07}$ & $1.40^{+0.11}_{-0.11}$ & -- & $0.66^{+0.44}_{-0.44}$ & $0.44^{+0.01}_{-0.01}$ & $2.79^{+0.92}_{-1.09}$ & $26.85^{+2.58}_{-2.58}$ \\
Abell0750 & $2.89^{+0.20}_{-0.20}$ & $3.95^{+0.49}_{-0.39}$ & $0.55^{+0.06}_{-0.06}$ & $2.08^{+0.30}_{-0.30}$ & $5.27^{+0.77}_{-0.76}$ & $7.85^{+0.38}_{-0.38}$ & $0.75^{+0.00}_{-0.00}$ & $19.73^{+1.92}_{-2.26}$ & $139.58^{+4.40}_{-4.40}$ \\
Abell0773 & $11.11^{+1.14}_{-1.14}$ & $7.50^{+1.58}_{-1.12}$ & $1.10^{+0.05}_{-0.05}$ & $7.46^{+1.39}_{-1.39}$ & $13.08^{+0.92}_{-0.91}$ & $12.33^{+0.46}_{-0.46}$ & $0.82^{+0.00}_{-0.00}$ & $22.02^{+2.04}_{-1.79}$ & $141.43^{+4.58}_{-4.58}$ \\
Abell0781 & $4.16^{+1.92}_{-1.92}$ & $5.92^{+2.40}_{-1.36}$ & $0.74^{+0.12}_{-0.12}$ & $4.45^{+1.69}_{-1.69}$ & -- & $9.58^{+0.71}_{-0.71}$ & $0.83^{+0.01}_{-0.01}$ & $16.58^{+3.43}_{-4.16}$ & $180.62^{+6.08}_{-6.08}$ \\
ZwCl0949.6+5207 & $4.52^{+0.99}_{-0.99}$ & $7.31^{+0.94}_{-0.89}$ & $0.40^{+0.04}_{-0.04}$ & -- & $3.22^{+0.69}_{-0.65}$ & $2.71^{+0.40}_{-0.40}$ & $0.80^{+0.00}_{-0.00}$ & $7.91^{+1.47}_{-1.51}$ & $44.37^{+3.38}_{-3.38}$ \\
Abell0907 & $5.91^{+0.22}_{-0.22}$ & $5.66^{+0.51}_{-0.41}$ & $0.93^{+0.06}_{-0.06}$ & $4.01^{+0.33}_{-0.33}$ & -- & $9.26^{+0.41}_{-0.41}$ & $0.60^{+0.00}_{-0.00}$ & $13.83^{+1.56}_{-1.71}$ & -- \\
Abell0963 & $7.89^{+0.29}_{-0.29}$ & $6.53^{+0.62}_{-0.50}$ & $0.80^{+0.05}_{-0.05}$ & $4.13^{+0.29}_{-0.29}$ & -- & $8.22^{+0.46}_{-0.46}$ & $1.29^{+0.00}_{-0.00}$ & $14.84^{+1.66}_{-1.78}$ & $65.01^{+3.66}_{-3.66}$ \\
ZwCl1021.0+0426 & $19.66^{+1.47}_{-1.47}$ & $9.04^{+1.51}_{-1.13}$ & $0.95^{+0.05}_{-0.05}$ & $10.80^{+2.50}_{-2.50}$ & $10.42^{+0.83}_{-0.82}$ & $9.81^{+0.60}_{-0.60}$ & $0.89^{+0.01}_{-0.01}$ & $9.27^{+1.87}_{-1.82}$ & $83.11^{+4.12}_{-4.12}$ \\
Abell1423 & $7.35^{+0.68}_{-0.68}$ & $8.20^{+1.54}_{-1.16}$ & $0.62^{+0.06}_{-0.06}$ & $6.42^{+1.46}_{-1.46}$ & $3.15^{+0.46}_{-0.47}$ & $7.61^{+0.40}_{-0.40}$ & $1.02^{+0.01}_{-0.01}$ & $9.90^{+1.22}_{-1.47}$ & $59.00^{+3.77}_{-3.77}$ \\
Abell1451 & $6.13^{+1.31}_{-1.31}$ & $8.87^{+1.45}_{-1.10}$ & $1.02^{+0.05}_{-0.05}$ & $7.57^{+1.07}_{-1.07}$ & $6.02^{+0.98}_{-1.01}$ & $11.52^{+0.49}_{-0.49}$ & $0.55^{+0.00}_{-0.00}$ & $18.77^{+2.11}_{-1.92}$ & -- \\
ZwCl1231.4+1007 & $7.87^{+0.66}_{-0.66}$ & $6.56^{+1.20}_{-0.89}$ & $0.69^{+0.11}_{-0.11}$ & $5.67^{+1.25}_{-1.25}$ & -- & $8.62^{+0.42}_{-0.42}$ & $0.99^{+0.00}_{-0.00}$ & $9.43^{+2.83}_{-1.82}$ & $93.13^{+4.41}_{-4.41}$ \\
Abell1682 & $4.99^{+2.00}_{-2.00}$ & $6.46^{+2.98}_{-1.49}$ & $0.84^{+0.04}_{-0.04}$ & $5.18^{+2.36}_{-2.36}$ & -- & $8.71^{+0.41}_{-0.41}$ & $1.04^{+0.00}_{-0.00}$ & $19.56^{+1.83}_{-2.15}$ & $118.56^{+4.51}_{-4.51}$ \\
Abell1689 & $15.81^{+0.55}_{-0.55}$ & $9.71^{+0.64}_{-0.51}$ & $1.31^{+0.05}_{-0.05}$ & $12.81^{+0.95}_{-0.95}$ & $27.55^{+2.27}_{-2.21}$ & $17.72^{+0.47}_{-0.47}$ & $0.74^{+0.00}_{-0.00}$ & $23.07^{+2.31}_{-2.51}$ & $163.62^{+4.13}_{-4.13}$ \\
Abell1763 & $15.20^{+1.56}_{-1.56}$ & $7.67^{+1.64}_{-1.32}$ & $1.61^{+0.09}_{-0.09}$ & $11.08^{+2.56}_{-2.56}$ & -- & $20.23^{+0.43}_{-0.43}$ & $1.17^{+0.01}_{-0.01}$ & $21.86^{+3.70}_{-3.18}$ & $172.16^{+5.30}_{-5.30}$ \\
Abell1835 & $22.22^{+0.79}_{-0.79}$ & $10.16^{+0.68}_{-0.55}$ & $1.43^{+0.07}_{-0.07}$ & $13.84^{+1.03}_{-1.03}$ & $22.26^{+1.60}_{-1.67}$ & $19.51^{+0.71}_{-0.71}$ & $1.29^{+0.00}_{-0.00}$ & $21.42^{+3.15}_{-2.75}$ & $134.55^{+4.89}_{-4.89}$ \\
Abell1914 & $17.08^{+1.36}_{-1.36}$ & $10.06^{+1.47}_{-1.22}$ & $1.11^{+0.07}_{-0.07}$ & $12.54^{+2.17}_{-2.17}$ & $21.10^{+2.71}_{-2.48}$ & $12.06^{+0.30}_{-0.30}$ & $0.96^{+0.00}_{-0.00}$ & $13.37^{+1.43}_{-1.71}$ & $110.67^{+3.61}_{-3.61}$ \\
ZwCl1454.8+2233 & $6.66^{+0.27}_{-0.27}$ & $4.74^{+0.42}_{-0.34}$ & $0.54^{+0.08}_{-0.08}$ & $3.10^{+0.36}_{-0.36}$ & $2.39^{+0.49}_{-0.52}$ & $6.21^{+0.60}_{-0.60}$ & $1.25^{+0.01}_{-0.01}$ & $6.64^{+1.71}_{-2.15}$ & $48.09^{+3.23}_{-3.23}$ \\
Abell2009 & $6.05^{+0.63}_{-0.63}$ & $7.44^{+1.56}_{-1.16}$ & $0.69^{+0.05}_{-0.05}$ & $4.72^{+1.03}_{-1.03}$ & $5.02^{+0.78}_{-0.80}$ & $4.44^{+0.38}_{-0.38}$ & $0.74^{+0.00}_{-0.00}$ & $9.51^{+1.78}_{-1.91}$ & $73.70^{+3.21}_{-3.21}$ \\
RXCJ1504.1-0248 & $16.65^{+1.86}_{-1.86}$ & $9.55^{+2.23}_{-1.52}$ & $1.06^{+0.08}_{-0.08}$ & $11.26^{+3.49}_{-3.49}$ & $12.17^{+1.26}_{-1.22}$ & $11.35^{+0.65}_{-0.65}$ & $0.97^{+0.00}_{-0.00}$ & $10.31^{+1.40}_{-1.56}$ & $61.06^{+3.79}_{-3.79}$ \\
Abell2111 & $5.93^{+2.76}_{-2.76}$ & $7.21^{+2.28}_{-1.52}$ & $0.68^{+0.08}_{-0.08}$ & $5.49^{+2.12}_{-2.12}$ & $5.58^{+0.76}_{-0.71}$ & $8.98^{+0.52}_{-0.52}$ & $0.64^{+0.00}_{-0.00}$ & $15.31^{+1.53}_{-1.80}$ & $138.66^{+4.96}_{-4.96}$ \\
Abell2204 & $15.84^{+0.66}_{-0.66}$ & $13.38^{+1.15}_{-0.76}$ & $1.23^{+0.08}_{-0.08}$ & $11.79^{+1.02}_{-1.02}$ & $17.71^{+1.77}_{-1.72}$ & $17.15^{+0.39}_{-0.39}$ & $0.65^{+0.00}_{-0.00}$ & $19.69^{+1.50}_{-1.27}$ & -- \\
Abell2219 & $32.91^{+2.60}_{-2.60}$ & $10.13^{+0.83}_{-0.70}$ & $1.68^{+0.11}_{-0.11}$ & $17.90^{+1.67}_{-1.67}$ & $18.42^{+1.37}_{-1.37}$ & $30.27^{+0.46}_{-0.46}$ & $1.04^{+0.01}_{-0.01}$ & $21.72^{+2.07}_{-1.83}$ & $169.10^{+5.10}_{-5.10}$ \\
RXJ1720.1+2638 & $9.63^{+0.57}_{-0.57}$ & $7.14^{+0.91}_{-0.73}$ & $0.71^{+0.07}_{-0.07}$ & $6.60^{+1.00}_{-1.00}$ & -- & $8.60^{+0.31}_{-0.31}$ & $1.01^{+0.00}_{-0.00}$ & $9.77^{+2.11}_{-1.36}$ & $63.89^{+2.97}_{-2.97}$ \\
Abell2261 & $13.04^{+1.12}_{-1.12}$ & $7.50^{+1.30}_{-1.09}$ & $1.23^{+0.06}_{-0.06}$ & $8.12^{+1.19}_{-1.19}$ & $12.36^{+1.52}_{-1.60}$ & $13.56^{+0.48}_{-0.48}$ & $1.78^{+0.01}_{-0.01}$ & $26.60^{+2.38}_{-3.64}$ & $142.94^{+4.89}_{-4.89}$ \\
RXCJ2102.1-2431 & $4.62^{+0.12}_{-0.12}$ & $5.32^{+0.46}_{-0.37}$ & $0.46^{+0.05}_{-0.05}$ & $2.29^{+0.18}_{-0.18}$ & -- & $4.00^{+0.36}_{-0.36}$ & $1.04^{+0.01}_{-0.01}$ & $7.77^{+0.87}_{-0.87}$ & -- \\
RXJ2129.6+0005 & $10.65^{+0.65}_{-0.65}$ & $5.94^{+0.75}_{-0.61}$ & $0.67^{+0.10}_{-0.10}$ & $5.47^{+0.95}_{-0.95}$ & $5.73^{+0.69}_{-0.89}$ & $5.76^{+0.48}_{-0.48}$ & $1.28^{+0.01}_{-0.01}$ & $7.53^{+1.60}_{-1.81}$ & $71.30^{+3.97}_{-3.97}$ \\
Abell2390 & $25.43^{+1.16}_{-1.16}$ & $10.79^{+0.95}_{-0.84}$ & $1.66^{+0.09}_{-0.09}$ & $16.91^{+1.57}_{-1.57}$ & $16.36^{+3.15}_{-3.07}$ & $24.07^{+0.52}_{-0.52}$ & $0.75^{+0.00}_{-0.00}$ & $17.44^{+2.02}_{-1.98}$ & $121.10^{+4.89}_{-4.89}$ \\
Abell2537 & $6.63^{+0.72}_{-0.72}$ & $9.93^{+3.73}_{-2.44}$ & $0.83^{+0.08}_{-0.08}$ & $6.30^{+2.30}_{-2.30}$ & $8.00^{+0.88}_{-0.86}$ & $9.77^{+0.63}_{-0.63}$ & $1.02^{+0.01}_{-0.01}$ & $19.48^{+2.65}_{-2.80}$ & $146.22^{+5.08}_{-5.08}$ \\
Abell2552 & $13.46^{+1.77}_{-1.77}$ & $9.69^{+2.75}_{-1.94}$ & $1.00^{+0.10}_{-0.10}$ & $9.22^{+2.89}_{-2.89}$ & $9.09^{+1.19}_{-1.19}$ & $11.66^{+0.63}_{-0.63}$ & $0.63^{+0.00}_{-0.00}$ & $19.51^{+4.15}_{-4.55}$ & $148.78^{+6.27}_{-6.27}$ \\
Abell2631 & $14.41^{+1.02}_{-1.02}$ & $6.91^{+1.18}_{-0.87}$ & $0.97^{+0.12}_{-0.12}$ & $6.69^{+1.21}_{-1.21}$ & $5.22^{+0.70}_{-0.83}$ & $11.95^{+0.57}_{-0.57}$ & $0.85^{+0.01}_{-0.01}$ & $13.75^{+2.31}_{-2.58}$ & $114.80^{+4.79}_{-4.79}$ \\
		      \hline
		\end{tabular}
	\end{center}
{\footnotesize}
\end{table*}

\section{Self-Similar Scaling}\label{sec:selfsimilar}

It is useful to review what might be expected for the outcome of our scaling relation constraints, and in this section we review predictions from self-similarity \citep{Kaiser1986}. The dominant force on the scale of galaxy clusters is gravity, which is scale invariant. This means that galaxy clusters, under the influence of gravity and shock heating only, are expected to be simply scaled versions of each other, with their properties determined only by their mass and redshift. 
Redshift determines the critical density

\begin{equation}
\rho_c(z) = E^2(z)\rho_{c,0},
\end{equation}
where the subscript $0$ refers to the present epoch.  

It is convention to define halo mass as that, centred on a local potential minimum, contained within a sphere of radius $r_{\Delta}$ encompassing an overdensity $\Delta$ relative to the critical density, thus
\begin{equation}
M_{\Delta} \ = \ \frac{4}{3} \pi  r_{\Delta}^3 \Delta \rho_c(z) \ \propto \ E^2(z) r_{\Delta}^3 .
\label{eqn:radius}
\end{equation}

Matter in self-similar, hydrostatic galaxy clusters satisfies the virial theorem between gravitational potential energy $U$ and kinetic energy $K$ ($\langle U \rangle = - 2\langle K \rangle$), leading to the expression for the circular velocity of the halo: $v_{\rm circ}^2 = M_{\Delta}/r_{\Delta}$. Combined with equation~(\ref{eqn:radius}), we see that the combination of mass and redshift sets the strength of the local gravitational potential:

\begin{equation}
v_{\rm circ}^3 \propto M_{\Delta}E(z).
\label{eqn:potential}
\end{equation}

This relation, which has been precisely calibrated by N-body simulations \citep{Evrard2008}, motivates our use of the effective potential well depth, $M_\Delta E(z)$, as the independent degree of freedom in the scaling relations we fit below. Note that we use the value $\Delta = 500$ in this work, because this is the radius which can be probed without extrapolation by all our measurements.

Applying the virial theorem to the ICM, the total kinetic energy can be written in terms of the average kinetic energy of the ICM particles, i.e. the cluster X-ray temperature $T_X$, leading to

\begin{equation}
T_X \propto \left[ M_{\Delta}E(z) \right]^{2/3}. \label{eqn:tx}
\end{equation}

The X-ray emission from the ICM is dominated by thermal bremsstrahlung emission, for which the resulting luminosity scales as $L_X \propto \rho_{\rm gas}^2 r^3 \Lambda(T_X)$, where there are two factors of the gas density $\rho_{\rm gas}$ because the radiation is produced by a two-body interaction, and $\Lambda(T_X)$ is the cooling function. In the soft-band range $\sim$[0.1$-$2.4]keV, the integral of the cooling function is nearly independent of $T_X$, while across the full energy range used for bolometric X-ray luminosity it scales with $T_X^{1/2}$. This leads to

\begin{equation}
\frac{L_{X, \rm soft}}{E(z)} \propto M_{\Delta}E(z),
\qquad
\frac{L_{X,\rm bol}}{E(z)} \propto \left[ M_{\Delta}E(z) \right]^{4/3}.
\end{equation}

As probes of the same thermal energy, $Y_{X}$ and $Y_{\rm SZ}$ have the same self-similar scaling, which can be derived from the product of $M_{\rm gas}$ and $T_X$:

\begin{equation}
YE(z) \propto \left[ M_{\Delta}E(z) \right]^{5/3},
\end{equation}

\noindent under the simple assumption of a constant gas fraction, $f_{\rm gas}$. We make the similar assumption of a constant stellar fraction, $f_{\star}$, giving

\begin{equation}
M_{\rm gas} = f_{\rm gas}M_{\Delta} \propto M_{\Delta},
\qquad
L_K \propto M_{\star} = f_{\star}M_{\Delta} \propto M_{\Delta},
\end{equation}

\noindent under the assumption that $L_K$ is a good indicator of the total stellar mass.

Finally, if we assume each cluster has a galaxy population drawn from a single luminosity function with some effective mean stellar mass, $m_{\star,\rm gal}$, we can also derive a relation between richness and mass:

\begin{equation}
\lambda = \frac{M_{\star}}{m_{\star,\rm gal}} \propto M_{\Delta}.
\end{equation}

\section{Linear Regression}\label{sec:regression}

We assume that scaling relations between observable properties and mass are described by power-law relations with constant slopes\footnote{While simulations suggest mass-dependent slope behaviour \citep{Farahi2018}, a constant slope is a good approximation for the narrow mass range probed by our sample.}.  We linearise the problem by using the natural log of the values and perform a Bayesian analysis to infer scaling parameters.  To do so correctly we have to take into account measurement errors, the halo mass function and the selection criteria.  Most commonly used regression methods (e.g. BCES, \citealt{Akritas1996}; and FITEXY, \citealt{Press1992,Tremaine2002}) can handle measurement errors, while methods from \citet{Kelly2007} and \citet{Mantz2016} also take into account the independent variable distribution by modelling it as a Gaussian mixture model inferred from the data.

However the selection function can still introduce significant biases, either directly when the selection variable is considered directly in the regression, or indirectly due to covariance between this selection variable and the observable of interest. We quantify this bias for the scaling relations presented in this paper by performing linear regression without correcting for selection effects.  Results are presented in Table~\ref{tab:bias} of Appendix \ref{sec:bias}. It is possible, in principle, to use the methods of \citet{Kelly2007} and \citet{Mantz2016} to correct for selection effects when the selection variable is on the dependent axis, by using upper limits and generating `censored' or missing data below the selection limit in an iterative process \citep{gelman2014bayesian}. However it is more complicated to correct for the bias caused by covariance with the selection variable, i.e. when considering a dependent variable which is not the selection variable, and this approach can be computationally challenging for a larger dataset.

We therefore develop a hierarchical Bayesian model similar to the methods of \citet{Kelly2007} and \citet{Mantz2016}, which simultaneously considers the selection variable alongside all other observables in order to explicitly model the property covariance, i.e. the intrinsic covariance between two observables at fixed halo mass, and correctly propagate selection effects.

\begin{table*}
\caption{Elements of the galaxy cluster observable vector.}\label{tab:MOR}
	\begin{center}
		\tabcolsep=0.8mm
		\begin{tabular}{ l l l }
            \hline
            Element, $S_{a}$ & Unit & Description \\
            \hline
\Lxrass$E(z)^{-1}$ & $10^{44} \, \rm erg/s$ & Selection variable: RASS soft-band X-ray luminosity \\
\Lx$E(z)^{-1}$ & $10^{44} \, \rm erg/s$ & Core-excised bolometric X-ray luminosity within $[0.15-1]r_{500,\rm WL}$\\
\Tx & $\rm keV$ & Core-excised ICM temperature within $[0.15-1]r_{500,\rm WL}$ \\ 
\Mgas$E(z)$ & $10^{14} \, M_{\odot}$ & ICM gas mass within $r_{500,\rm WL}$ \\ 
\Yx$E(z)$ & $10^{14} \, M_{\odot}\rm keV$ & Spherical ICM X-ray thermal energy \\ 
\Ysza$E(z)$ & $10^{-5} \, \rm Mpc^2$ & Spherical ICM SZ thermal energy within $r_{500,\rm WL}$ \\ 
\Ypl$E(z)$ & $10^{-5} \, \rm Mpc^2$ & Cylindrical ICM SZ thermal energy within $r_{500,\rm WL}$ \\ 
\Lbcg$E(z)$ & $10^{12} \, L_{\odot}$ & BCG $K$-band luminosity \\ 
\Lk$E(z)$ & $10^{12} \, L_{\odot}$ & Total $K$-band luminosity within $r_{500,\rm WL}$\\ 
\richness$E(z)$ & none & redMaPPer richness (count of galaxies) \\
\Mwl$E(z)$ & $10^{14} \, M_{\odot}$ & Weak-lensing $M_{500}$ mass \\
            \hline
        \end{tabular}
	\end{center}
{\footnotesize}
\end{table*}

\subsection{Hierarchical Bayesian Model}\label{sec:maxlik}

We define log-space variables, $\mu \equiv \ln(M)$ and $\textbf{s} \equiv \ln(\textbf{S})$, where $M$ is the total halo mass and $\textbf{S}$ the vector of observables given in Table~\ref{tab:MOR}.  In practice we normalize mass using the median weak-lensing mass of the sample.  
At a fixed redshift, the joint probability that there exists a cluster with given observables and mass can be written as the product
\begin{equation}
P(\textbf{s}, \mu \, | \, \boldsymbol{\theta} , \boldsymbol{\psi}) = P(\textbf{s} \, | \, \mu  , \boldsymbol{\theta})  P(\mu \, | \, \boldsymbol{\psi}),
\label{eq:jointLikeModel}
\end{equation}
\noindent where $\boldsymbol{\theta}$ is the set of parameters that characterise the scaling relation of observable properties with mass, and $\boldsymbol{\psi}$ characterises the distribution of the independent variable, in this case the cosmological mass function of halos.  For the analysis presented here, we simplify the latter term by assuming a fixed cosmology and use the second-order mass function model of \citet{Evrard2014} at redshift 0.22. Since the mass function shape has only a modest effect on the posterior scaling parameter constraints, we do not attempt to marginalize over cosmology and so drop $\boldsymbol{\psi}$ from the equations below.

We note that the mass discussed above is the true unobserved halo mass which we marginalize over. The small sample size and limited set of observables force us to make the simplifying assumption that weak-lensing mass is an unbiased measure of true halo mass, albeit with non-zero scatter of $\sim$20 per cent \citep[e.g.][]{Becker&Kravtsov2011,Oguri2011,Bahe2012}. We retain weak-lensing mass, $M_{\rm WL}$, in the vector of observables $\textbf{s}$, and treat it in a special way to avoid severe parameter degeneracies of the type discussed in \citet{Penna-Lima:2017}.

We model $P(\textbf{s} \, | \, \mu  , \boldsymbol{\theta})$, the first term in the joint probability distribution in equation~(\ref{eq:jointLikeModel}), as a log-normal distribution, 
\begin{equation}
P( \textbf{s} \, | \, \mu , \boldsymbol{\theta}) \propto {\rm det}(\Sigma)^{-\frac{1}{2}} \exp \left\{ -\frac{1}{2} ( \textbf{s}  - \langle\textbf{s}\rangle )^{T} \Sigma^{-1}  ( \textbf{s}  - \langle\textbf{s}\rangle ) \right\},
\end{equation}
\noindent where $\langle\textbf{s}\rangle = \boldsymbol{\alpha} \mu + \boldsymbol{\pi}$ and the model parameters, $\boldsymbol{\theta} = \left\{ \boldsymbol{\pi} , \boldsymbol{\alpha} , \Sigma \right\}$, include the intercepts $\boldsymbol{\pi}$ and slopes $\boldsymbol{\alpha}$ of the log-mean behaviour, as well as the property covariance matrix $\Sigma$ of Gaussian deviations about the log-mean. Each diagonal element of the covariance matrix specifies the variance of a property, while the off-diagonal elements are the property covariance, all at fixed true halo mass. Except for the parameters connected to weak-lensing mass, which are fixed as explained below, the remainder are unknown parameters to be constrained. Parameter priors are uninformative, as specified in Table \ref{tab:prior}. 

We impose a strict prior on the scaling of $M_{\rm WL}$ that assumes unit slope and intercept with true mass, and a fixed log-normal scatter of $0.2$. We tested values for the scatter of $0.1$ and $0.3$, finding that our results and inferred parameters are insensitive to this choice. We assume zero {\sl intrinsic} correlation between weak-lensing mass and all other observable properties ($r_{M_{\rm WL},S_a} = 0$ for all properties, $S_a$). 
We include the correlation of weak-lensing mass measurement uncertainty with the other observables defined within the weak-lensing radius \citep[so-called `aperture bias', e.g.][]{Okabe2010}.  

In the likelihood below, true masses of all clusters are treated as extra degrees of freedom, or hyperparameters, with posteriors shaped primarily by the input weak-lensing mass measurements and secondarily by collective distance from the mean property scaling relations.  Because of the relatively narrow mass range probed by the LoCuSS sample, the assumed form of the mass function is not very important.  Because our focus is on scaling relation model parameters, the likelihood does not contain explicit terms relating to the size of the selected sample.  In other words, the sample volume is not a factor in our model.

In practice we do not measure the true values of $\textbf{s}$; our measurements, $\textbf{s}_{o}$, include observational uncertainties.  We again assume a log-normal form for the measurement errors,  
\begin{equation}
   P(\textbf{s}_{o}  | \textbf{s}) \propto {\rm det}(\Sigma_{\rm err})^{-\frac{1}{2}} \exp \left\{ -\frac{1}{2} ( \textbf{s}_{o}  - \textbf{s}  )^{T} \Sigma_{\rm err}^{-1}  ( \textbf{s}_{o}  - \textbf{s} ) \right\},
\end{equation}
where $\Sigma_{\rm err}$ is the measurement error covariance. This matrix includes both diagonal elements given by the square of the fractional errors in each cluster's measured properties, and off-diagonal `aperture bias' terms for \Mgas, \Lk \ and \Ysza \ properties measured within the characteristic radius inferred from weak-lensing mass. The aperture bias contributions are the fraction of an observable's uncertainty that is due to the radial error, calculated by remeasuring the observable within $r_{500,\rm WL} \pm \delta r$ to propagate the radial uncertainty, where $\delta r$ is $\sim$50$-$130$\rm kpc$, or $\sim$4$-$15 per cent of $r_{500,\rm WL}$. The propagated aperture uncertainties are added in quadrature with the observables' other statistical uncertainty. While most other observables are measured within the weak-lensing radius, they are largely unaffected by small radial changes and so don't require these off-diagonal terms.

The probability of measuring the observable properties, $\textbf{s}_{o,i}$, of a specific cluster, $i$, is found by marginalizing over the true quantities, \textbf{s}, resulting in
\begin{align}
P( \textbf{s}_{o,i} | \mu_i , \boldsymbol{\theta}) & \propto {\rm det}(\Sigma_{{\rm tot},i})^{-\frac{1}{2}} \times \\
   & ~~ \exp \left\{ -\frac{1}{2} ( \textbf{s}_{o,i}  - \langle {\textbf{s}}_{o}\rangle_i )^{T} \Sigma_{{\rm tot},i}^{-1} ( \textbf{s}_{o,i}  - \langle {\textbf{s}}_{o}\rangle_i  ) \right\}, \nonumber
\end{align} 
where $\langle {\textbf{s}}_{o}\rangle_i  = \boldsymbol{\alpha} \mu_i + \boldsymbol{\pi}$, with $\mu_i$ the unobserved true halo mass of the $i^{\rm th}$ cluster, and $\Sigma_{{\rm tot},i} = \Sigma + \Sigma_{{\rm err},i}$.  We make a similar log-normal assumption about the weak-lensing mass measurements -- which is an element in $\textbf{s}_{o}$ -- and include the measurement error and its aperture-driven covariance with other measured property uncertainties in the regression analysis. 

Finally, we are able to account for the effect of sample selection, as the vector of observables includes the selection variable \citep{gelman2014bayesian,Kelly2007}. Our selection function is simply a redshift dependent \Lxrass \ threshold (see Fig.~\ref{fig:sample}), which is taken into account using a redshift dependent step function.  Letting $y \equiv \ln L_{X,\rm RASS}$ and denoting the $z$-dependent threshold luminosity as $y_t(z)$, the odds of selection given a true mass, $\mu_i$, and model parameters, $\boldsymbol{\theta}$, are
\begin{equation}
\Phi_i(\mu_i, \boldsymbol{\theta}) = 
   \int {\rm d} y \ \Theta(y - y_t(z_i)) \ P(y  \, | \mu_i \,, \boldsymbol{\theta}),
\label{eq:sel-exp}
\end{equation}
where $\Theta(z)$ is the Heaviside function. With the assumed log-normal form, the integral yields a complementary error function that is evaluated for each cluster at each step in the MCMC analysis.

The expression in equation~(\ref{eq:sel-exp}) is used to re-normalize the contribution of each cluster to the likelihood. The likelihood of the observed sample properties is then
\begin{equation} \label{eq:sele-func}
  \mathcal{L}  = \prod\limits_{i \in \mathcal{C}} \,\, \Phi_i^{-1}(\mu_i, \boldsymbol{\theta}) \ P({\bf s}_{o,i}  \,|\, \mu_i \,, \boldsymbol{\theta} ),
\end{equation}
where $\mathcal{C}$ is the cluster sample.  Compared to a selection-unweighted likelihood (see Appendix \ref{sec:bias}), the odds factor adds support in regions where the $L_{X,\rm RASS}$--$M$ relation has a lower mean amplitude, steeper slope, and larger variance.

We consider the set of 41 true halo masses as additional degrees of freedom and perform the MCMC analysis in this space joined with 75 model degrees of freedom consisting of slope, normalization, and variance for 10 properties, and 45 correlation coefficients.  Uninformative priors, $P(\boldsymbol{\theta})$, on the latter parameters are specified in Table \ref{tab:prior} and the halo mass function, $P(\mu_i)$, is used as a prior on cluster true masses.  At every iteration of the MCMC analysis, the likelihood is renormalized according to equation~(\ref{eq:sel-exp}), and the resulting posterior probability distribution in the full model parameter space is
\begin{equation}\label{eq:sele-func-2}
  P( \boldsymbol{\theta}, \mu_i \,| \, {\bf s}_{o,i} ) \propto \left[ \prod\limits_{i \in \mathcal{C}} \,\, \Phi_i^{-1}(\mu_i, \boldsymbol{\theta}) \ P({\bf s}_{o,i}  \,|\, \mu_i \,, \, \boldsymbol{\theta} ) \right] P(\mu_i \,, \boldsymbol{\theta}),
\end{equation}
where $P(\mu_i \,, \boldsymbol{\theta}) = P(\mu_i) P(\boldsymbol{\theta})$ is the prior distribution. 

We then determine the model parameter constraints, $P(\boldsymbol{\theta} \,| \, {\bf s}_{o,i})$, by marginalizing over the posterior distributions of the 41 halo masses. In Section \ref{sec:massest}, we perform the complementary marginalization and present posterior estimates of true mass for the 41 LoCuSS clusters.

The MCMC algorithm is based on the \texttt{PyMC} library \citep{pymc2010} and proceeds as follows.  For each iteration, a mass is assigned to each cluster drawn randomly from the halo mass function, i.e. the prior distribution. Then a new set of model parameters, $\boldsymbol{\theta}$, are drawn randomly from the prior distribution specified in Table \ref{tab:prior}.  With the assigned cluster masses and chosen set of parameters, the selection function is evaluated and the likelihood evaluated. The initial seeds are adapted in a way to minimize the number of steps needed to reach equilibrium. We choose the central value of the weak-lensing masses as the initial seed for each unobserved halo mass, $\mu_i$, and the scaling parameters are initialized with the estimates from the uncorrected fit in Appendix \ref{sec:bias}. This choice of initial seeds allows us to reach equilibrium faster and does not have an effect on the posterior distribution. The performance of this method is demonstrated and compared with other methods in Appendix \ref{sec:maxliktests}.

Our method is able to handle missing data, meaning systems for which not all elements of the data vector are available. We marginalize over these missing quantities by setting the missing values to the median of that observable quantity and assuming a large error, 999 in the natural log, on the missing value. 

\begin{table}
\centering
\caption{Prior distributions of the scaling relation parameters for any property, $a$, other than weak-lensing mass. The same priors are used for all properties and pairwise combinations, $a$,$b$.}
\label{tab:prior}
\begin{tabular}{|l|l|l|}
\hline
Parameter                       & Description             & Prior    \\ \hline
$\pi_a$          & Intercept               & $\mathcal{N}(0, 100)$ \\ 
$\alpha_a$       & Slope                   & $\mathcal{N}(0, 100)$ \\ 
$\sigma_{a\,|\,\mu}$       & Scatter               & $\mathcal{U}(0, 5)$   \\ 
$r_{a,b\,|\,\mu}$  & Correlation coefficient & $\mathcal{U}(-1, 1)$   \\ \hline
\end{tabular}
\end{table}

\section{Results}\label{sec:results}

In this section we apply the hierarchical Bayesian method described in Section \ref{sec:maxlik} to the LoCuSS data described in Section \ref{sec:data}. We discuss the resulting scaling relation parameters below, focusing on the individual properties in turn. Constraints on property covariances are presented in a companion paper \citep{Farahi:inprep}.  

In order to characterise the scaling relations between cluster observables and mass, we use a fixed mass pivot defined by the sample average, $M_p = 7.41\times 10^{14}M_{\odot}$, and fit the log-mean behaviour of property $a$ to the form,
\begin{align}
s_a = \alpha_a (\mu + e(z))  + \pi_a,
\end{align}

\noindent where $\mu = \ln(M_{\rm halo}/M_p)$, $e(z) = \ln{E(z)}$, and the normalization is in the natural log using units given in Table~\ref{tab:MOR}.  We remind the reader that one of the elements of the observable vector, ${\bf s}_{o}$, is the weak-lensing mass, which is assumed to be an unbiased estimator of true mass with fixed slope $\alpha_{\ln M_{WL}}=1$ and normalization $\pi_{\ln M_{WL}} = 0$.  Since our method constrains the covariance between observables at fixed mass, we use the same independent variable, $\mu + e(z)$, for all properties.  Where this is not the natural independent variable derived in Section \ref{sec:selfsimilar} (i.e. for \Mgas, $L_K$ and \richness) we include an additional factor of $e(z)$ on the dependent axis, as listed in Table \ref{tab:MOR}.

As a check, we also perform the fits with $\mu$ as the independent variable and appropriately modified $e(z)$ factors on the dependent axes. As expected within such a narrow redshift range, the results are consistent.

\subsection{Scaling Relations Parameters}\label{sec:relations}

\begin{figure*}
  \centering
  \includegraphics[width=0.35\linewidth]{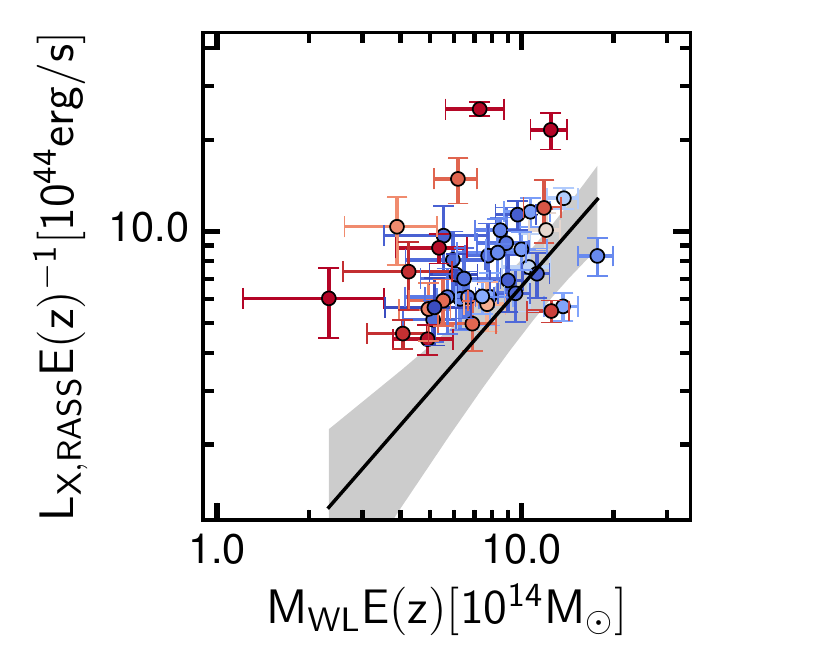}
  \hspace{0.03\linewidth}
  \includegraphics[width=0.35\linewidth]{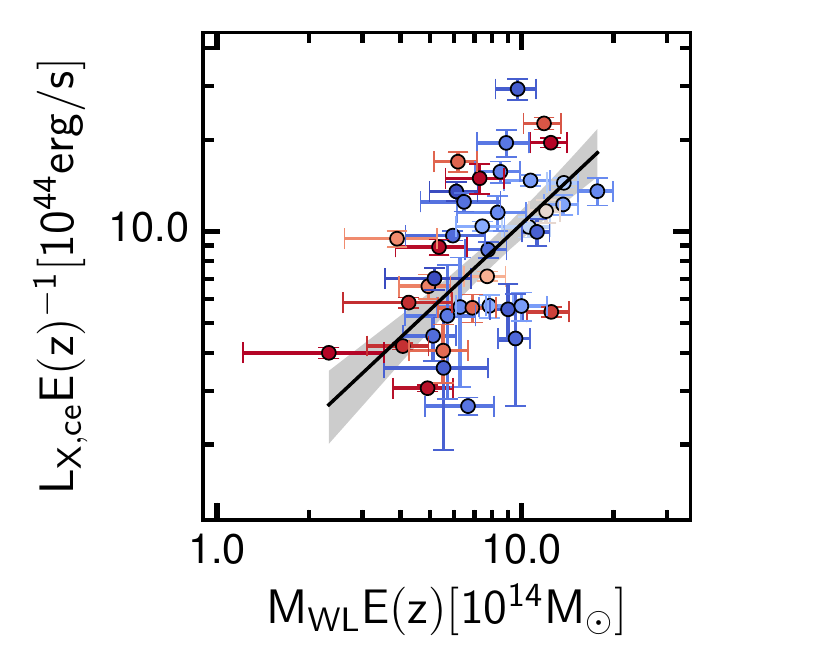}
  \hspace{0.03\linewidth}
  \includegraphics[width=0.35\linewidth]{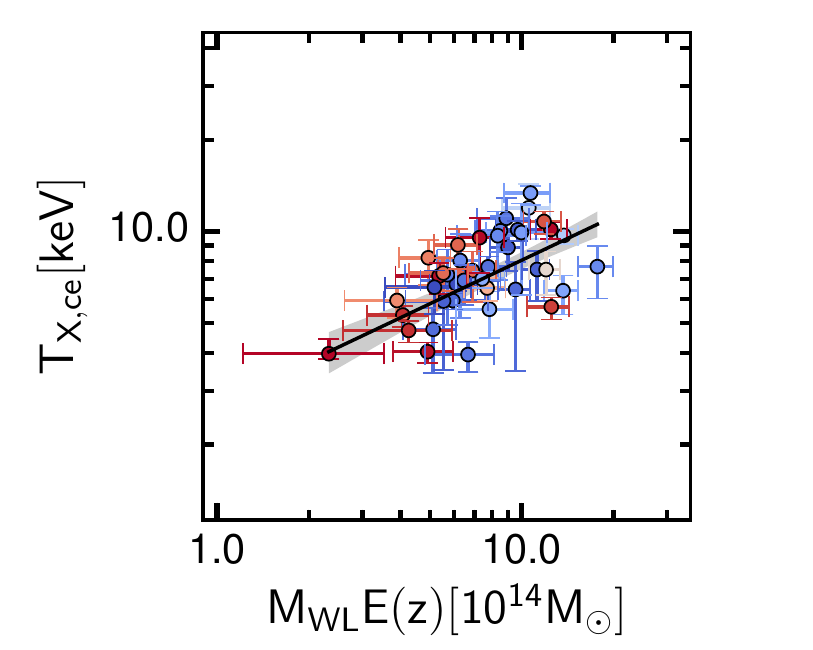}
  \hspace{0.03\linewidth}
  \includegraphics[width=0.35\linewidth]{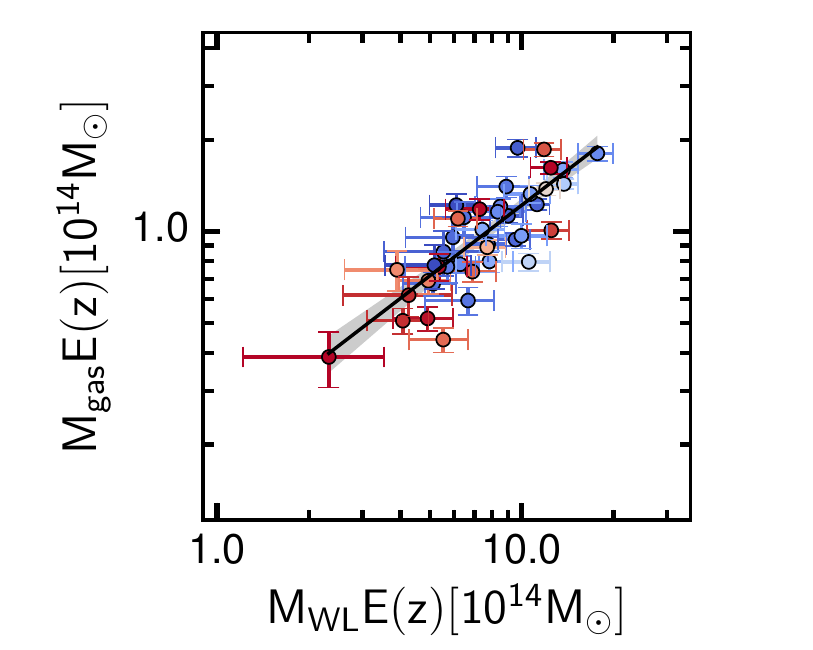}
  \hspace{0.03\linewidth}
  \includegraphics[width=0.35\linewidth]{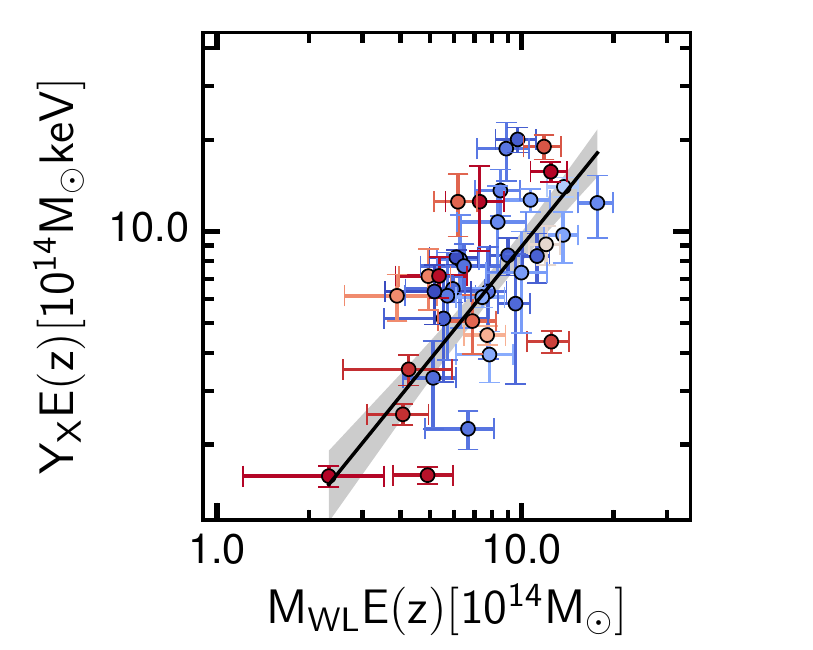}
  \hspace{0.03\linewidth}
  \includegraphics[width=0.35\linewidth]{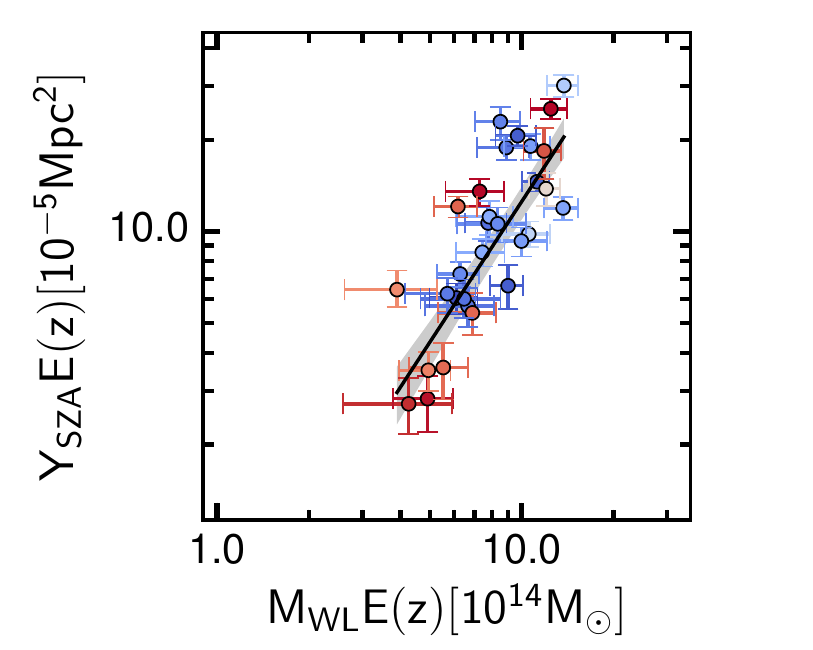}
  \hspace{0.03\linewidth}
  \includegraphics[width=0.35\linewidth]{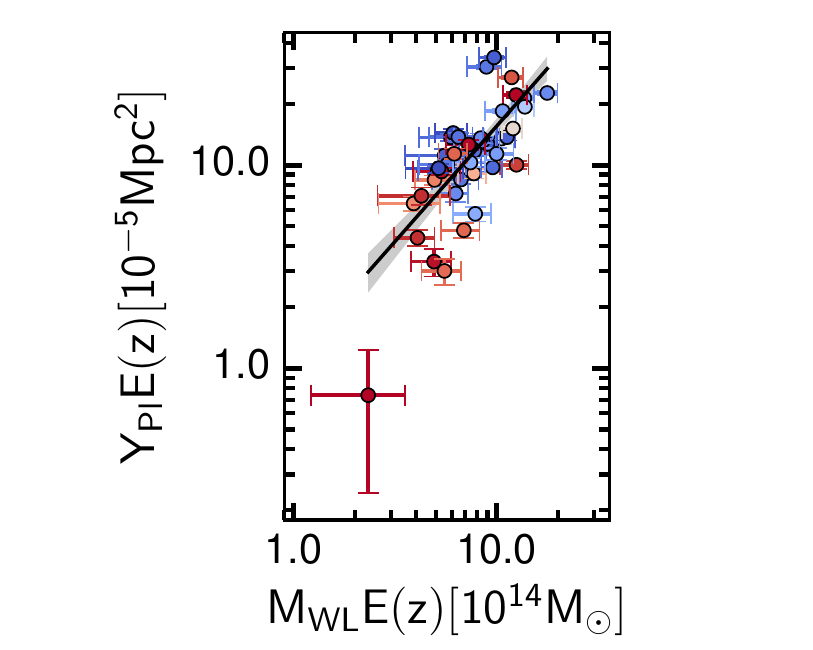}
  \hspace{0.03\linewidth}
  \includegraphics[width=0.35\linewidth]{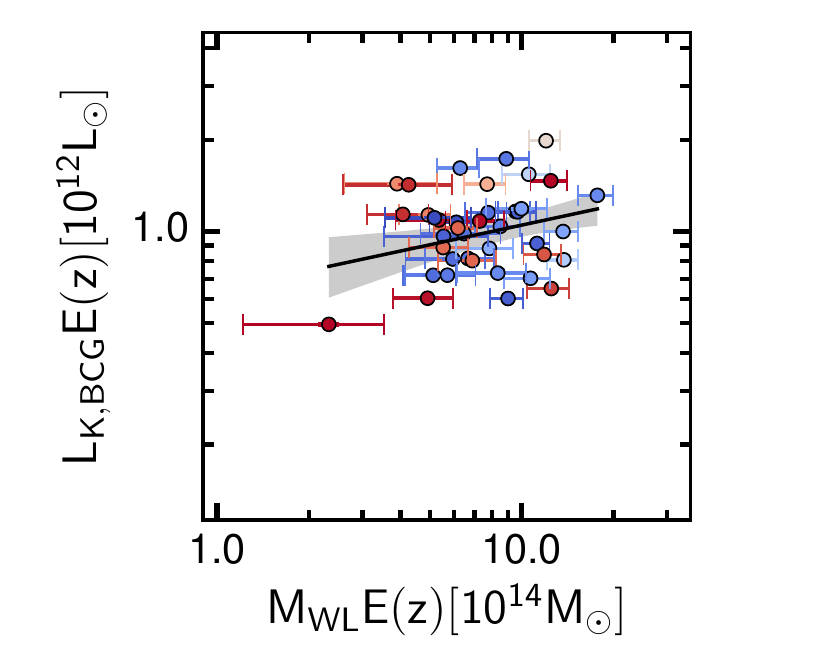}
  \hspace{0.03\linewidth}
  \includegraphics[width=0.35\linewidth]{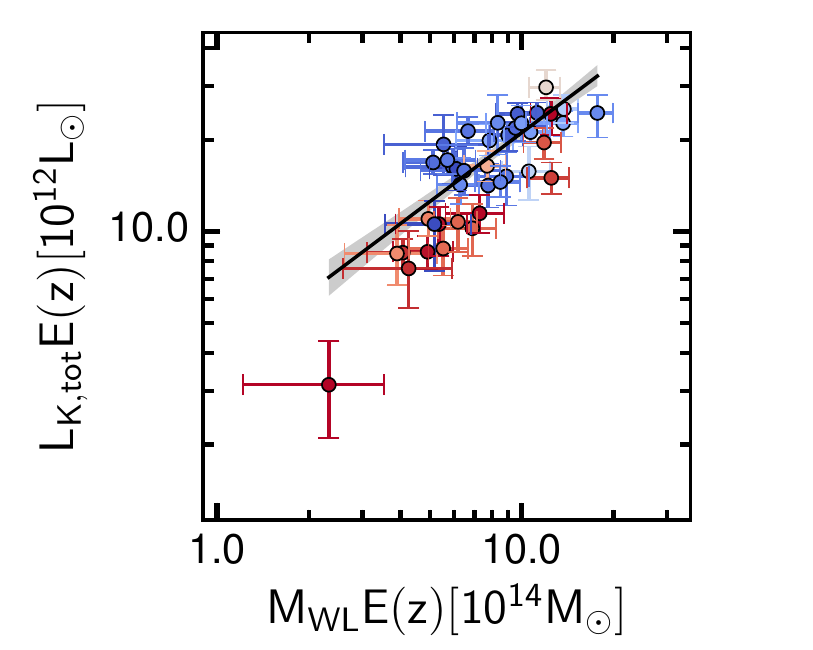}
  \hspace{0.03\linewidth}
  \includegraphics[width=0.35\linewidth]{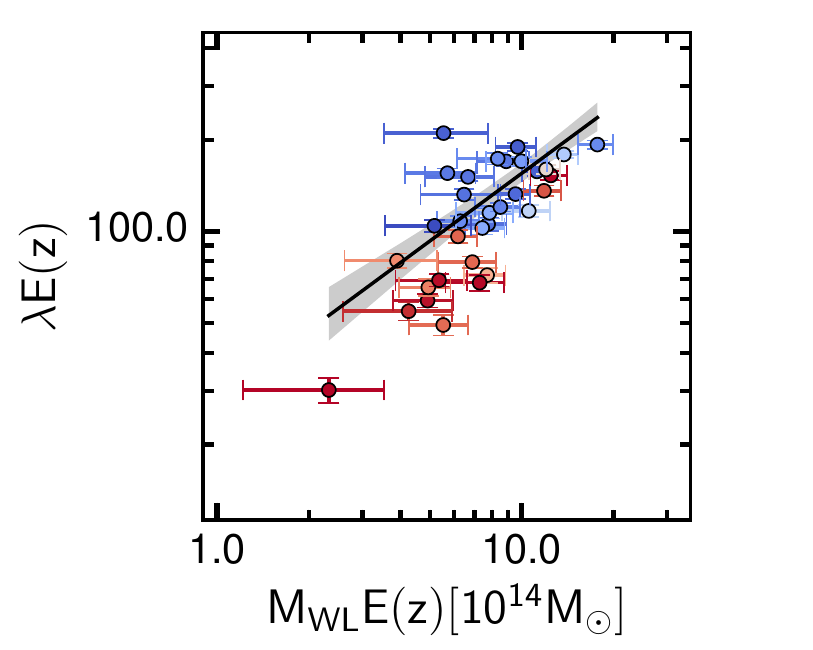}
  \caption{Scaling relations between cluster observable properties and potential well depth, $M_{\rm WL}E(z)$. Individual cluster points with error bars are shown, while the hierarchical Bayesian fits and 68 per cent confidence regions of the mean behaviours are given by solid lines and grey-scales, respectively. The colour scale indicates the central entropy $K(<20\rm kpc)$, with red being lower entropy, cool-core clusters and blue being higher entropy, non cool-core clusters.}
  \label{fig:fits}
\end{figure*}

The resulting posterior estimates of the scaling relation parameters are summarized in Table~\ref{tab:fits}, shown in Fig.~\ref{fig:fits}, and discussed below.  In ensuing subsections, we begin by presenting results for the selection variable, \Lxrass,  then proceed to examine hot gas and stellar scaling behaviours.  Subsequent sections discuss intrinsic property variance and the physical origins of deviations about the mean relations.

\begin{table}
\caption{Scaling relation parameters constrained by our heirarchical Bayesian method. See Table~\ref{tab:MOR} for intercept units.}\label{tab:fits}
	\begin{center}
		\tabcolsep=0.8mm
		\begin{tabular}{ l c c c c }
            \hline
            Observable & Intercept & Slope & Scatter & Self-Similar \\
            $S_a$ & $\exp(\pi_a)$ & $\alpha_a$ & $\sigma_{a | \mu}$ & Slope \\
            \hline
\vspace{3.0truept}
\Lxrass  & $ 4.70 ^{+ 1.65 }_{- 1.28 } $  &  $ 1.15 ^{+ 0.37 }_{- 0.42 } $  &  $ 0.54 ^{+ 0.11 }_{- 0.17 } $ & 1.00 \\ 
\vspace{3.0truept}
\Lx  & $ 8.01 ^{+ 0.85 }_{- 0.81 } $  &  $ 0.94 ^{+ 0.19 }_{- 0.21 } $  &  $ 0.38 ^{+ 0.04 }_{- 0.05 } $ & 1.33 \\ 
\vspace{3.0truept}
\Tx  & $ 6.98 ^{+ 0.46 }_{- 0.43 } $  &  $ 0.47 ^{+ 0.10 }_{- 0.11 } $  &  $ 0.20 ^{+ 0.03 }_{- 0.04 } $ & 0.66 \\ 
\vspace{3.0truept}
\Mgas  & $ 0.97 ^{+ 0.05 }_{- 0.05 } $  &  $ 0.77 ^{+ 0.10 }_{- 0.10 } $  &  $ 0.16 ^{+ 0.03 }_{- 0.03 } $ & 1.00 \\ 
\vspace{3.0truept}
\Yx  & $ 6.18 ^{+ 0.65 }_{- 0.65 } $  &  $ 1.23 ^{+ 0.19 }_{- 0.20 } $  &  $ 0.34 ^{+ 0.05 }_{- 0.05 } $ & 1.66 \\ 
\vspace{3.0truept}
\Ysza  & $ 7.93 ^{+ 1.06 }_{- 0.96 } $  &  $ 1.53 ^{+ 0.20 }_{- 0.22 } $  &  $ 0.31 ^{+ 0.07 }_{- 0.08 } $ & 1.66 \\ 
\vspace{3.0truept}
\Ypl  & $ 11.10 ^{+ 0.92 }_{- 0.93 } $  &  $ 1.14 ^{+ 0.15 }_{- 0.16 } $  &  $ 0.29 ^{+ 0.04 }_{- 0.04 } $ & 1.66 \\ 
\vspace{3.0truept}
\Lbcg  & $ 0.98 ^{+ 0.09 }_{- 0.09 } $  &  $ 0.21 ^{+ 0.15 }_{- 0.16 } $  &  $ 0.34 ^{+ 0.04 }_{- 0.05 } $ & -- \\ 
\vspace{3.0truept}
\Lk  & $ 16.85 ^{+ 0.73 }_{- 0.79 } $  &  $ 0.75 ^{+ 0.10 }_{- 0.10 } $  &  $ <0.16^* $ & 1.00 \\ 
\vspace{3.0truept}
\richness  & $ 124.49 ^{+ 8.49 }_{- 11.25 } $  &  $ 0.74 ^{+ 0.14 }_{- 0.13 } $  &  $ 0.24 ^{+ 0.04 }_{- 0.05 } $ & 1.00 \\ 
            \hline
        \end{tabular}
	\end{center}
{\footnotesize $^*$ The \Lk \ scatter is not bounded from below (see Fig.~\ref{fig:pdf}), so the value quoted is the 95th percentile.}
\end{table}

\subsubsection{Selection variable}

The posterior parameter constraints on the scaling of \Lxrass \ with mass, listed in the first row of Table \ref{tab:fits}, entail large uncertainties that are driven by significant sample incompleteness as a function of mass.  The upper left panel of Fig.~\ref{fig:fits} shows that all but 4 of the 41 clusters lie above the best-fit underlying scaling relation; the selection skims off only the brightest systems as a function of mass.  This behaviour is a textbook example of Malmquist bias \citep{Allen2011,Mantz2016,Giles2017}.

While the inferred slope of $1.15^{+0.37}_{-0.42} $ agrees with the self-similar expectation, the $35$ per cent uncertainty in slope dilutes the impact of this statement.  The intrinsic scatter (in natural log) of $0.54^{+0.11}_{-0.17}$ is higher than the $0.38^{+0.04}_{-0.05}$ seen for the core-excised counterpart \Lx, which we interpret as the consequence of including the core.  We have also performed analysis using \emph{Chandra}/\emph{XMM-Newton} luminosities that include the core, finding an intrinsic scatter of $0.51^{+0.08}_{-0.08}$, consistent with the \Lxrass \ value. 

The relatively large uncertainty in the \Lxrass \ scaling parameters allows only weak estimates of the correlation coefficients between \Lxrass \ luminosity and other cluster properties.  The largest coefficients, with values between 0.4 and 0.6 and uncertainties of roughly 0.2, are with follow-up X-ray measures and \Ysza.  The full set of coefficients, provided in Table~\ref{tab:rlxrass} of Appendix~\ref{sec:bias}, includes hint of an anti-correlation between hot gas mass and stellar mass discussed further in the companion paper \citep{Farahi:inprep}.

\subsubsection{X-ray Observables}
For the X-ray properties (rows 2 through 5 of Table~\ref{tab:fits}), posterior constraints on the slopes of the scaling relations are consistently shallower than self-similar model expectations at the $\sim$1$-$2$\sigma$ level, with uncertainties ranging from 0.1 (\Mgas \ and \Tx) to 0.2 (\Lx \ and \Yx).  The shallow behaviour for \Mgas \ is unexpected, as previous studies covering a wider dynamic range in cluster mass have found that mean gas mass increases with halo mass in a super-linear fashion, \Mgas \ $\propto M^{1.2}$ \citep[e.g.][]{Pratt2009}. However, as discussed below, the slope we find is only in $\sim$1.5$\sigma$ tension with the Weighing the Giants study of \citet{Mantz_wtg}, who find a slope of $1.004 \pm 0.014$ for a high-mass sample of clusters.  
A trend toward self-similar behaviour in the highest halo masses is seen in recent hydrodynamical simulations that include AGN heating \citep{Farahi2018}.

We highlight that there is a degeneracy between the posterior slope of a property and the covariance between that property and the selection variable, \Lxrass.  Physically, we expect a positive correlation between \Mgas \ and \Lxrass \ residuals, but find the correlation coefficient to be only $0.24^{+0.21}_{-0.24}$.  If this value were constrained higher, the slope of the \Mgas \ relation would also increase. To demonstrate this, we perform the analysis with a uniform prior between 0.7 and 1 on this correlation coefficient, finding the slope of the \Mgas \ relation increases $\sim$1.5$\sigma$, from $0.77^{+0.10}_{-0.10}$ to $0.90^{+ 0.11}_{-0.11}$, consistent with both the self-similar prediction and the Weighing the Giants result.

\subsubsection{SZ Observables}\label{sec:szresults}

We find that the slopes of the two SZ-derived $Y$ relations are consistent with each other, with \Ysza \ being steeper than \Ypl \ at the level of $1.5 \sigma$.  \Ysza \ is within $1 \sigma$ of the self-similar slope of $5/3$, and the two SZ values bracket the \Yx \ slope of $1.23^{+0.19}_{-0.20}$.

Regarding normalization, the cylindrical measurement of \Ypl \ can be converted to a spherical estimate by dividing by a factor $Y_{\rm cyl}/Y_{\rm sph} = 1.2$ \citep{Arnaud2010}. When we apply this conversion factor to the \Ypl \ intercept, the resulting value of $9.25^{+0.77}_{-0.78}$ is consistent with the \Ysza \ value of $7.93^{+1.06}_{-0.96}$. To compare to the X-ray normalization, we follow \citet{Arnaud2010} and apply a conversion factor,
\begin{align}
C_{\rm XSZ} = \frac{\sigma_T}{m_ec^2} \frac{1}{\mu_em_p} = 1.416 \times 10^{-19} \frac{\rm Mpc^2}{\rm M_\odot keV},
\end{align}
\noindent giving a $Y_X$ intercept of $8.75^{+0.92}_{-0.92}$.  To summarize, we find good agreement between the normalizations of all three relations that measure the electron thermal energy content.

While the \Ysza \ slope is in agreement with the self-similar relation, the \Ypl \ slope is shallower. The \Ypl \ measurement errors for the low mass clusters are large, so they don't have a strong influence on the fit. The fit parameters are largely constrained by the intermediate and high mass clusters, and an increase in the \Ypl \ measurement of intermediate mass clusters would act to shallow the fitted slope. Indeed we find the highest ratios of \Ypl \ to \Ysza \ in low and intermediate mass clusters.

We note that the \Ysza \ relation is constrained using a subsample of 33 clusters, due mostly to contamination as detailed in Section \ref{sec:sza}. If there was correlation between cluster mass and the extended sources that lead to contamination, this could lead to a bias in the constrained relation. We refit all scaling relations using only this subsample of 33 systems, finding the results largely consistent within errors.

\subsubsection{Stellar Observables}\label{sec:stellar}

The measures of galactic stellar content, $L_{K{,\rm BCG}}$, \Lk \ and $\lambda$, provide complementary insights into the star formation history of high mass halos. Both \Lk \ and \richness \ attempt to measure the total stellar content of a cluster, but they differ in detail. The total K-band luminosity, \Lk, is a background-corrected estimate that uses all member galaxies within the weak-lensing estimate of $r_{500}$, whereas \richness \ is a red-sequence weighted estimate determined within an aperture scaling as $\lambda^{0.2}$. The former is luminosity-weighted while the latter is number-weighted. We highlight that any interpretation of the stellar content derived from these galaxy observable scaling relations relies on the assumption that they are reliable tracers of the stellar mass. This is likely sensitive to the details of the measurement, and determining the best stellar mass estimate would require further study.

Despite their differences, the slopes of the \Lk \ and \richness \ scaling relations are consistent, and in both cases shallower than the self-similar prediction. As both measures scale with total stellar mass, this is consistent with a stellar fraction that decreases with increasing halo mass, implying that star-forming efficiency is a decreasing function of halo mass \citep{Gonzalez2007,Lagana2011}.  
This result is supported by abundance matching arguments \citep{Behroozi2013,Kravtsov2013}, and AGN-based feedback scenarios in cosmological hydrodynamics models are tuned to produce this feature \citep{Croton2006, Delucia2007, Planelles2013, Pillepich2018, Farahi2018}.
Both weak-lensing \citep{Simet2017} and ensemble spectroscopic \citep{Farahi2016} mass estimate methods find mean mass scaling behaviour, $M \propto \lambda^{1.3}$, consistent with our findings. 

The close agreement in the \Lk \ and \richness \ slope values may be somewhat fortuitous. The radius within which \richness \ is measured scales more slowly ($\lambda^{0.20}$) than the halo radius implied from its scaling with weak-lensing mass ($\lambda^{0.45}$), within which \Lk  \ is measured. While this could potentially lead to proportionally smaller increases in \richness \ compared to \Lk \ as halo mass increases, a secondary factor such as a decreasing star forming fraction in higher mass halos may compensate for the scale mismatch effect. We note that the correlation coefficient between \Lk \ and \richness \ at fixed \Mwl, presented in \citet{Farahi:inprep}, is near unity: $ 0.77 ^{+0.16 }_{-0.27 }$.

The \Lbcg \ scaling relation is very shallow, almost consistent with zero, demonstrating that the luminosity of the BCG is not a strong function of mass for clusters in this mass range. As halo mass increases, so does the galaxy velocity dispersion, and accretion onto the BCG slows relative to the total mass growth of the cluster. As these two processes are largely uncoupled it leads to large scatter in the relation, consistent with our finding that the \Lbcg \ relation has a larger intrinsic scatter than the \Lk \ relation.

The normalizations of the BCG and total $L_K$ relations provide a simple estimate of the fraction of stellar mass associated with the BCG.  We find a value of $5.8 \pm 0.5$ per cent, with the uncertainty dominated by the error in the BCG normalization. A comparison to the literature is difficult to do homogeneously, as the precise values will rely on the method used for BCG and intracluster light separation, as well as background subtraction.

\citet{Ziparo2016} applied very similar methods to ours to a sample of clusters from the XXL survey with weak-lensing masses between $10^{14}$ and $10^{15}$, finding $L_{K,\rm BCG}/L_{K,500}$ between 3.5 and 20 per cent. Using slightly different methodology but again finding consistent results, \citet{Lin2004b} found $L_{K,\rm BCG}/L_{K,200}$ ranged from 3 to $\sim$18 per cent, again for clusters with masses similar to our sample. These values, calculated using $L_{K,200}$, provide a lower limit on $L_{K,\rm BCG}/L_{K,500}$.

Halo occupation distribution models also enable calculation of the BCG/total stellar fraction. For instance, \citet{Leauthaud2012} use lensing, clustering and stellar masses to parameterise the occupation of halos. Although these models are often driven by galaxies halos with masses less than clusters, the parameterisation do allow calculations at all masses. In the lowest redshift bin ($z \sim 0.3$), \citeauthor{Leauthaud2012} found that halos with masses greater than $10^{14}$ had BCG/total stellar fraction below 10 per cent.

\begin{figure}
  \centering
  \includegraphics[width=\linewidth]{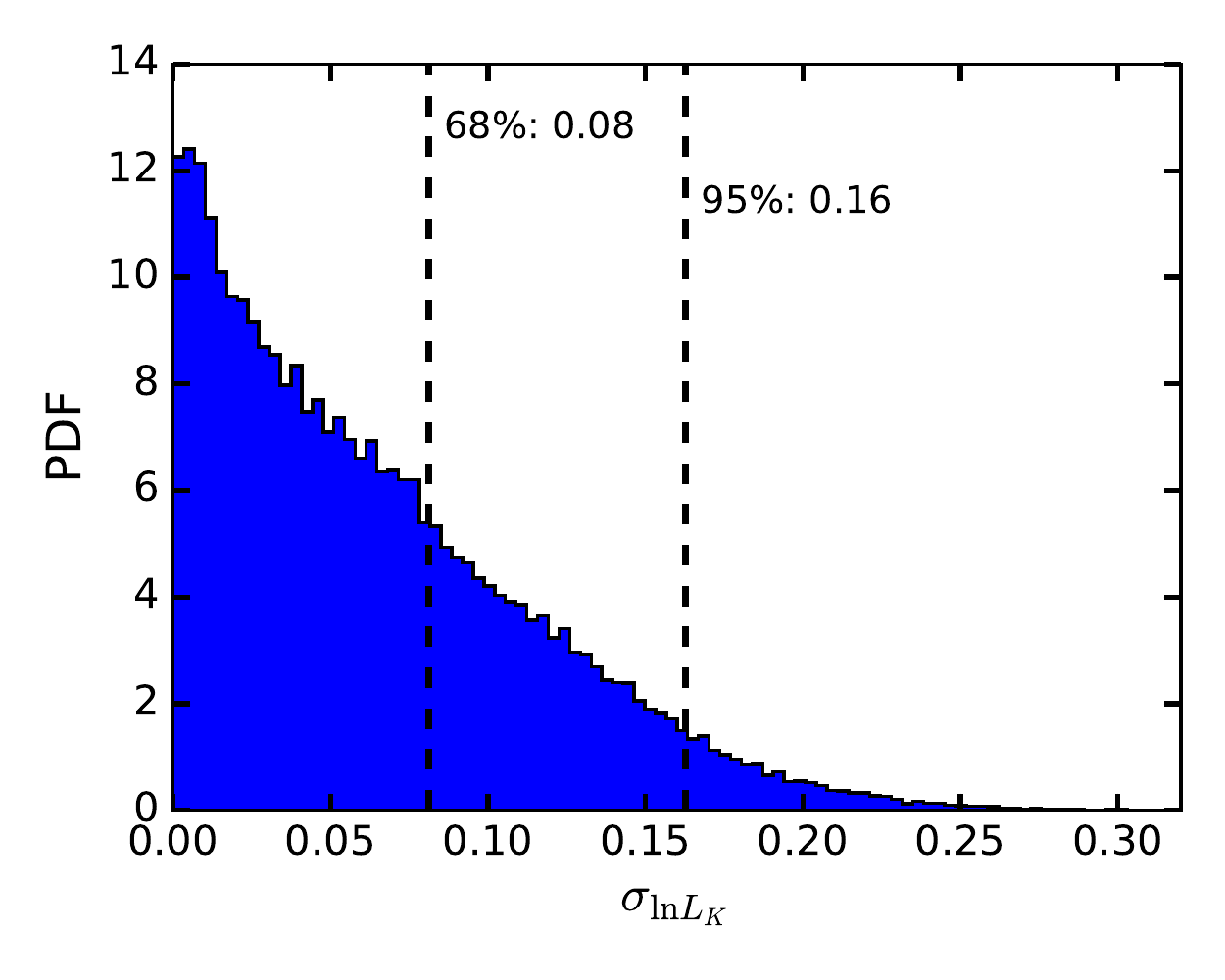}
  \caption{Posterior PDF of the scatter in total K-band luminosity, $\sigma_{\ln L_K}$, with the 68th and 95th percentiles indicated.}
  \label{fig:pdf}
\end{figure}

\subsection{Intrinsic Variance}\label{sec:intrinsicVariance}

Knowledge of the intrinsic variance in cluster properties is important for precise cosmological studies with the population, but empirical estimates of the full covariance matrix, including both on-diagonal scatter and off-diagonal pair correlations have only recently begun to emerge  \citep{Okabe2010,Maughan2014,Mantz2016}. 

Caution is required when estimating the covariance of sample properties, as the statistical (measurement) errors must be accurately determined and the selection model must be correctly described.  Considerable interest lies in the intrinsic scatter  of an individual property, $\sigma_{a|\mu}$, and its related scatter in halo mass. 

The effect of including sample selection has a significant effect on the posterior intrinsic scatter estimates.  The `naive' regression model (see Appendix \ref{sec:bias}) produces scatter estimates that differ significantly from Table~\ref{tab:fits} for several X-ray properties, including the \Lxrass \ selection variable. Note, however, that the intrinsic scatter constraints on \Mgas \ and \Tx, as well as all of the SZ and optical properties, are consistent between the two treatments.  

Since the model that includes selection effects should be closer to unbiased, we employ the values in Table~\ref{tab:fits} as our primary results, with a note of caution that posterior scatter constraints for \Lx \ and \Yx \ appear to be most sensitive to the selection model.  

Reviewing the intrinsic scatter values, we note that \Mgas \ and \Lk \ have the lowest values, while the \Lxrass \ selection variable is highest.  The posterior in \Lk \ scatter has no finite lower bound.  As shown in Fig. \ref{fig:pdf}, the posterior probability distribution function (PDF) of the intrinsic scatter in the \Lk \ relation is not well fit by a Gaussian, so we quote 68th and 95th percentiles of 0.08 and 0.16 respectively. The 95th percentile is below the central value of the intrinsic scatter in the \richness \ relation, $0.24^{+0.04}_{-0.05}$. We note that the definition of membership for the two observables is different and therefore recalculate \Lk \ using membership as determined in the \richness \ calculation, finding the result unchanged. We interpret this as an indication that \Lk, as a tracer of the stellar mass, is a slightly better proxy for cluster mass than the richness.

We find good agreement between the intrinsic scatter of $\sim$0.3 for all three $Y$ relations.  

From Table~\ref{tab:fits} we can estimate the mass proxy power using the inferred scatter in mass $\sigma_{\mu |a} = \sigma_{a|\mu}/\alpha_a$. BCG K-band luminosity is by far the least effective, with a wide scatter of $1.6$ in logarithmic mass.  Total K-band light, on the other hand, is much more tightly correlated, with an upper limit of $\sim$20 per cent. Gas mass provides $\sim$0.20$\pm$0.05 fractional accuracy in mass, similar to all measures of $Y$.  We find no evidence that $Y$ is the lowest scatter mass proxy.  We stress that these estimates are with respect to the weak-lensing mass values, and the inference with respect to true mass is dependent on our simplifying assumptions discussed in Section~\ref{sec:regression}. Larger homogeneous samples of the type used here are needed to provide more accurate estimates of the intrinsic property covariance. 

\subsection{Origin of Scatter}\label{sec:scatter}

\begin{figure*}
  \centering
  \includegraphics[width=0.4\linewidth]{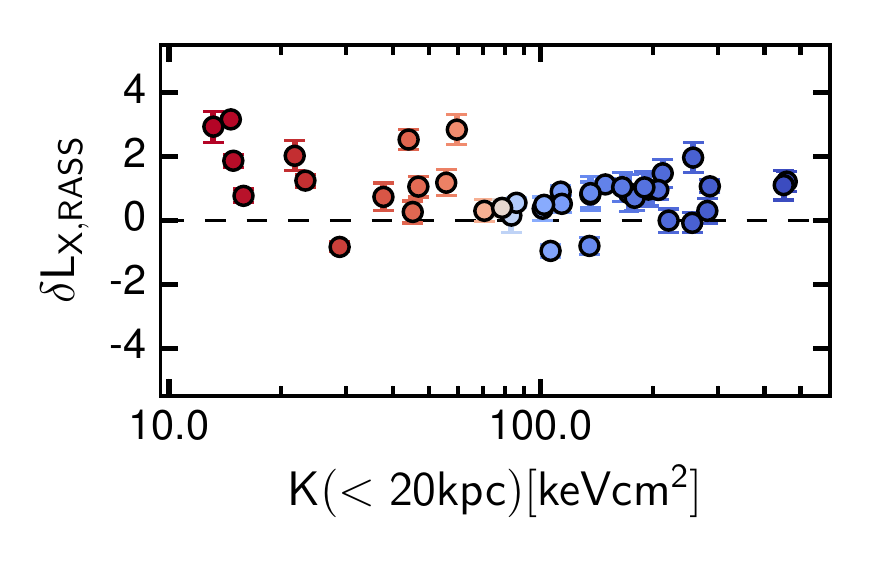}
  \hspace{0.05\linewidth}
  \includegraphics[width=0.4\linewidth]{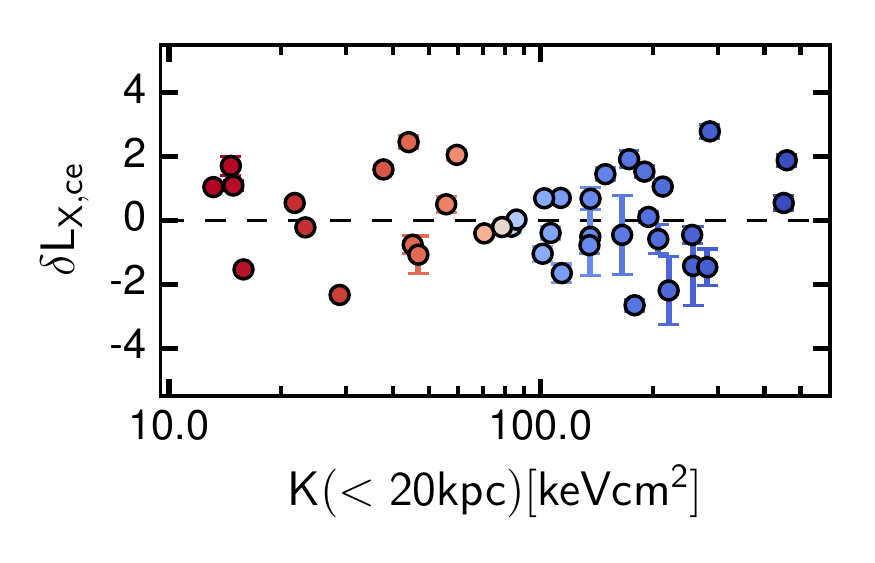}
  \includegraphics[width=0.4\linewidth]{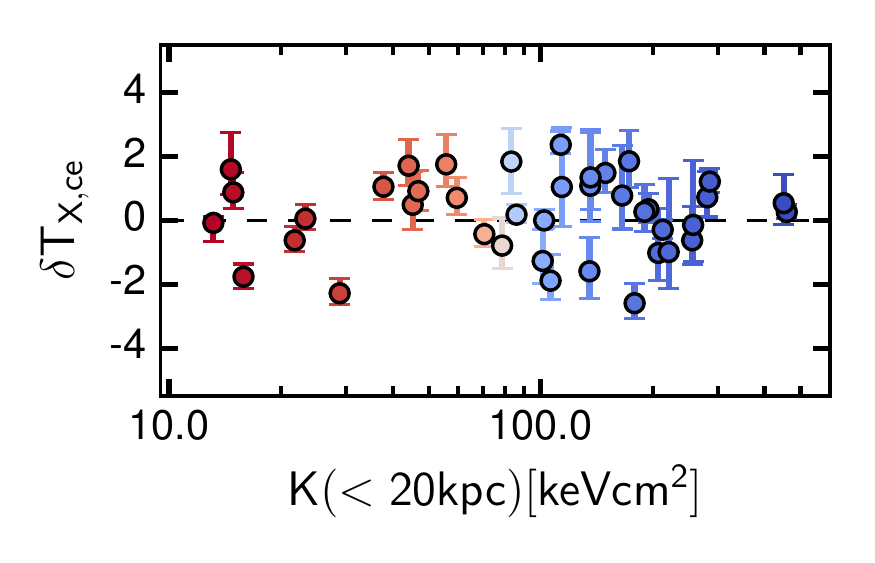}
  \hspace{0.05\linewidth}
  \includegraphics[width=0.4\linewidth]{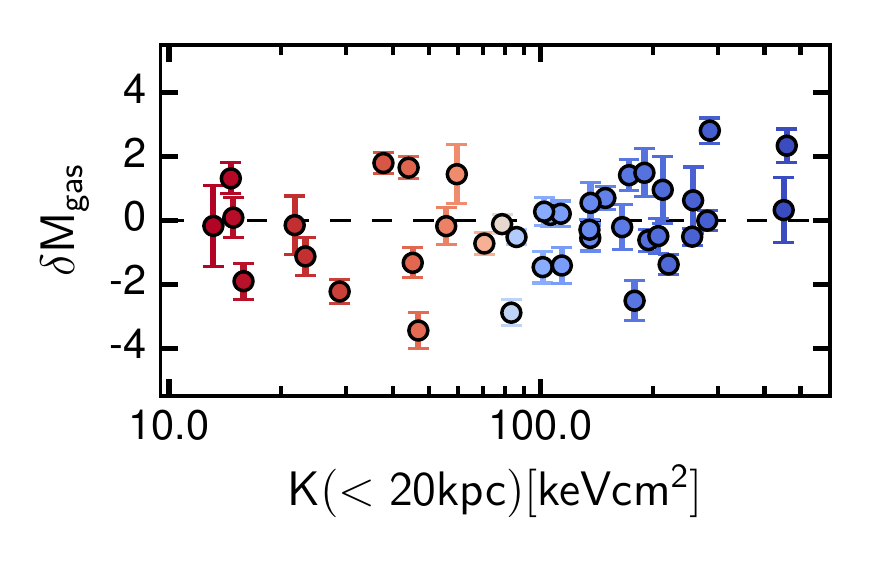}
  \includegraphics[width=0.4\linewidth]{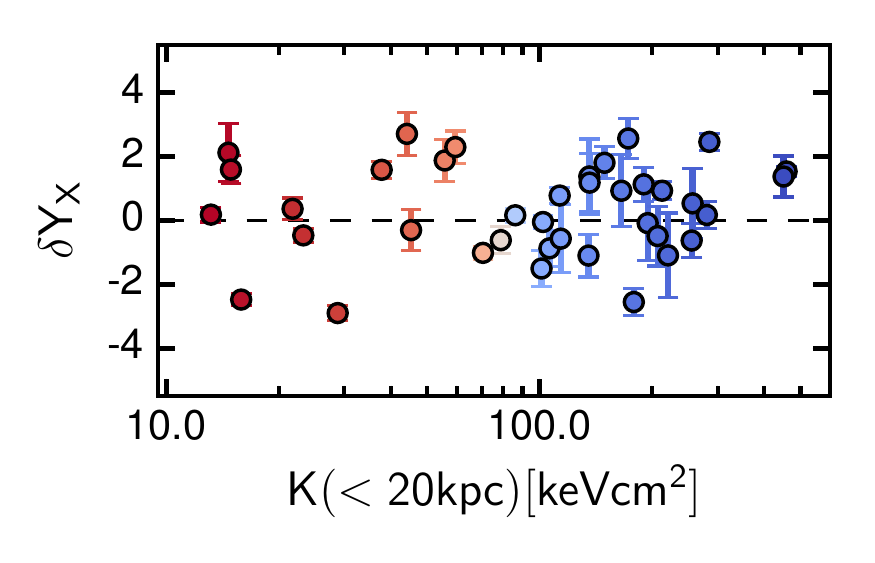}
  \hspace{0.05\linewidth}
  \includegraphics[width=0.4\linewidth]{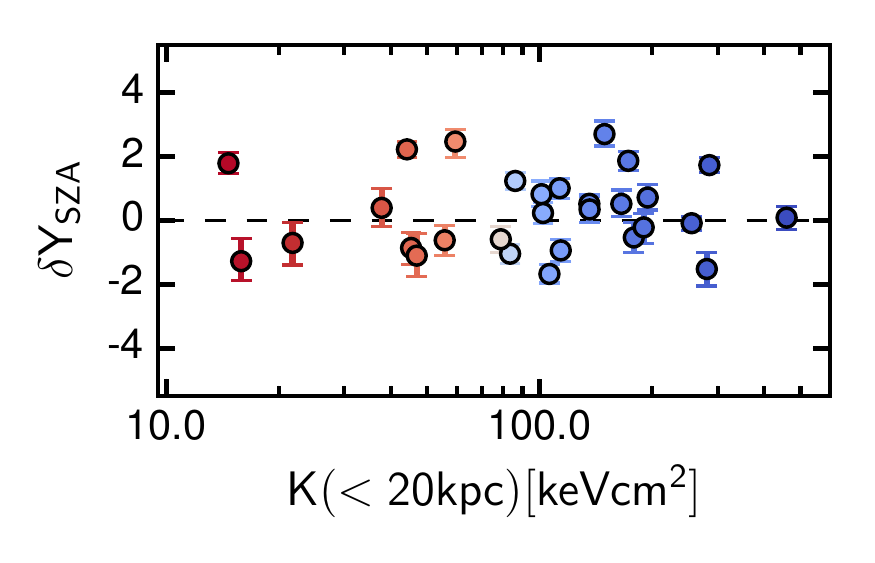}
  \includegraphics[width=0.4\linewidth]{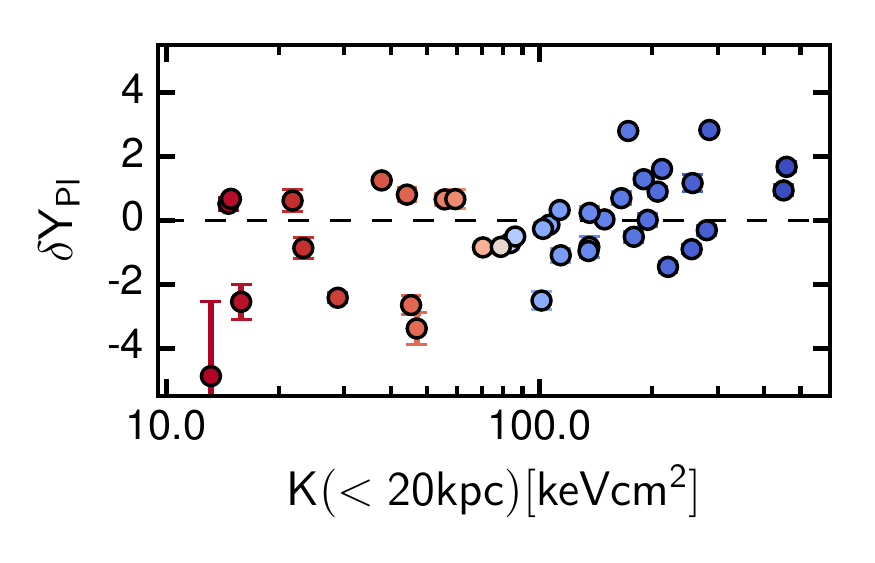}
  \hspace{0.05\linewidth}
  \includegraphics[width=0.4\linewidth]{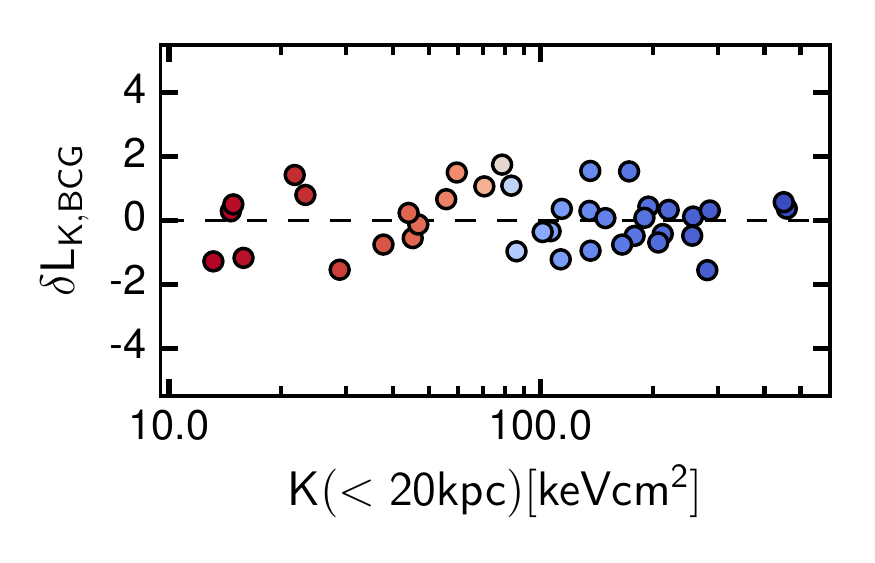}
  \includegraphics[width=0.4\linewidth]{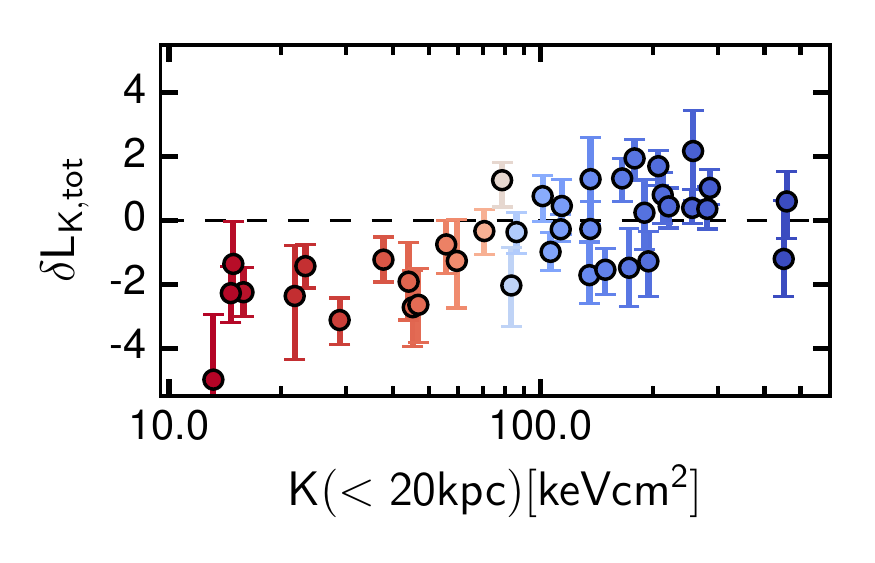}
  \hspace{0.05\linewidth}
  \includegraphics[width=0.4\linewidth]{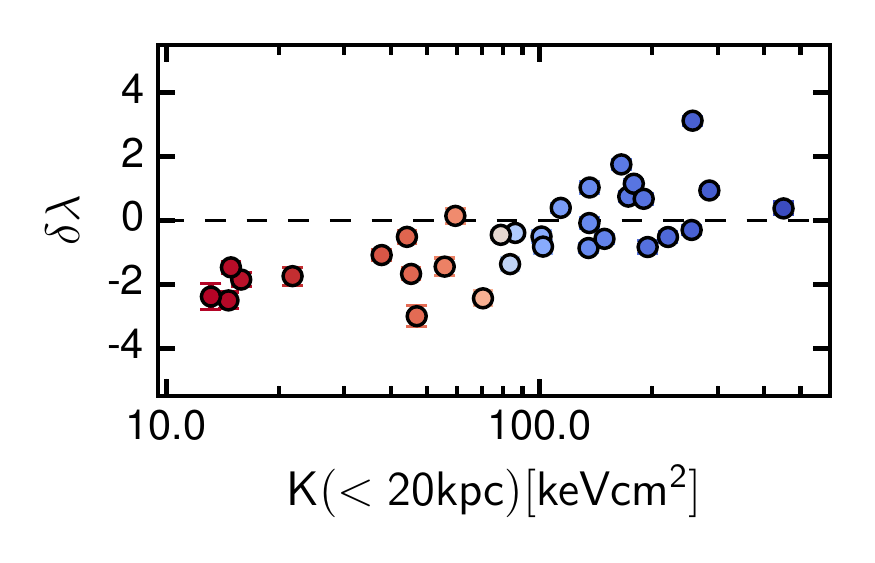}
  \caption{Normalized residuals from each scaling relation, defined in equation~(\ref{eq:residual}), as a function of entropy in the central $20\rm kpc$ of the cluster. Colours indicate $K(<20\rm kpc)$, as in Fig.~\ref{fig:fits}.}
  \label{fig:residuals}
\end{figure*}

To motivate exploration of potential physical origins of the scatter in the scaling relations, in Fig.~\ref{fig:residuals} we compare the residuals in each property with the central entropies of the clusters. The central entropy, $K(<20 \rm kpc)$, measured in the inner $20 \rm kpc$ \citep{Sanderson2009} is an indicator of the formation history of the cluster, with a lower entropy suggesting a less disturbed cluster with a cool core, and thus earlier formation epoch and/or less rich recent merger history \citep{Rasia2015,Hahn2017}.

In Appendix \ref{sec:indicators} we consider multiple other indicators of the level of disturbance in the cluster -- central surface brightness, centroid shift, BCG/centroid separation and magnitude gap -- finding results consistent with those of the central entropy described below.

We define the residual, $\delta a_i$, in property $a$ as the vertical distance in logarithmic space between the $i^{\rm th}$ cluster's measurement and the posterior mean scaling relation, normalized by the intrinsic scatter of that relation:
\begin{equation}\label{eq:residual}
\delta a_i = \frac{ s_{a,i} - (\hat{\pi}_a + \hat{\alpha}_a \mu_i)}{\hat{\sigma}_a} ,
\end{equation} 
where the hatted quantities are the posterior central estimates of the scaling law parameters for property $a$, and $\mu_i$ is the weak-lensing mass of the $i^{\rm th}$ cluster. We use the 95th percentile of 0.16 for $\sigma_{L_K}$.  We highlight that the residuals from a given scaling relation don't necessarily average to zero, due to sampling biases introduced by the selection model. This effect is strongest in the \Lxrass \ selection variable, but translates to other observables through non-zero covariance.

The \Lxrass \ measurement contains the core, which will contribute more to the signal for clusters with cool cores than those without. We therefore expect large positive residuals in the low entropy clusters, as we see clearly in the top left panel of Fig.~\ref{fig:residuals}. In the \emph{Chandra}/\emph{XMM-Newton} X-ray observables, we see no clear trend in the residuals with cluster entropy.

While we find no trend in \Ysza \ (or \Yx) residuals, we do find a trend in \Ypl \ of more positive (negative) residuals in higher (lower) entropy clusters. This could suggest that a fixed \citet{Arnaud2010} pressure profile performs less well in non cool-core clusters, as a boosted signal in the outskirts would increase the \Ypl \ measurement and produce a positive residual. This interpretation is supported by the results in Appendix \ref{sec:indicators}, where we find the same trend in indicators sensitive to the gas morphology.

The clearest trends we find in Fig.~\ref{fig:residuals} are in the lower two panels, showing residuals of the total cluster optical observables -- \Lk \ and \richness \ -- with more positive (negative) deviations in higher (lower) entropy clusters. This trend is reproduced in most structural indicators in Appendix \ref{sec:indicators}. The trend is also seen clearly in the two lower panels of Fig.~\ref{fig:fits} and discussed further in Section \ref{sec:residuals}.

\subsection{Posterior Distribution on True Halo Mass}\label{sec:massest}

\begin{figure*}
  \centering
  \includegraphics[width=0.66\linewidth]{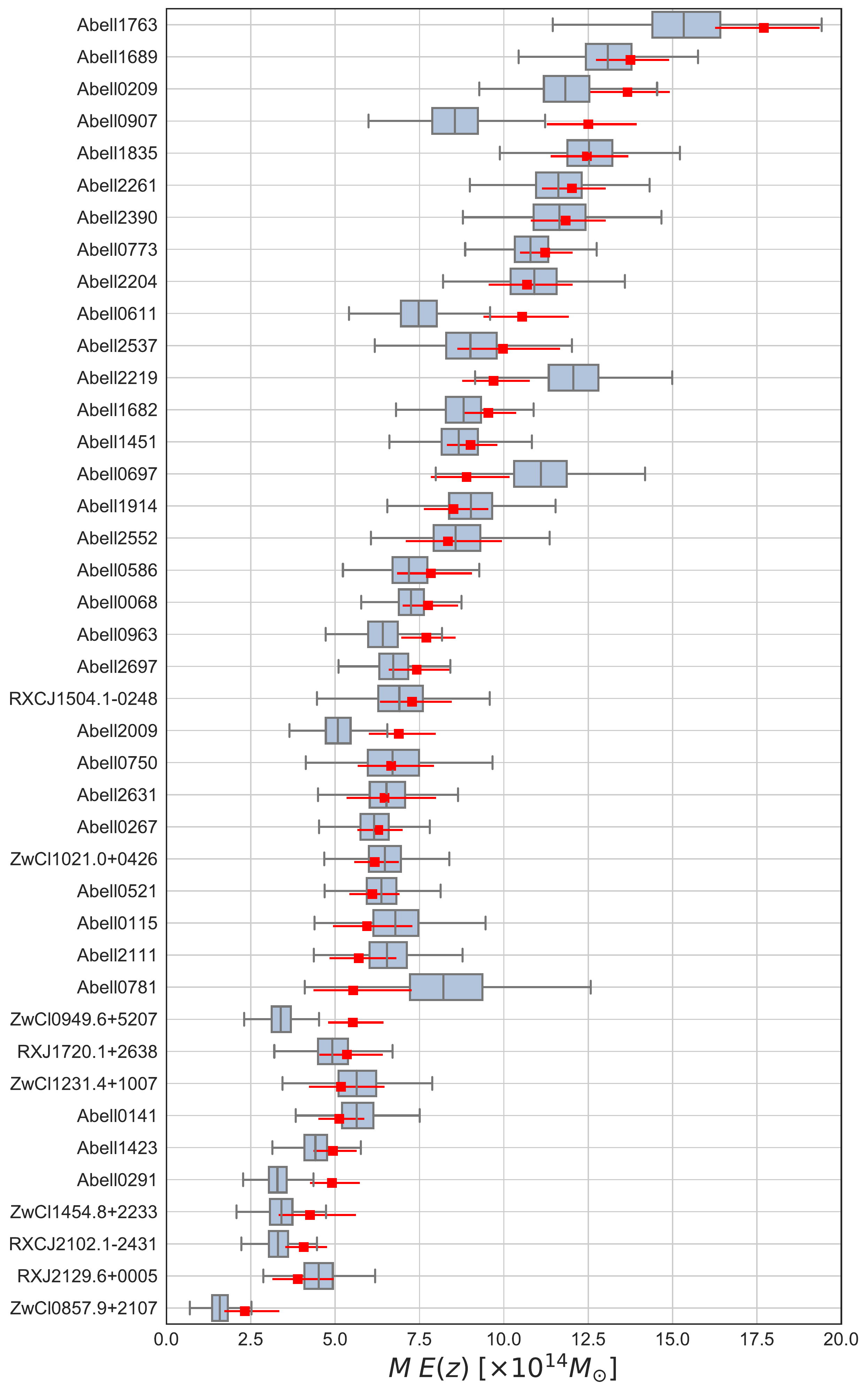}
  \caption{The posterior constraints on true halo mass from the hierarchical Bayesian fit in grey, alongside the measured weak-lensing cluster masses in red. The grey box plots and whiskers show the [25$-$75] and [0.3$-$99.7] percentile ranges, respectively, while the errors on the red points show the 25th and 75th percentiles according to the measurement errors on the weak-lensing measurements. The data points are ordered by weak-lensing mass.}
  \label{fig:posterior_mass}
\end{figure*}

Our model fits for the cluster halo mass, and so generates a posterior distribution for the true mass of each cluster. We report these posterior constraints in the final column of Table \ref{tab:sample}, and display them next to our weak-lensing mass estimates in Fig.~\ref{fig:posterior_mass}. Any differences are due to a combination of two effects -- the mass function favouring low mass systems, and the scaling relations favouring systems that lie near the expectation value. The latter effect can be seen by considering Fig.~\ref{fig:posterior_mass} alongside Fig.~\ref{fig:all_residuals}. Clusters with negative residuals from the scaling relations tend to have posterior masses smaller than their weak-lensing masses (e.g. Abell0907 and Abell0291), while those with positive residuals have the opposite (e.g. Abell2219 and Abell0697).

\section{Discussion}\label{sec:disc}

\subsection{Scaling Relations in the Literature}

To obtain robust scaling relations requires an unbiased measurement of the true mass, an understanding of and correction for the selection of the sample, and a method which allows for the covariance between the selection variable and the observable property. Mainly due to the paucity of high signal-to-noise, uniform, multiwavelength data for well-defined cluster samples, the number of studies in the literature which meet all of these criteria is small. We will largely restrict ourselves to these studies for comparison.

The most similar study to our own is that of \citet{Mantz_wtg}, who use weak-lensing measurements and gas mass as estimators of the true mass, and attempt to model the selection of their clusters. For the ICM properties, they also allow for the covariance of those properties with the selection variable. Their sample includes 27 clusters with weak-lensing masses and a larger sample with gas mass measurements, and span a slightly wider redshift range than ours. In mild conflict with our results, \citeauthor{Mantz_wtg} report that the core-excised gas temperature and the gas mass agree with the self-similar predictions. They find a \Tx \ relation slope of 0.62$\pm$0.04, consistent with the self-similar expectation of 2/3 but only $\sim$1$\sigma$ discrepant with our estimate of $0.47^{+0.10}_{-0.11}$. Their estimate of the \Mgas \ relation slope is 1.007$\pm$0.012, in agreement with unity and again marginally consistent with our estimate of $0.77^{+0.10}_{-0.10}$. It is unclear what causes the differences in our results, however given our method, selection and data analysis are all different from \citeauthor{Mantz_wtg}, a difference of this magnitude is not unexpected.

Similar to our results, \citeauthor{Mantz_wtg} also find that the soft-band X-ray luminosity is steeper than the self-similar expectation, and suggest that this is due to non-gravitational heating and cooling processes in cluster cores.

Our study is the first to look at the simultaneous scaling of X-ray, SZ and optical properties, and so there are few results to compare to the SZ and optical properties. \citeauthor{Mantz_wtg} provide an empirical scaling (without modelling the covariance and correcting for sample selection) and find a shallower $Y_{\rm SZ}$ slope than self-similarity would predict (1.31 $\pm$ 0.03). Note that this measurement is using \Mgas \ as the mass parameter, but \citeauthor{Mantz_wtg} find a one-to-one relation between \Mgas \ and \Mwl. This result is bracketed by our \Ypl \ and \Ysza \ slopes.

Although not corrected for selection effects, studies have placed constraints on the optical scaling relations of \Lk \ \citep[e.g.][]{Lin2003,Lin2004a,Mulroy2014,Mulroy2017} and \richness \ \citep[e.g.][]{Rykoff2012,Mantz_wtg,Simet2017,Melchior2017}, finding the slopes to be shallower than the self-similar predictions, consistent with our results.  

Results from recent numerical simulations indicate that AGN heating produces departures from self-similar scaling relations. Several independent groups find that galactic physics with AGN feedback steepens the ICM scaling relations \citep{Planelles2013,LeBrun2017,Hahn2017,Pillepich2018}, in moderate tension with our X-ray findings.  The overall star formation efficiency declines with increasing halo mass in these simulations, producing stellar mass scaling relations that are sub-linear with M, in agreement with the LoCuSS behaviour. We caution that a concern when making sample comparisons is the possibility that the scaling relation slopes run with halo mass and, to a lesser extent, redshift \citep{Farahi2018}.

\subsection{Cluster Residuals}\label{sec:residuals}

In this section we consider the trends observed in Section \ref{sec:scatter} in more detail.

\begin{figure*}
  \centering
  \includegraphics[width=0.8\linewidth]{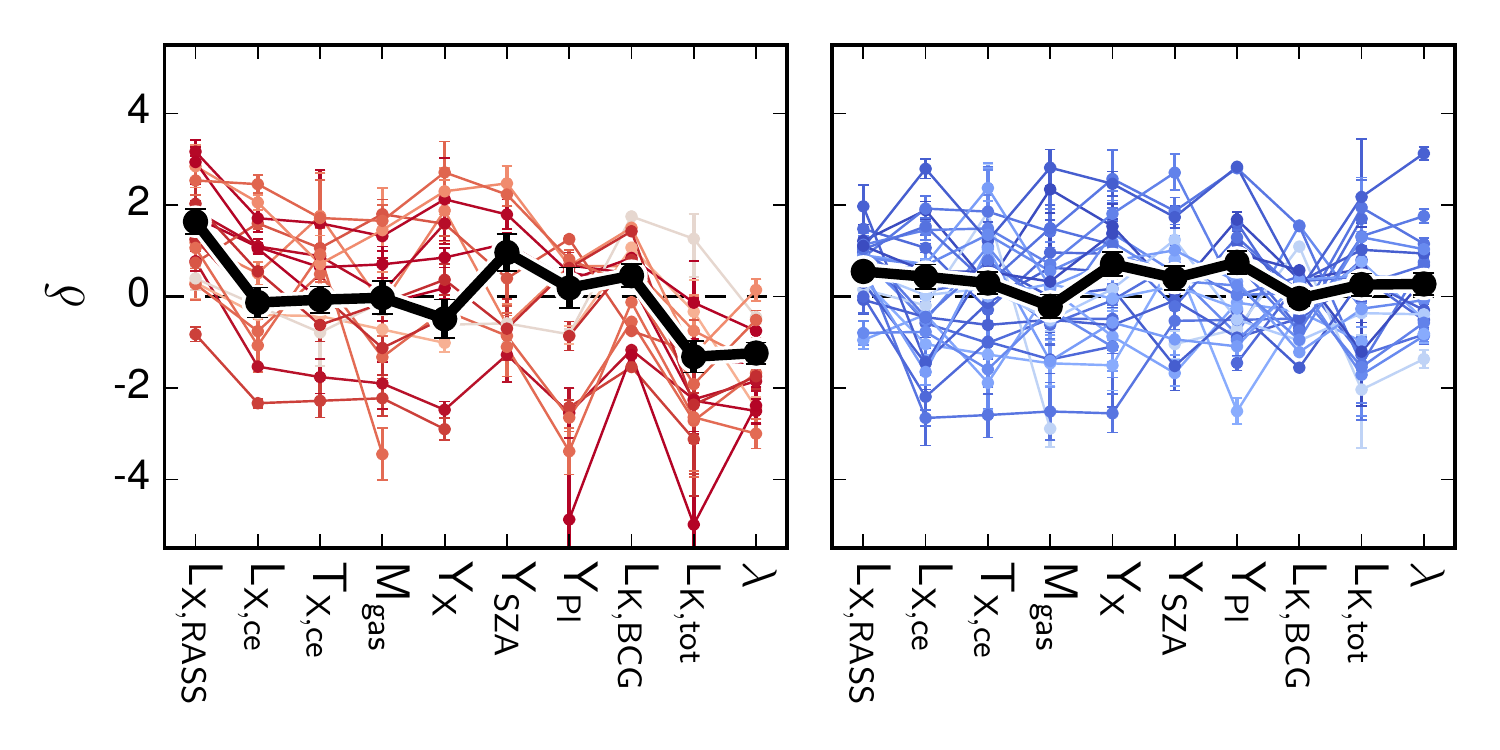}
  \caption{Normalized residuals from the scaling relations, defined in equation~(\ref{eq:residual}), for low-entropy ($K<80$, left) and high-entropy ($K>80$, right) subsamples, with $K(<20\rm kpc)$ determined by \citet{Sanderson2009_bimodality} in units of keV-cm$^2$.  Colours indicate $K(<20\rm kpc)$, as in Fig.~\ref{fig:fits}. Ensemble average values are shown in black, inversely weighted by the square of the measurement uncertainty, with error bars displaying the standard deviation in the mean.}
  \label{fig:stacked_residuals}
\end{figure*}

In Fig.~\ref{fig:stacked_residuals} we split the clusters into low- ($K<80$) and high- ($K>80$) entropy subsamples using central entropy, $K(<20\rm kpc)$ \citep{Sanderson2009_bimodality}, and show combined residuals from each scaling relation for clusters within each subsample. Ensemble average values are inversely weighted by the square of the uncertainty in that measurement. The $x$-axis order starts with X-ray measurements before progressing to SZ and optical. Lines are coloured by central entropy as in Fig.~\ref{fig:fits}. For completeness and additional clarity, we show the data for individual clusters in Appendix \ref{sec:individual_residuals}.

Except for the \Lxrass \ selection variable, residuals of the remaining gas observables average to near zero for both subsamples, indicating that both high- and low-entropy clusters follow similar mean scaling relations between these gas observables and mass. 

Surprisingly, the residuals in the total cluster optical content, \Lk \ and \richness, show a clear difference between the two subsamples. Interpreting them as a measure of stellar content, this suggests that at fixed mass, lower entropy clusters contain a smaller stellar mass and a smaller number of galaxies than higher entropy clusters.

This finding may be a signal of bias connected to halo formation epoch, if high central entropy is an indicator of a later formation epoch. The majority of star formation in the Universe took place at high redshift (z$\sim$1$-$3), and it is well known that galaxies in the field are more star forming than their cluster satellite counterparts \citep{Wetzel2012,Haines2015}. Galaxies in later forming clusters may be able to form more stellar mass because the progenitor halos spend more time in the field during this epoch of cosmic star formation before being quenched in the cluster environment. Conversely, early forming clusters would quench their galaxies earlier, and the massive galaxies would undergo more merging than their field counterparts. The net result would be both a lower stellar mass and a lower richness in older, lower entropy clusters.

It is important to note, however, that the \Lxrass \ selection criteria may contribute to the trend we see. This selection favours detection of brighter, cool-core clusters, with lower central entropy.  The low mass end of our sample is certainly incomplete, and potentially the absent systems are preferentially non cool-core clusters. Rather than the non cool-core clusters containing a systematically higher stellar fraction than the cool-core clusters, it is consistent with Fig.~\ref{fig:fits} that these non cool-core clusters are simply missing from the lower mass end of our sample. Inclusion of these missing clusters could possibly drive up the intrinsic scatter constraints in \Lk \ and \richness. Studies based on optically-selected samples will shed light on this issue \citep{Rykoff2014, Rykoff2016}.

\section{Summary}\label{sec:summary}

The task of constraining scaling relations is complicated by the effects of the selection function and covariance. In this paper we have presented a new multivariate approach to correct for these effects, and applied it to a multiwavelength observational dataset for which the selection function is well defined. For the first time, we have provided well-constrained scaling relation parameters with mass for a range of galaxy cluster observables, and our main results are as follows:

\begin{enumerate}
\item We find that the ICM scaling relations are shallower than the self-similar expectations at the 1$-$2$\sigma$ level.
\item The results of the integrated optical observables, \Lk \ and \richness, are in good agreement, with slopes of $\sim$0.75 suggesting that star forming efficiency is a decreasing function of cluster halo mass.
\item We find no distinction between the core-excised X-ray or high-resolution SZ relations of clusters of different central entropy.
\item Clusters with low central entropy have negative residuals from the integrated optical scaling relations, suggesting that early forming clusters have a lower stellar fraction than their younger counterparts.
\end{enumerate}

Following conclusion (iii), selecting based on core-excised X-ray or high-resolution SZ may lead to a more dynamically diverse sample of clusters since neither property's scaling relation is impacted by the presence of a cool core. Further investigation with samples including lower mass clusters is needed to fully understand any dependence of the cluster stellar fraction on its dynamical state. While our results in this work are limited by the low number of observed clusters, our method will be applicable to future surveys and will lead to excellent constraints on the physics of clusters and the cosmological parameters.

\section*{Acknowledgements}

We thank Arif Babul and members of the LoCuSS collaboration for their support and assistance. SLM and GPS acknowledge support from the STFC. AF and AEE acknowledge support from NASA \textit{Chandra} Grant G06-17116B. AF acknowledges support from a McWilliams Postdoctoral Fellowship. GPS acknowledges support from the Royal Society. CO, DPM, ZA, JEC, and CARMA operations were supported by NSF grant AST-1140019. CPH acknowledges support from PRIN INAF 2014. This paper is dedicated to the memory of Dr. Yu-Ying Zhang.

\appendix

\section{Selection Bias}\label{sec:bias}

Table \ref{tab:bias} shows the scaling relation parameters inferred from the \citet{Kelly2007} method, without correcting for selection effects. Comparison with the constraints from our hierarchical Bayesian method, shown in Table \ref{tab:fits}, quantifies the bias from the selection function and the importance of accounting for it. The bias in the \Lxrass \ parameters are largest, as expected for the selection variable. The magnitude of the bias in other observables is consistent with the magnitude of that observable's covariance with \Lxrass, shown in Table~\ref{tab:rlxrass}.

\begin{table}
\caption{Scaling relation parameters inferred from the \citet{Kelly2007} method without correcting for selection effects.}\label{tab:bias}
	\begin{center}
		\tabcolsep=0.8mm
			\begin{tabular}{ l c c c }
            \hline
            Observable & Intercept & Slope & Scatter \\
            $S_a$ & $\exp(\pi_a)$ & $\alpha_a$ & $\sigma_{a | \mu}$ \\
            \hline
\vspace{3.0truept}
\Lxrass  & $ 7.61 ^{+ 0.52 }_{- 0.56 } $  &  $ 0.47 ^{+ 0.23 }_{- 0.23 } $  &  $ 0.37 ^{+ 0.05 }_{- 0.06 } $  \\ 
\vspace{3.0truept}
\Lx  & $ 8.08 ^{+ 0.68 }_{- 0.75 } $  &  $ 1.02 ^{+ 0.29 }_{- 0.30 } $  &  $ 0.48 ^{+ 0.06 }_{- 0.08 } $  \\ 
\vspace{3.0truept}
\Tx  & $ 7.03 ^{+ 0.33 }_{- 0.33 } $  &  $ 0.55 ^{+ 0.14 }_{- 0.15 } $  &  $ 0.22 ^{+ 0.04 }_{- 0.04 } $  \\ 
\vspace{3.0truept}
\Mgas  & $ 0.90 ^{+ 0.04 }_{- 0.04 } $  &  $ 0.99 ^{+ 0.13 }_{- 0.14 } $  &  $ 0.17 ^{+ 0.04 }_{- 0.05 } $  \\ 
\vspace{3.0truept}
\Yx  & $ 6.43 ^{+ 0.60 }_{- 0.65 } $  &  $ 1.31 ^{+ 0.30 }_{- 0.30 } $  &  $ 0.47 ^{+ 0.08 }_{- 0.09 } $  \\ 
\vspace{3.0truept}
\Ysza  & $ 8.01 ^{+ 0.81 }_{- 0.83 } $  &  $ 1.91 ^{+ 0.33 }_{- 0.36 } $  &  $ 0.30 ^{+ 0.09 }_{- 0.09 } $  \\ 
\vspace{3.0truept}
\Ypl  & $ 10.00 ^{+ 0.76 }_{- 0.77 } $  &  $ 1.37 ^{+ 0.23 }_{- 0.25 } $  &  $ 0.35 ^{+ 0.06 }_{- 0.07 } $  \\ 
\vspace{3.0truept}
\Lbcg  & $ 1.00 ^{+ 0.05 }_{- 0.05 } $  &  $ 0.18 ^{+ 0.18 }_{- 0.18 } $  &  $ 0.32 ^{+ 0.04 }_{- 0.04 } $  \\ 
\vspace{3.0truept}
\Lk  & $ 14.99 ^{+ 0.70 }_{- 0.70 } $  &  $ 0.97 ^{+ 0.13 }_{- 0.14 } $  &  $ 0.12 ^{+ 0.05 }_{- 0.06 } $  \\ 
\vspace{3.0truept}
\richness  & $ 100.82 ^{+ 6.16 }_{- 6.44 } $  &  $ 1.17 ^{+ 0.18 }_{- 0.19 } $  &  $ 0.20 ^{+ 0.07 }_{- 0.07 } $  	\\
            \hline
	    \end{tabular}
	\end{center}
{\footnotesize}
\end{table}

\begin{table}
\caption{The covariance between \Lxrass \ and the observables, constrained by our hierarchical Bayesian method.}\label{tab:rlxrass}
	\begin{center}
		\tabcolsep=0.8mm
			\begin{tabular}{ l c }
            \hline
            Observable & Correlation coefficient \\
            $S_a$ & $r_{a,L_{X,\rm RASS}}$ \\
            \hline
\vspace{3.0truept}
\Lx         & $ 0.43^{+0.15}_{-0.19} $ \\
\vspace{3.0truept}
\Tx         & $ 0.33^{+0.21}_{-0.25} $ \\
\vspace{3.0truept}
\Mgas       & $ 0.24^{+0.21}_{-0.24} $ \\
\vspace{3.0truept}
\Yx         & $ 0.44^{+0.16}_{-0.21} $ \\
\vspace{3.0truept}
\Ysza       & $ 0.57^{+0.17}_{-0.24} $ \\
\vspace{3.0truept}
\Ypl        & $ 0.18^{+0.20}_{-0.23} $ \\
\vspace{3.0truept}
\Lbcg       & $ 0.12^{+0.21}_{-0.23} $ \\
\vspace{3.0truept}
\Lk         & $ -0.07^{+0.58}_{-0.47} $ \\
\vspace{3.0truept}
\richness   & $ -0.30^{+0.24}_{-0.21} $ \\
            \hline
	    \end{tabular}
	\end{center}
{\footnotesize}
\end{table}

\section{Performance of Hierarchical Bayesian Method}\label{sec:maxliktests}

\begin{figure*}
  \centering
  \includegraphics[width=0.3\linewidth]{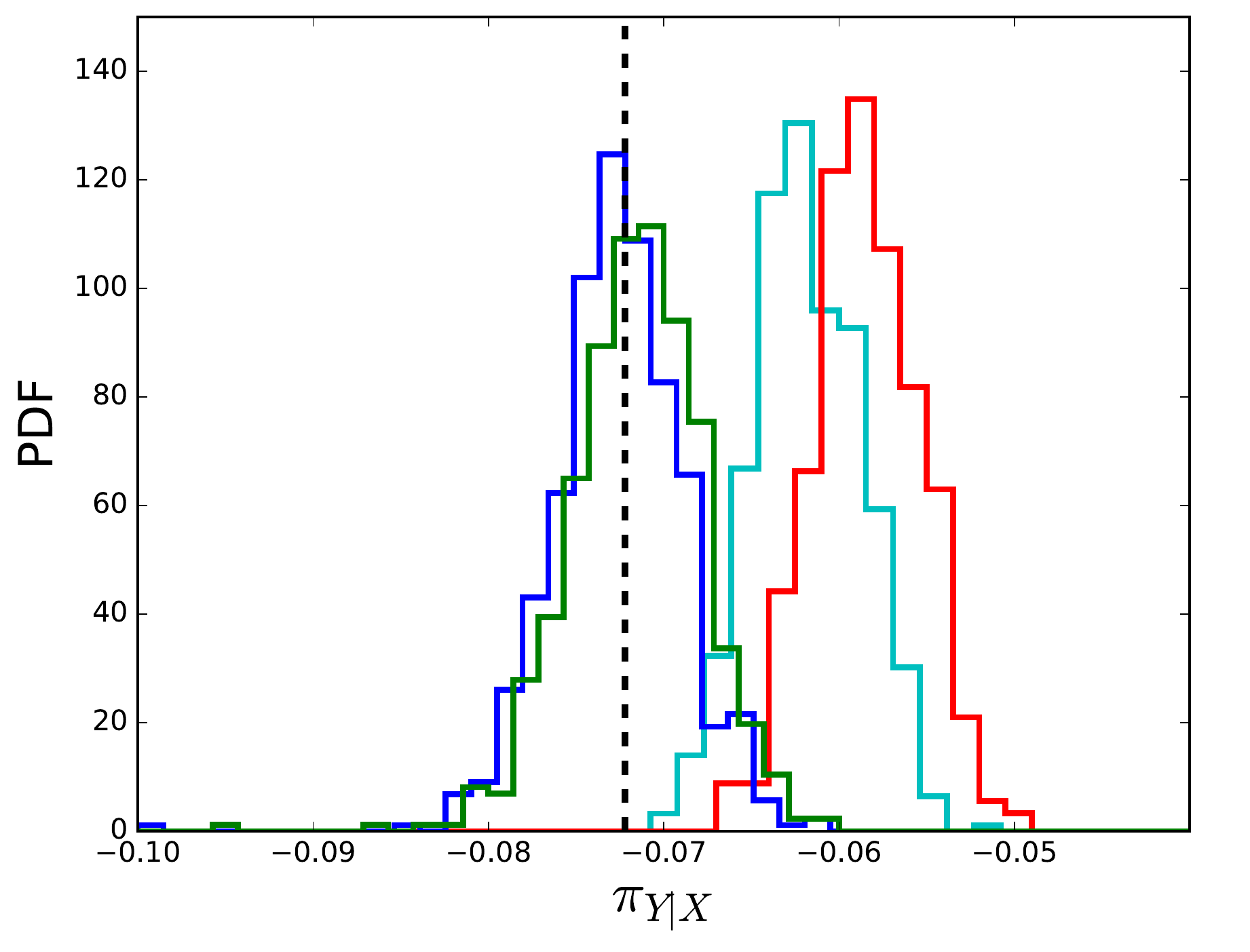}
  \hspace{0.01\linewidth}
  \includegraphics[width=0.3\linewidth]{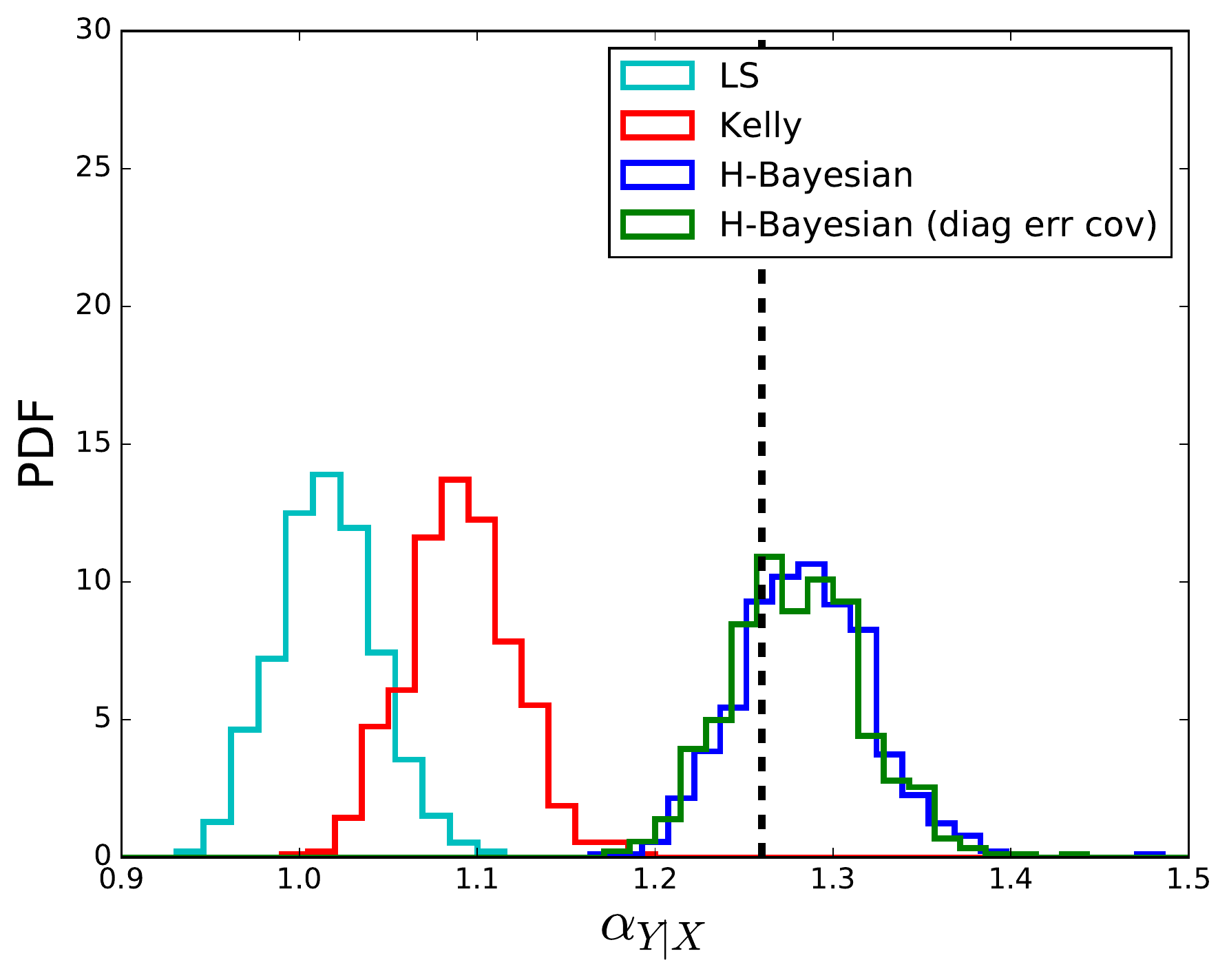}
  \hspace{0.01\linewidth}
  \includegraphics[width=0.3\linewidth]{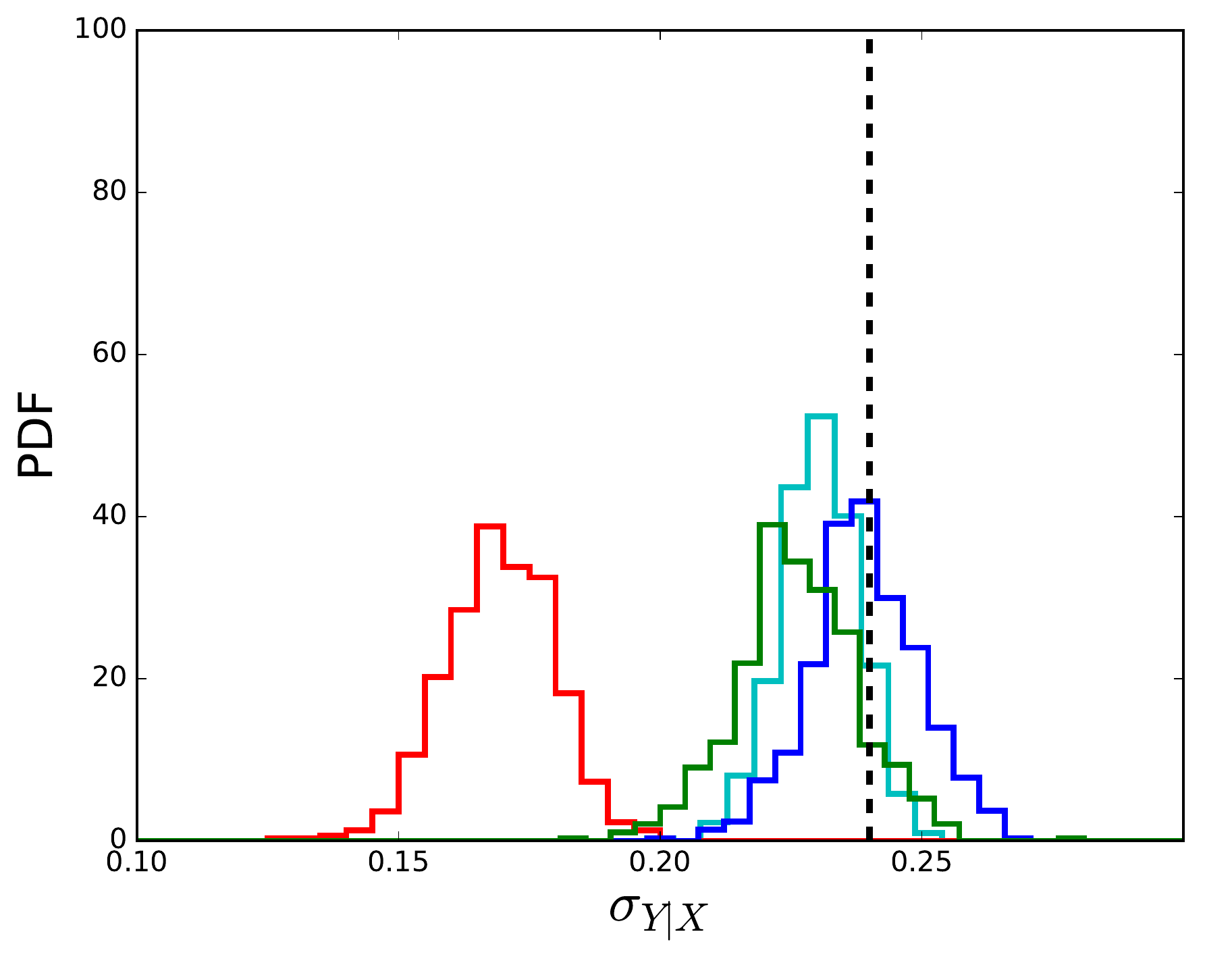}
  \includegraphics[width=0.3\linewidth]{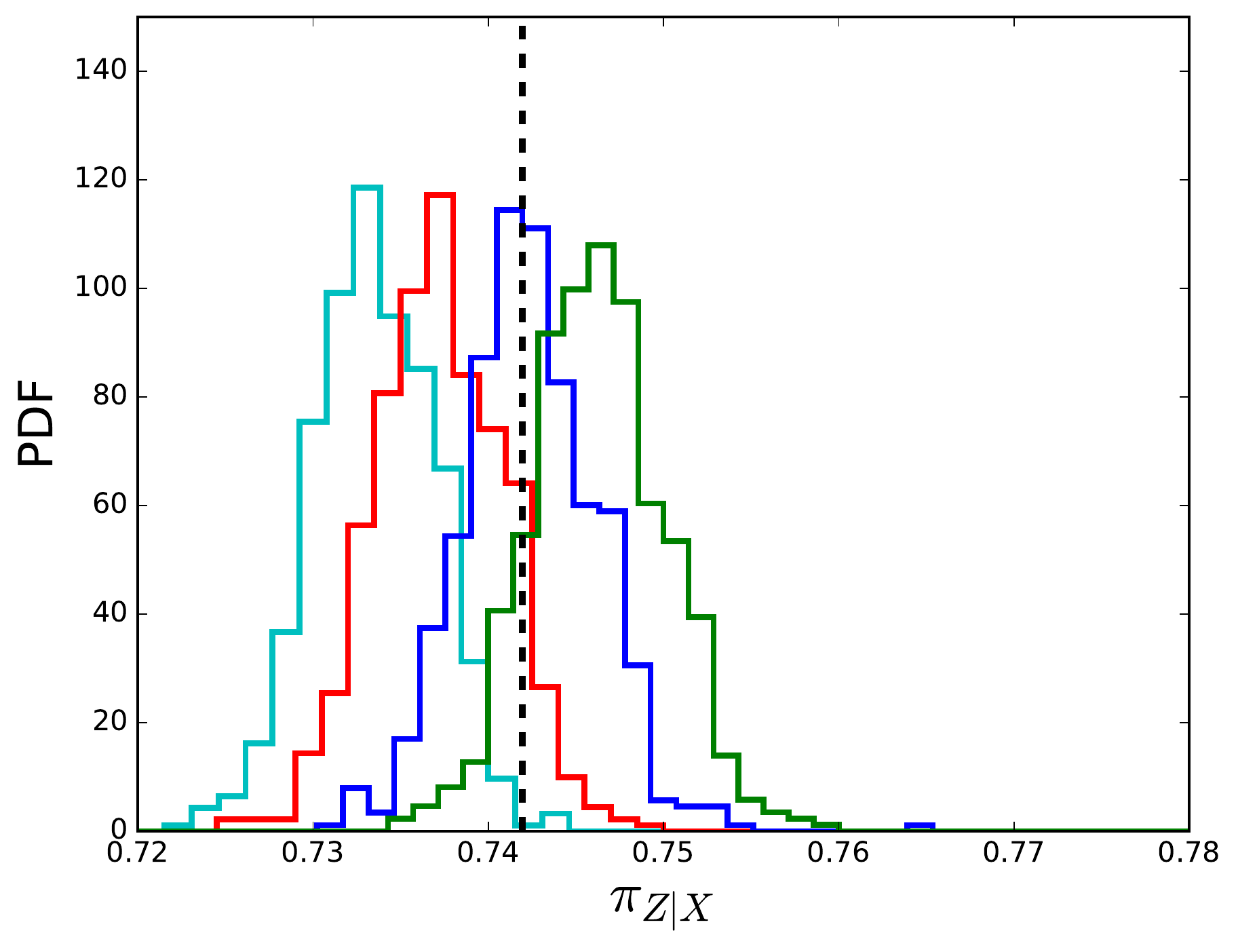}
  \hspace{0.01\linewidth}
  \includegraphics[width=0.3\linewidth]{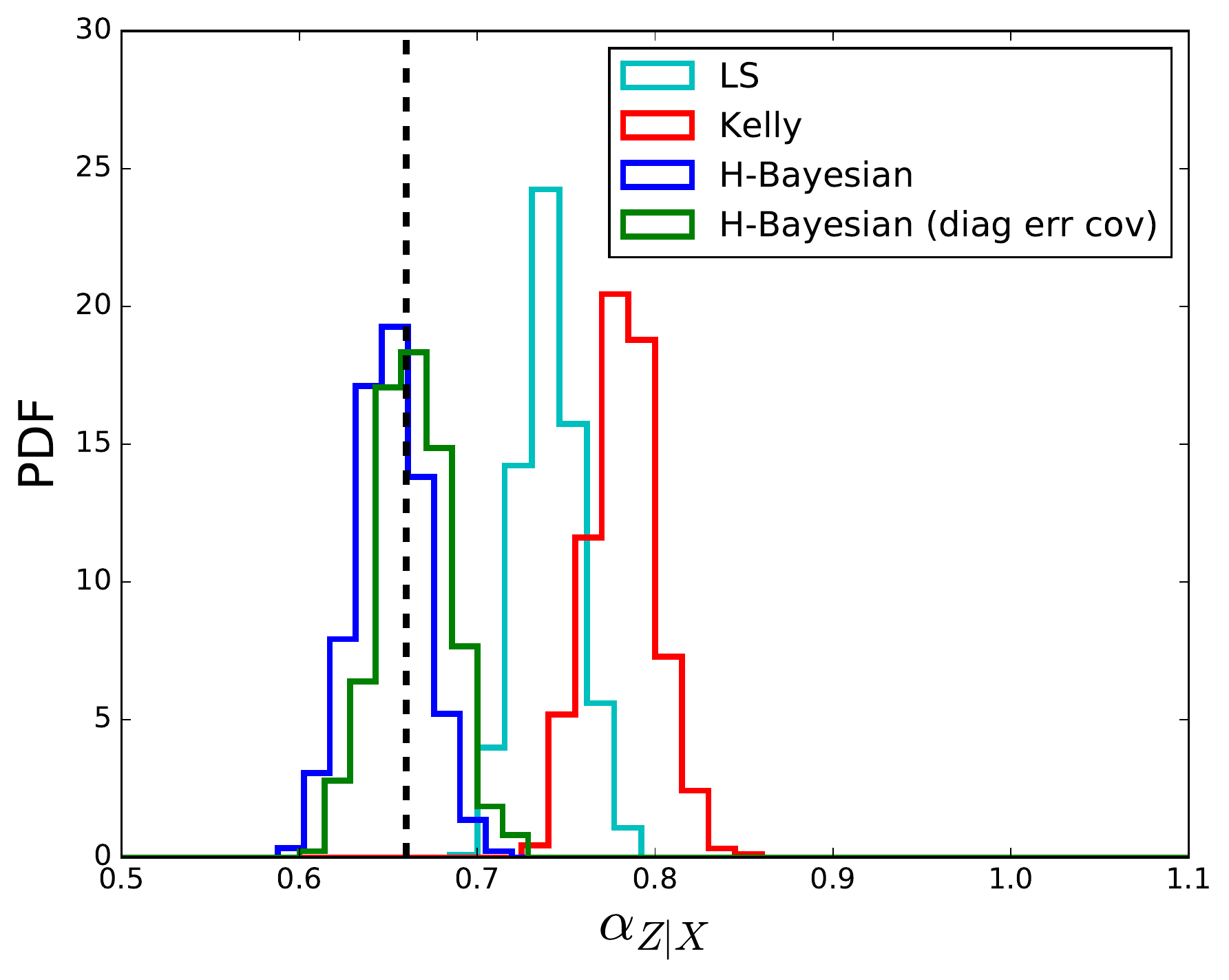}
  \hspace{0.01\linewidth}
  \includegraphics[width=0.3\linewidth]{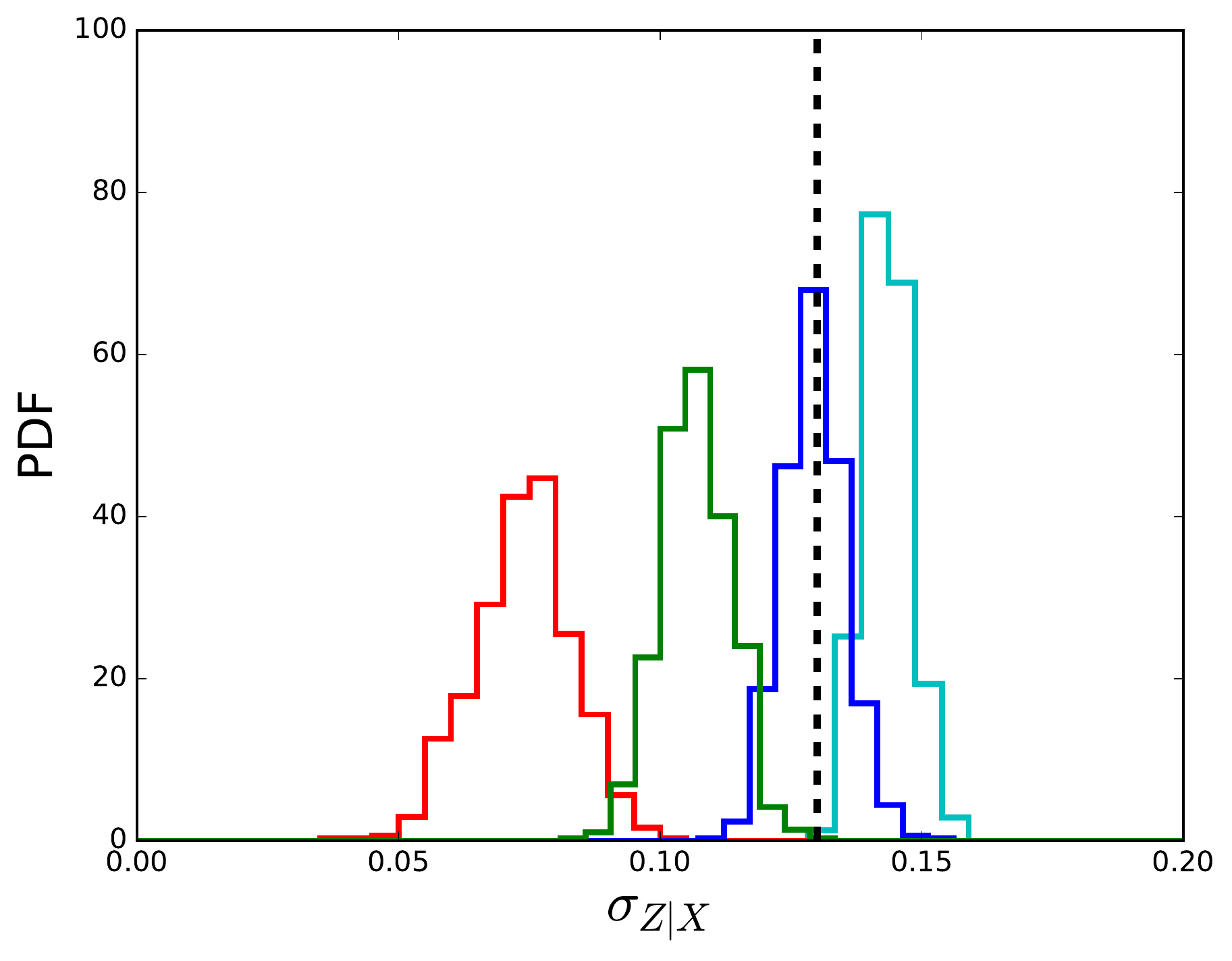}
\caption{Distribution of the best fit parameters for 1000 mock datasets, constrained by four different methods:
LS - Ordinary Least Squares (cyan); \citeauthor{Kelly2007} - the method of \citet{Kelly2007}, without correcting for selection effects (red); H-Bayesian - the hierarchical Bayesian model presented in Section \ref{sec:maxlik} (blue); H-Bayesian (diag err cov) - the same model, without modelling the non-diagonal elements of the error covariance (green). The dashed lines show the input values.}
  \label{fig:test}
\end{figure*}

We test the performance of the hierarchical Bayesian method on 1000 mock datasets, generated using the following steps:

\begin{enumerate}
\item Generate X values assuming a mass function using the \texttt{hmf} code \citep{Murray2013}.
\item Generate Y values assuming a Y-X scaling relation.
\item Generate Z values assuming a Z-X scaling relation and a correlation coefficient of -0.7 between Y and Z at fixed X.
\item Apply correlated measurement errors with variance 0.01 to X, Y and Z values with a correlation coefficient of 0.7 at fixed X.
\item Select those above a Y limit.
\end{enumerate}
After applying the Y selection, each dataset contains $\sim$50 objects, similar to our LoCuSS sample. We calculate the best fit parameters for each dataset, and show the distribution of these parameters in Fig.~\ref{fig:test}, finding all parameters to be well constrained.

We compare the best fit parameters calculated using different methods:

\begin{enumerate}
\item LS: Ordinary Least Squares.
\item Kelly: the method of \citet{Kelly2007}, without correcting for selection effects.
\item H-Bayesian: the hierarchical Bayesian model presented in Section \ref{sec:maxlik}.
\item H-Bayesian (diag err cov): the same model, without modelling the non-diagonal component of error covariance.
\end{enumerate}

As expected, the methods that do not consider the selection function (LS and \citeauthor{Kelly2007}) constrain a shallower slope (and higher intercept) for selection variable Y and a steeper slope (and lower intercept) for Z due to its negative covariance with Y. This leads the \citeauthor{Kelly2007} method to underestimate the intrinsic scatter in both relations, while a simple LS method is more accurate. We note that while both H-Bayesian methods are accurate in the Y relation where modelling full error covariance is unimportant, the H-Bayesian method that does not model full error covariance is less accurate in the Z relation. This figure illustrates the importance of modelling both the selection function and the error covariance on the inferred parameter, particularly for the scatter parameter of the non-selection variables.

\section{Other Structural Indicators}\label{sec:indicators}

\begin{figure*}
  \centering
  \includegraphics[width=0.4\linewidth]{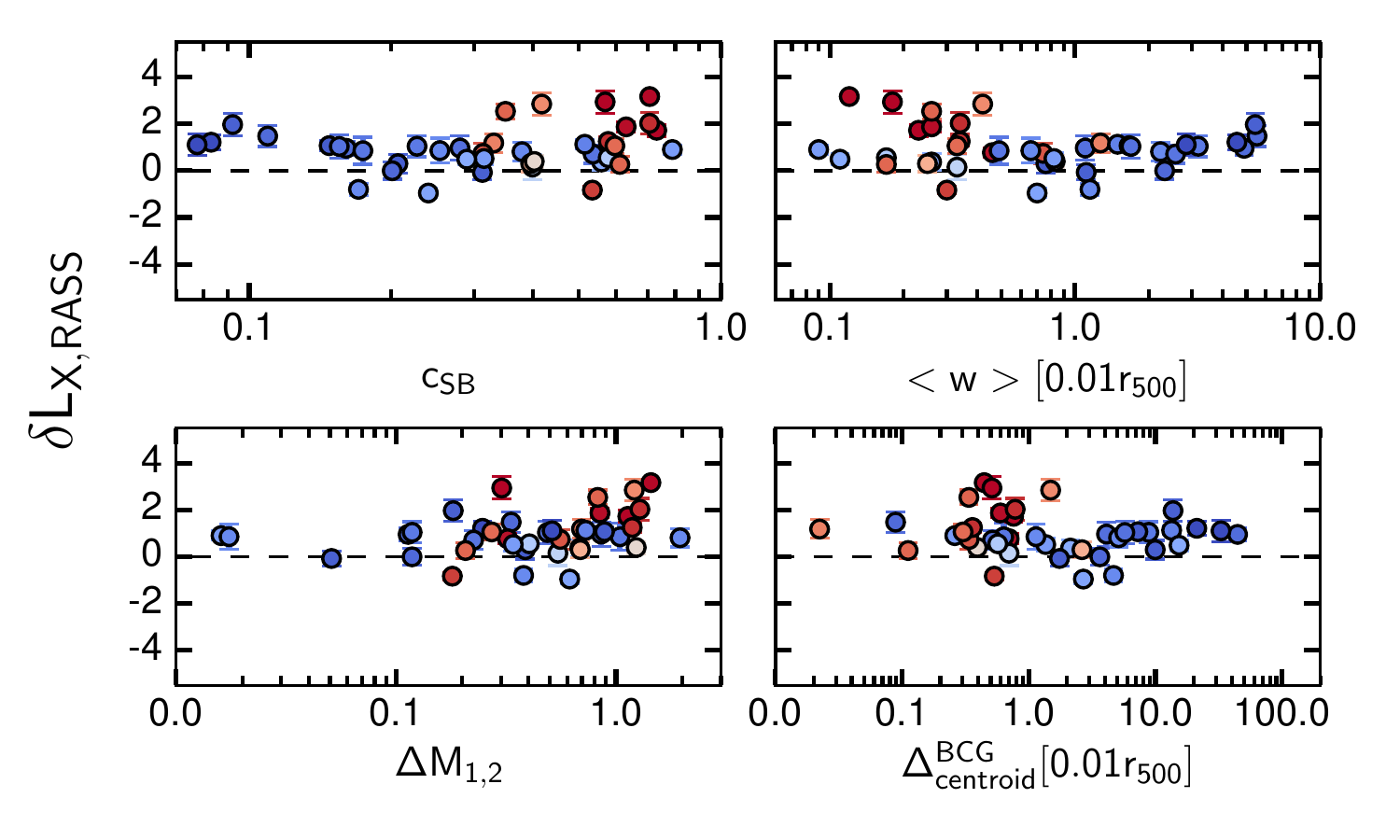}
  \hspace{0.05\linewidth}
  \includegraphics[width=0.4\linewidth]{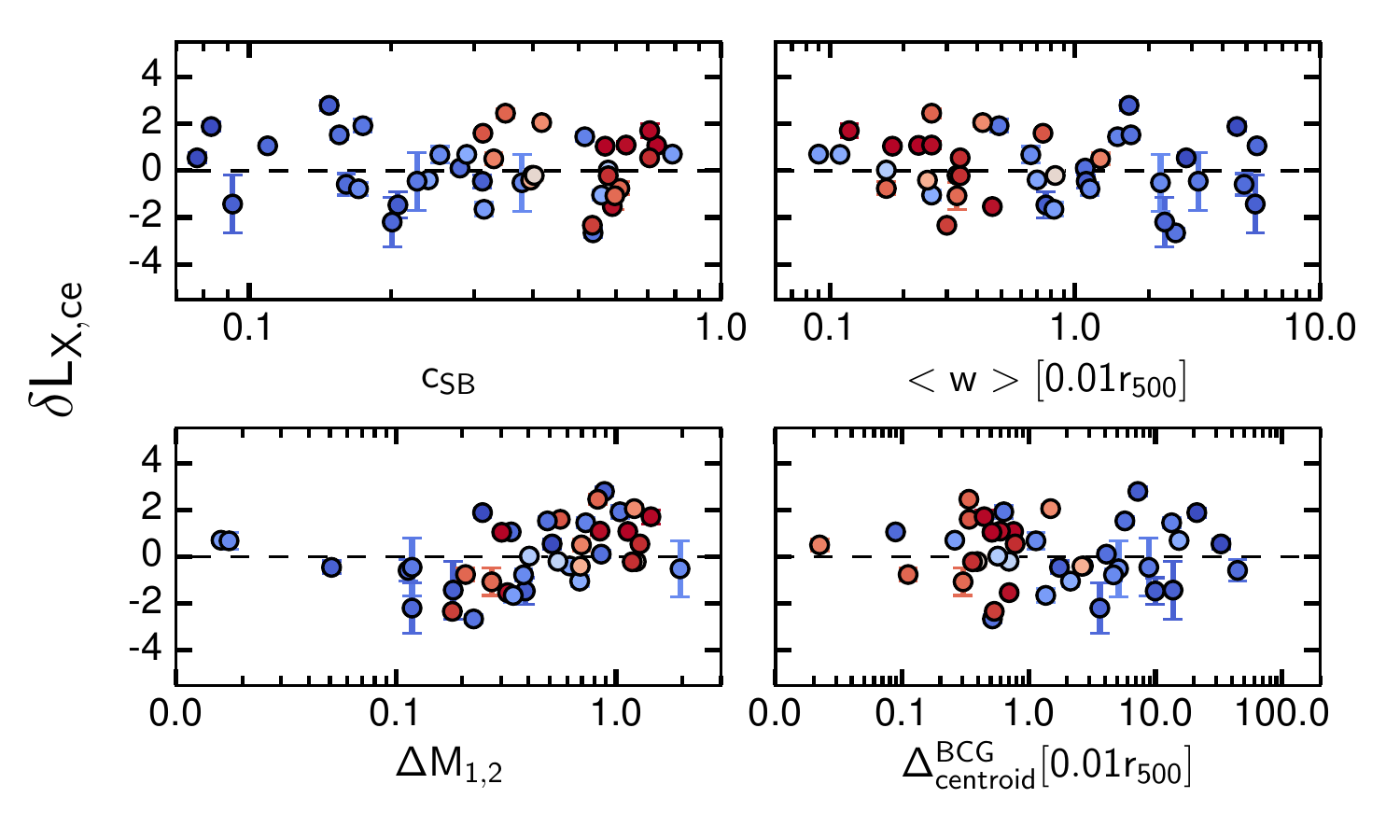}
  \includegraphics[width=0.4\linewidth]{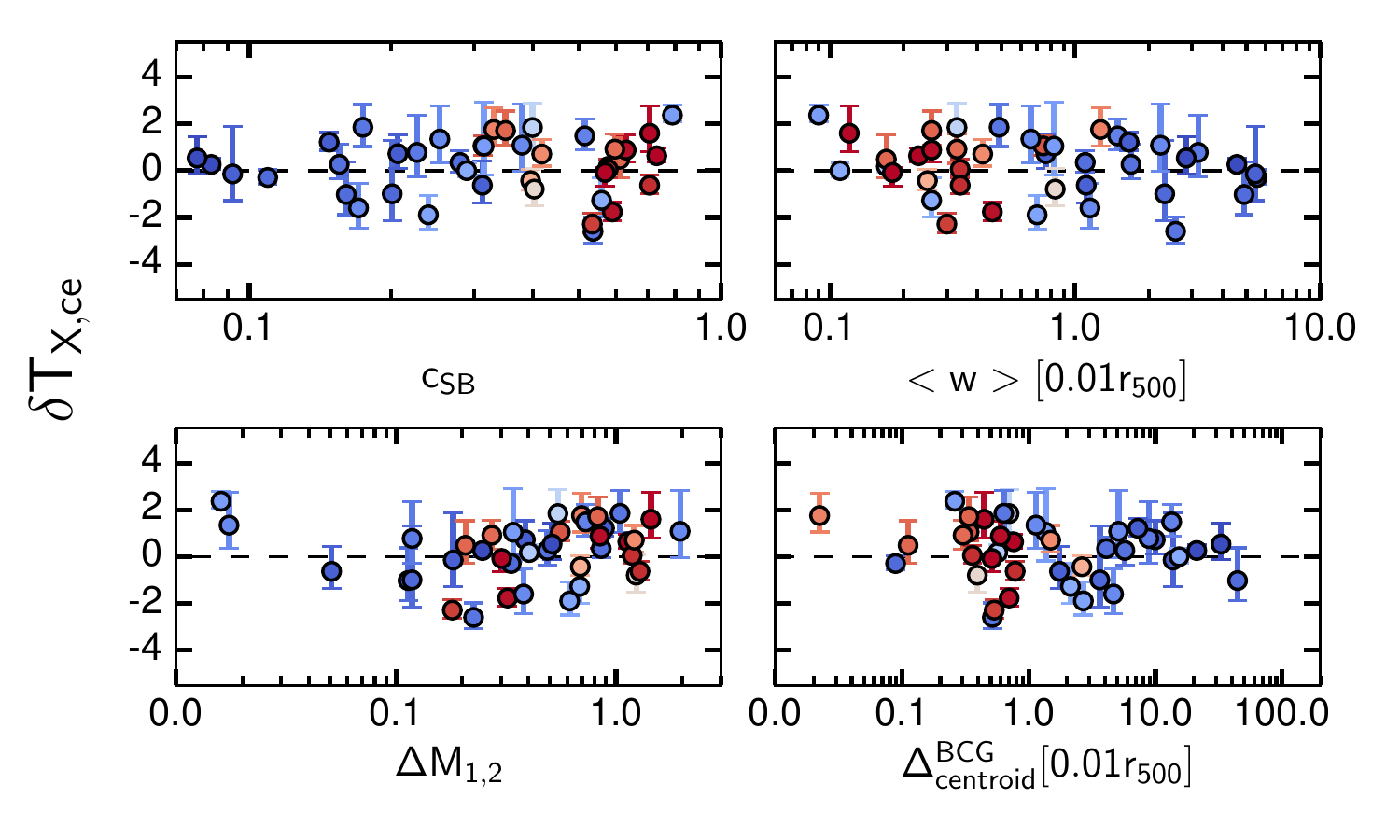}
  \hspace{0.05\linewidth}
  \includegraphics[width=0.4\linewidth]{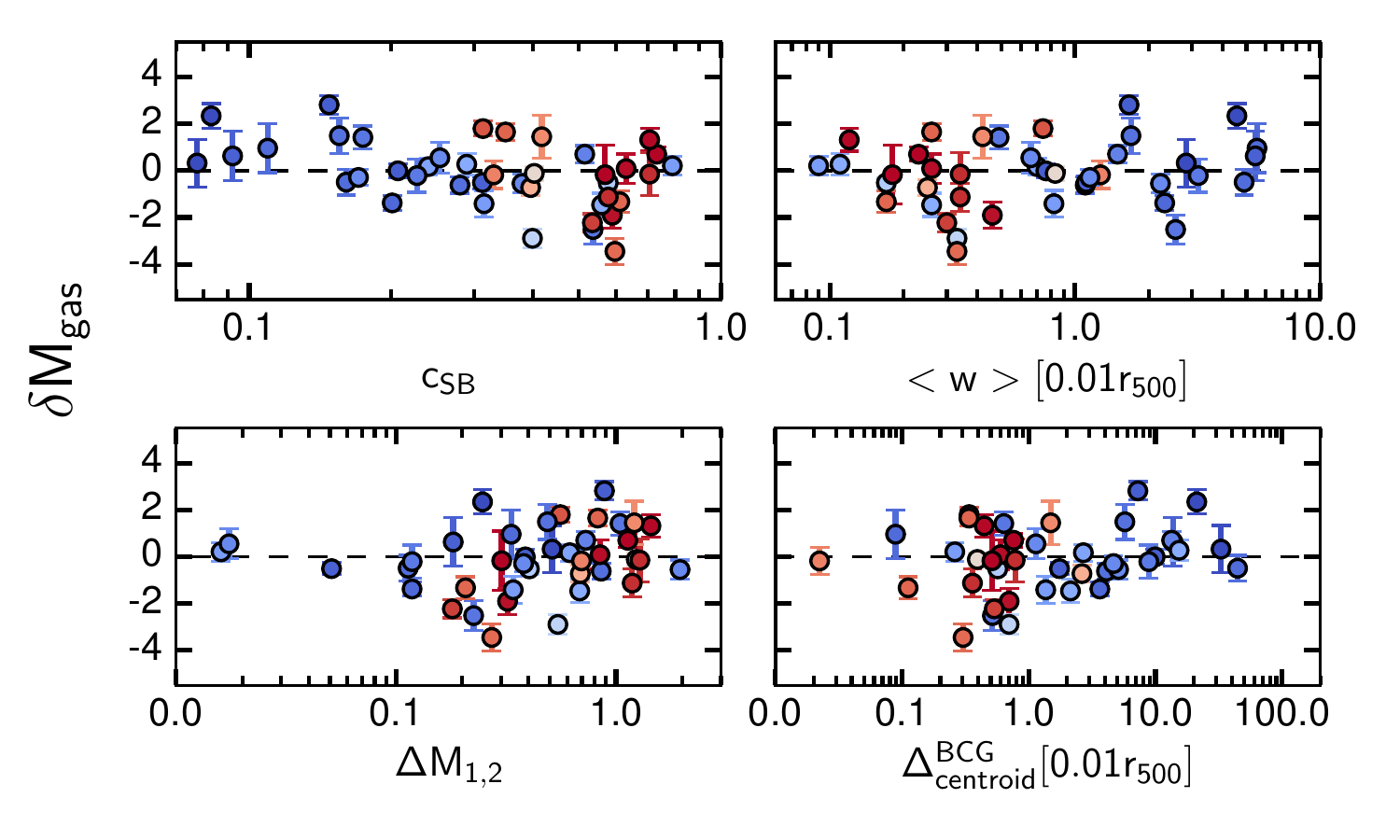}
  \includegraphics[width=0.4\linewidth]{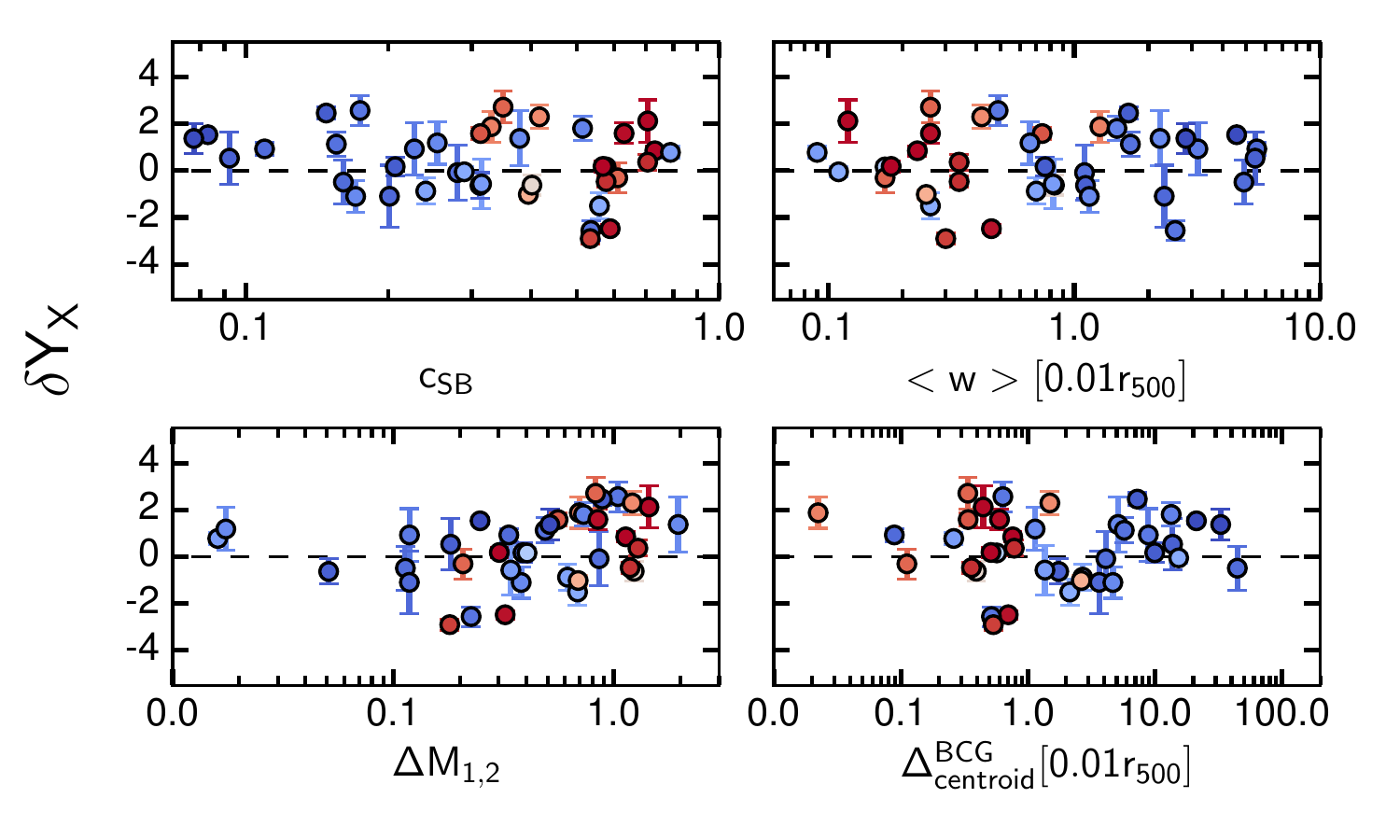}
  \hspace{0.05\linewidth}
  \includegraphics[width=0.4\linewidth]{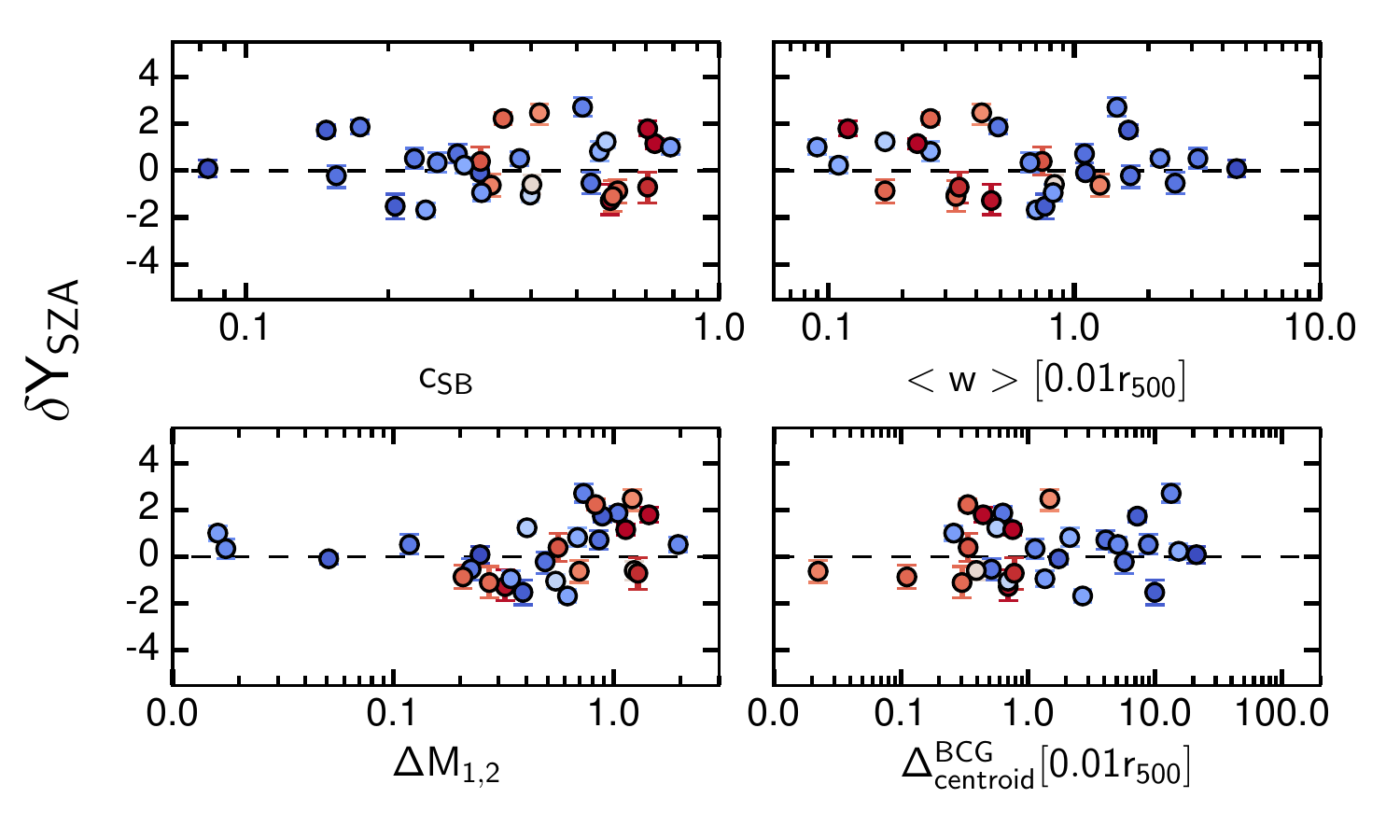}
  \includegraphics[width=0.4\linewidth]{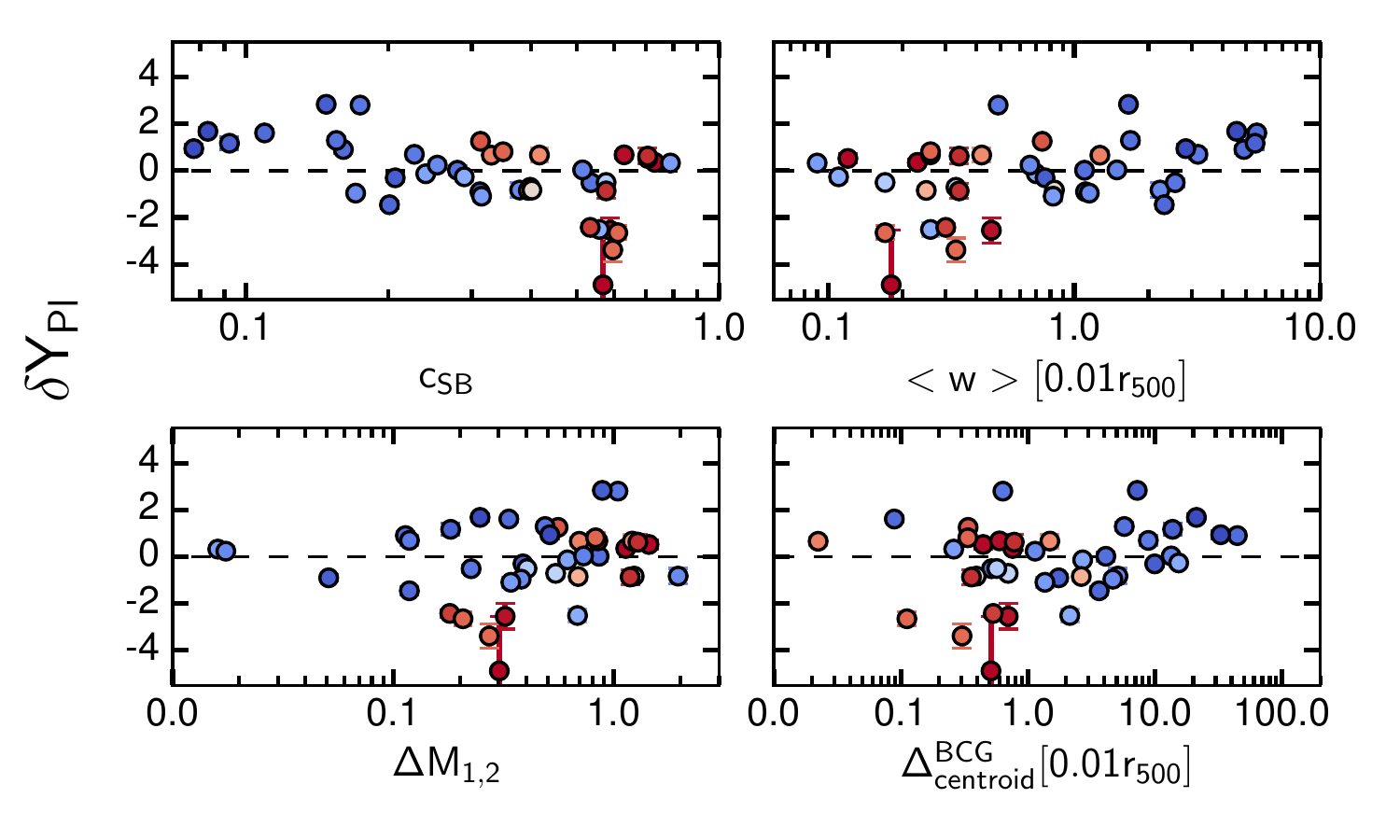}
  \hspace{0.05\linewidth}
  \includegraphics[width=0.4\linewidth]{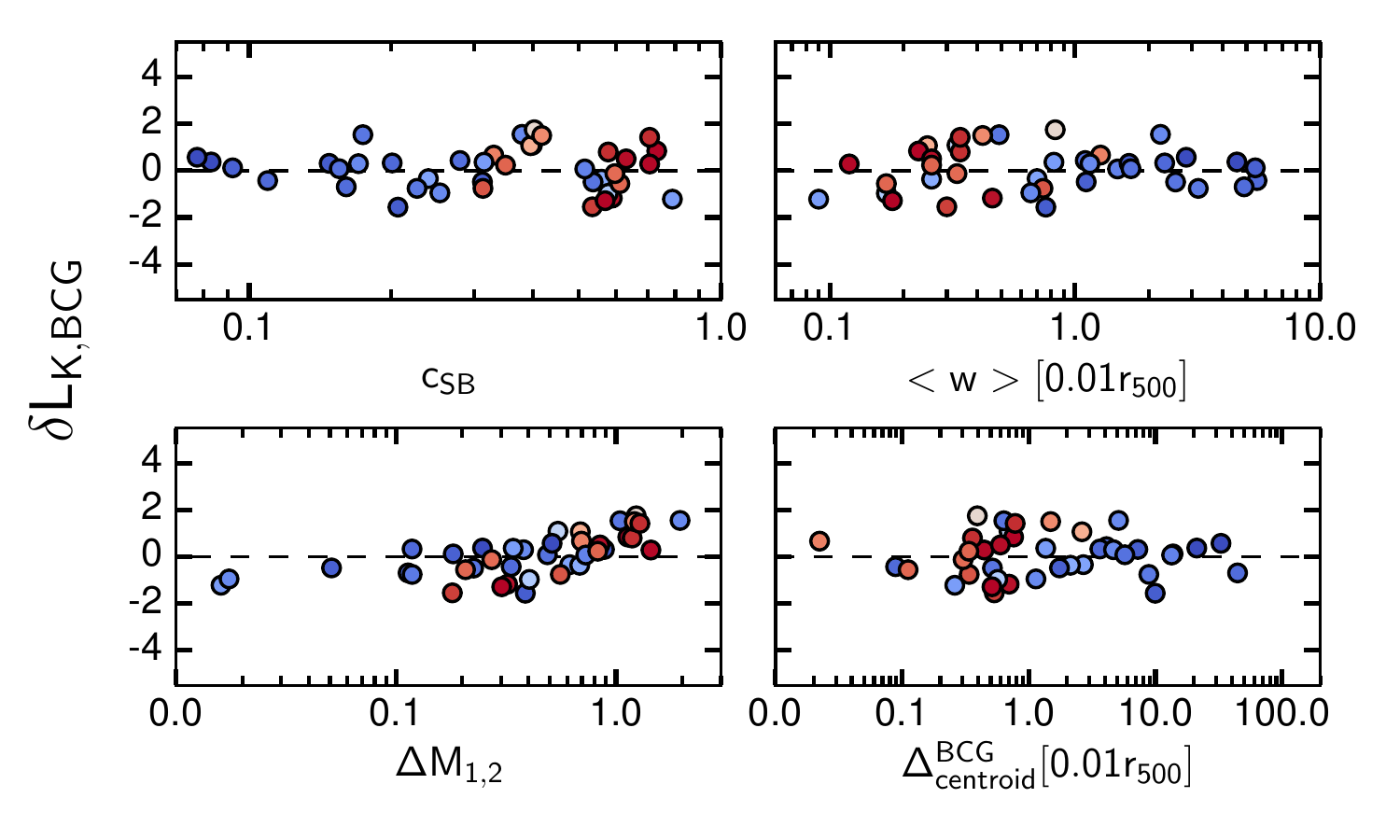}
  \includegraphics[width=0.4\linewidth]{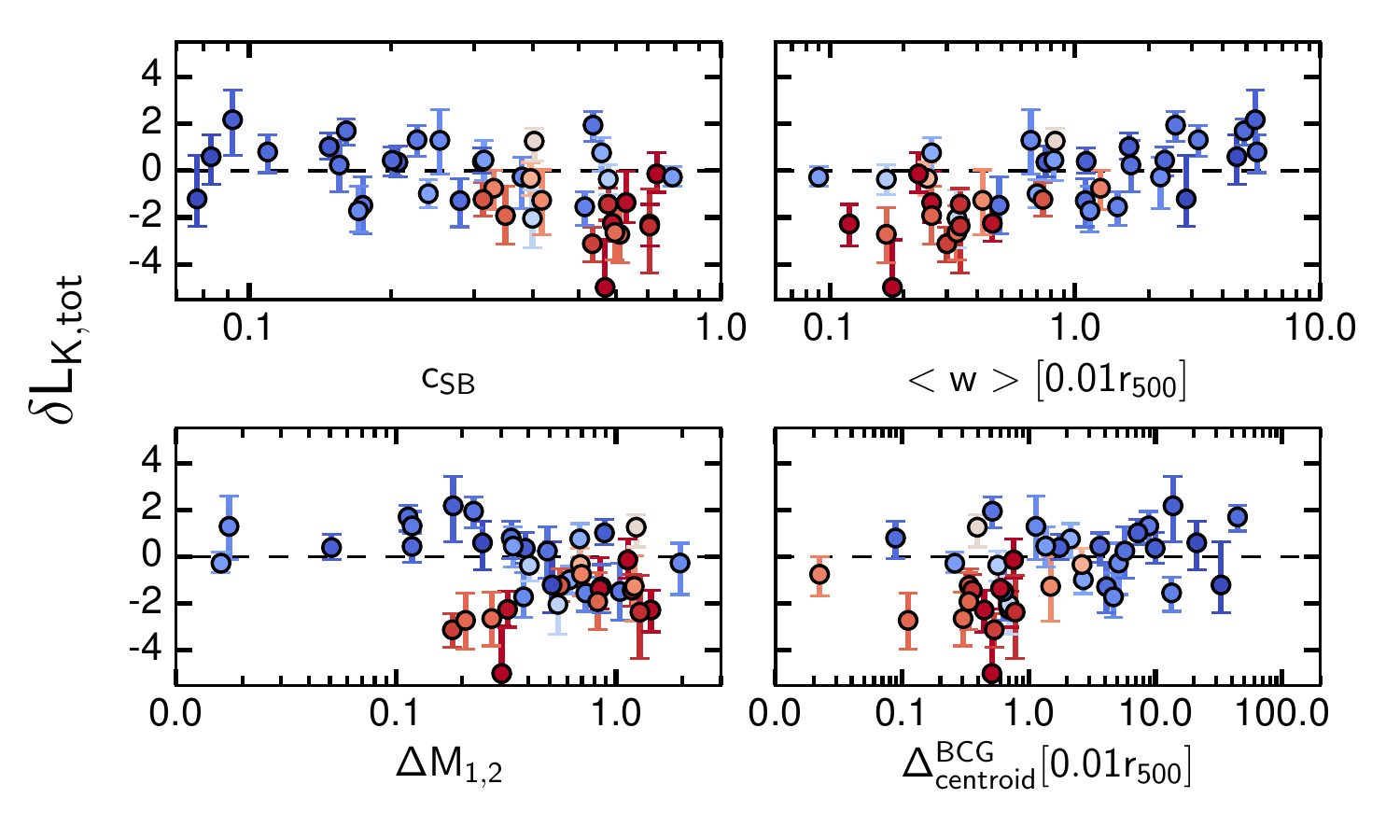}
  \hspace{0.05\linewidth}
  \includegraphics[width=0.4\linewidth]{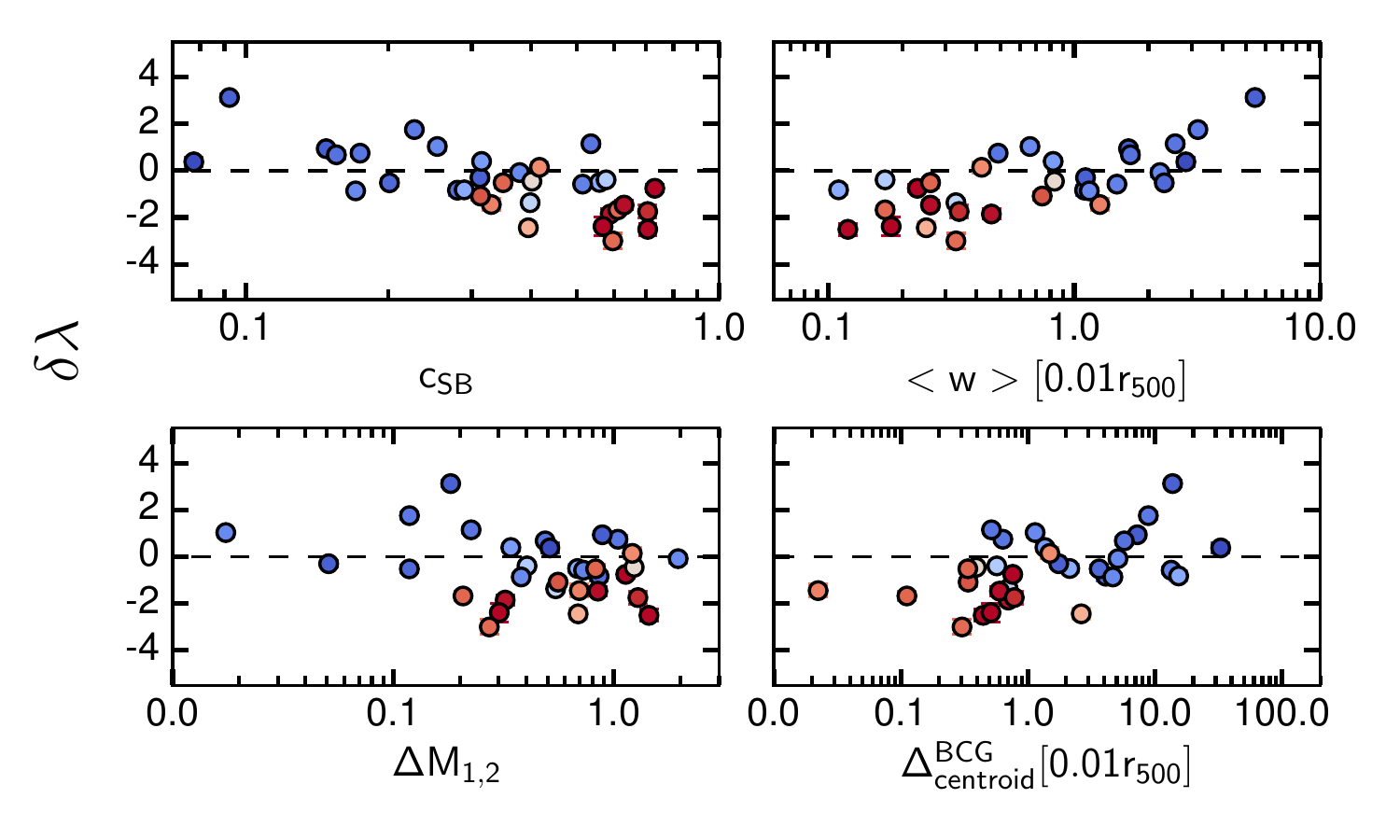}
  \caption{Normalized residuals from scaling relations, defined in equation~(\ref{eq:residual}), as a function of (clockwise) surface brightness concentration, centroid shift, BCG/centroid separation and magnitude gap. Colours indicate central entropy $K(<20\rm kpc)$, as in Fig.~\ref{fig:fits}.}
  \label{fig:residuals_appendix}
\end{figure*}

In this section, we compare residuals from the scaling relations against several structural indicators of the cluster, and display the results in Fig.~\ref{fig:residuals_appendix}.

The surface brightness concentration, $c_{\rm SB}$, is defined as the ratio of the central surface brightness within $40 \rm kpc$ and the ambient surface brightness within $400 \rm kpc$. A large $c_{\rm SB}$ suggests the presence of a cool core, and therefore a less dynamically disturbed cluster. The centroid shift, $\langle w \rangle$, taken from \citet{Martino2014}, is the standard deviation of the projected separation between the X-ray peak and the X-ray centroid calculated in circular apertures in the range $[0.05 - 1]r_{500}$. We also consider the projected separation between the X-ray centroid and the BCG, $\Delta^{\rm BCG}_{\rm centroid}$. Both projected separation parameters ($\langle w \rangle$ and $\Delta^{\rm BCG}_{\rm centroid}$) are sensitive to the dynamical state of the cluster, with a large value suggesting a more disturbed cluster. Finally, we include the magnitude gap, $\Delta M_{1,2}$, between the two brightest galaxies within $0.5r_{\rm vir}$. A larger magnitude gap suggests that bright galaxies have had time since the last major merger to accrete onto the BCG, therefore suggesting a less disturbed cluster.

The trends seen in Section \ref{sec:scatter} in residuals from the integrated optical observables (\Lk \ and \richness) as a function of central entropy $K(<20\rm kpc)$ are reproduced strongly in the structural indicators sensitive to gas morphology. They are less clear in the indicators sensitive to the galaxies. The $K(<20\rm kpc)$ trend in \Ypl \ is reproduced by indicators sensitive to the gas morphology, consistent with the explanation that measurements of more disturbed non cool-core clusters are overestimated by the assumption of an \citet{Arnaud2010} profile.

We find positive correlation between $\Delta M_{1,2}$ and residuals from \Lbcg, as expected, with a larger $\Delta M_{1,2}$ suggesting a brighter BCG.

\section{Individual Cluster Residuals}\label{sec:individual_residuals}

\begin{figure*}
  \centering
  \includegraphics[width=\linewidth]{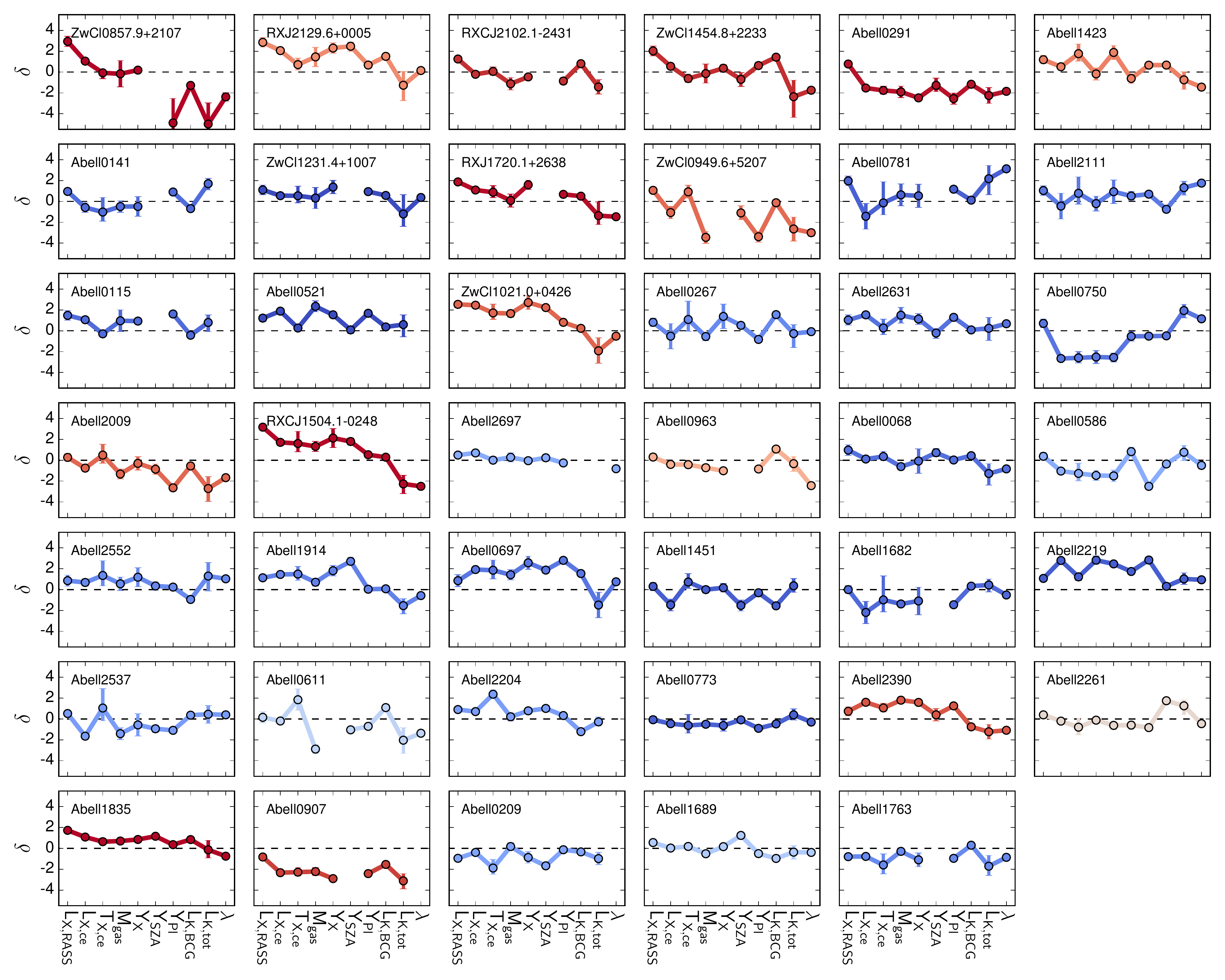}
  \caption{Normalized residuals from the scaling relations, defined in equation~(\ref{eq:residual}), for all clusters. The panels are ordered by increasing \Mwl, and colours indicate the cluster central entropy $K(<20\rm kpc)$, as in Fig.~\ref{fig:fits}.}
  \label{fig:all_residuals}
\end{figure*}

In Fig.~\ref{fig:all_residuals} we present the unstacked cluster residuals discussed in Section \ref{sec:residuals}. The panels are ordered by increasing \Mwl, and colours indicate $K(<20\rm kpc)$, as in Fig.~\ref{fig:fits}.

\bibliographystyle{mnras}
\bibliography{mulroybib}

\label{lastpage}
\end{document}